\documentclass[letter]{aa} % for the letters
\usepackage{graphicx}
\usepackage{txfonts}
\usepackage{natbib}
\bibpunct{(}{)}{;}{a}{}{,}
\begin{document}
   \title{SPITZER survey of dust grain processing in stable discs around binary post-AGB stars.
\thanks{Based on observations obtained at the European Southern Observatory (ESO),
La Silla, observing program 072.D-0263, on observations made with the 1.2\,m Flemish Mercator telescope
at Roque de los Muchachos, Spain, the 1.2\,m Swiss Euler telescope at La Silla,
Chile and on observations made with the SPITZER Space Telescope (program id 3274), which is operated 
by the Jet Propulsion Laboratory, California Institute of Technology under a contract with NASA.}
}

   \author{
          C. Gielen\inst{1}
          \and
          H. Van Winckel\inst{1}
          \and
          M. Min \inst{2}
          \and
          L.B.F.M. Waters \inst{1,2}
          \and
          T. Lloyd Evans \inst{3}
          }

%  \offprints{C. Gielen}

   \institute{Instituut voor Sterrenkunde,
              Katholieke Universiteit Leuven, Celestijnenlaan 200D, 3001 Leuven, Belgium\\
	      \email{clio.gielen@ster.kuleuven.be}
              \and
              Sterrenkundig Instituut 'Anton Pannekoek', 
              Universiteit Amsterdam, Kruislaan 403, 1098 Amsterdam, The Netherlands
              \and
              SUPA, School of Physics and Astronomy, 
              University of St Andrews, North Haugh, St Andrews, Fife KY16 9SS, United Kingdom
            }

   \date{Received ; accepted }

% \abstract{}{}{}{}{}
% 5 {} token are mandatory
  \abstract
   {}
   {We investigate the mineralogy and dust processing in the circumbinary discs of binary post-AGB stars using high-resolution TIMMI2 and SPITZER infrared spectra.
     }
   {We perform a full spectral fitting to the infrared spectra using the most recent opacities of amorphous and crystalline dust species.
 This allows for the identification of the carriers of the different emission bands. Our fits also constrain the physical properties of different dust species and grain sizes responsible for the observed emission features.  }
   {In all stars the dust is oxygen-rich: amorphous and crystalline silicate dust species prevail and no
features of a carbon-rich component can be found, the exception being EP\,Lyr, where a mixed chemistry of both oxygen- and carbon-rich species is found.
Our full spectral fitting indicates a high degree of dust grain processing. 
The mineralogy of our sample stars shows that the dust is constituted of irregularly shaped and relatively large grains, with typical grain sizes larger than 2\,$\mu$m. 
The spectra of nearly all stars show a high degree of crystallinity, where magnesium-rich end members of olivine and pyroxene silicates dominate.
Other dust features of e.g. silica or alumina are not present at detectable levels.
Temperature estimates from our fitting routine show that a significant fraction of grains must be cool, significantly cooler than the glass temperature. This shows that radial mixing is very efficient is these discs and/or indicates different thermal conditions at grain formation. 
Our results show that strong grain processing is not limited to young stellar objects and that the physical processes occurring in the discs are very similar to those
in protoplanetary discs.}
   {}
   \maketitle
%
%________________________________________________________________

\section{Introduction}

Post-AGB stars are low- and intermediate-mass stars that evolve rapidly from the Asymptotic Giant Branch (AGB) towards
the Planetery Nebula (PNe) phase, before cooling down as white dwarves.
During the previous AGB phase, the star has undergone severe mass loss, leaving behind a slowly expanding circumstellar dust shell.
Depending on the dilution in the line-of-sight, the expected spectral energy distribution (SED) is double-peaked,
with the first UV-visible peak indicative of the central star and the second infrared peak coming from thermal emission of the cool dust in the circumstellar environment (CE).
The release of the IRAS All-Sky Survey mission made a large-scale identification of post-AGB stars on the basis of their
infrared colours possible \citep{lloydevans85,hrivnak89,oudmaijer92}. More recent surveys on post-AGB stars include
\citet{szczerba07}, who present the latest catalogue of Galactic post-AGB stars with almost 400 objects.

IRAS colour-colour diagrams revealed a large sample of stars that did not show the expected double-peaked SED \citep{lloydevans85,deruyter06}.
Instead, they show a strong near-IR excess, pointing to the presence of hot dust in the system,
while the central stars are currently too hot to have an ongoing dusty mass loss.
Correlations with optical databases of variable stars point to the presence of pulsating stars that also show
this specific IR-excess. These RV\,Tauri stars are luminous evolved stars that cross the Population II Cepheid instability strip 
and populate a well-defined part of the IRAS colour-colour diagram \citep{lloydevans85,raveendran89,lloydevans99}.
Recent interferometric studies \citep{deroo07b,deroo07a} prove that the circumstellar emission originates from
a very compact region and that the SEDs of these stars are well modelled with a passive
2D disc model \citep{dullemond04,deroo07b,gielen07}. The spectral slope of the submillimetre SED points to the presence of 
large, up to $\mu$m and cm sized, grains in these discs \citep{deruyter06,gielen07}, which have settled to form a cool disc midplane.
The inner radius of these discs is determined by the dust sublimation radius. The hot inner rim is puffed up by gas presssure from the central star and
radiates mainly in the near-IR, while the outer parts of the disc can be strongly flared and are responsible for
the strong absorption and re-radiation of the stellar light.
In a few cases the direct confirmation of the Keplerian kinematics of the circumstellar discs comes from resolved interferometric CO measurements \citep{bujarrabal88,bujarrabal05,bujarrabal07}.
Recent studies have shown that these post-AGB stars are very likely all binaries \citep{maas02,maas03,vanwinckel99,vanwinckel07,gielen07}.

Famous examples of these objects include HD\,44179, the central star of the Red Rectangle. It is a carbon-rich post-AGB object for
which the binarity was proven by \citet{vanwinckel95}. The star is surrounded by an oxygen-rich disc \citep{waters98} and is resolved 
in ground-based high spatial resolution imaging \citep{roddier95,menshchikov02} as well as in HST optical images \citep{cohen04}.
The disc is also resolved in interferometric CO measurements, where the Keplerian velocity of the disc was detected \citep{bujarrabal05}.

Infrared spectroscopy is the ideal tool to study the physico-chemical characteristics of the
circumstellar material, since it samples resonances of the most dominant dust species and important ro-vibrational bands of dominant molecules.
Spectroscopic data obtained with the Infrared Space Observatory (ISO) allowed for the first time to study the mineralogy of circumstellar environments
of objects in different evolutionary stages \citep[e.g.][]{waters96,waelkens96}.
More advanced studies of the CE of young stellar objects \citep[e.g.][]{vanboekel05,lisse07}
and mass-losing objects \citep{molster02a,molster02b,molster02c} further identified the dominant dust species and grain sizes of the circumstellar dust. 

Infrared spectral studies of post-AGB objects \citep[][and references therein]{vanwinckel03} allowed for the detection 
of different CE chemistries. The more typical C-rich post-AGB stars are characterised by a strong amorphous SiC feature at 11.3\,$\mu$m.
In these stars, a strong unidentified 21\,$\mu$m feature can sometimes be found \citep{kwok89}.
In addition to the detection of O-rich post-AGB stars, dominated by silicate species, surprisingly also mixed chemistries were found, showing features
of both O-rich and C-rich dust species.

Young stellar objects are the natural environment to study circumstellar disc physics, but
previous studies of the mineralogy of discs around the few brightest infrared evolved objects  with ISO \citep{molster99,molster02a,molster02b,molster02c}
show strong dust processing, with oxygen-rich and highly crystalline dust.
\citet{molster02a,molster02b,molster02c} found that a high degree of crystallisation is indicative for the
presence of a stable disc, and not a dusty outflow.  

In our pilot study \citep{gielen07}, based on SPITZER-IRS and ISO data, we investigate the mineralogy and spectral energy distribution of two post-AGB stars, RU\,Cen and AC\,Her.
The spectra of these stars are extremely similar and show a strong resemblance to the infrared spectrum of the solar-system comet Hale-Bopp
and young stellar objects, such as HD\,100546. The observed crystalline emission features are well modelled using magnesium-rich crystalline silicates.
The grains are irregularly shaped, with grain sizes of 0.1\,$\mu$m and 1.5\,$\mu$m. Both hot and cool grains are necessary
to reproduce the observed spectrum. The spectral energy distributions of both stars were fitted using a 2D passive disc model \citep{dullemond04}.
The discs surrounding these objects start at 35\,AU from the central star, which is well beyond the dust sublimation radius. The outer radius is less well constrained,
but extends till about 300\,AU. The discs are strongly flared.

In this work we study the mineralogy of the discs in a large sample of post-AGB binaries and perform a detailed fitting of the observed emission features.
We have observed 21 binary post-AGB objects with the SPITZER-IRS spectrograph and look for relations between the dust parameters
and stellar characteristics. 

The outline of this paper is as follows: in Sects.~\ref{programstars} and ~\ref{observations} we introduce our programme stars and the different observation and reduction
strategies used. Section~\ref{sed} contains the construction of the spectral energy distributions and the total extinction determination.
In Sect.~\ref{general} we give a general overview of the spectra and the observed emission features. The profile and full spectral fitting
is presented in Sects.~\ref{profilefit} and ~\ref{fullfit}. The discussion and conclusion of our analysis are given in Sects.~\ref{discussion} and \ref{conclusions}.

\section{Programme stars}
\label{programstars}

From the total sample of 51 published binary post-AGB stars likely surrounded by a stable disc \citep{deruyter06}, we selected
the 21 stars with fluxes below the saturation limit of the SPITZER-IRS spectrograph. Spectral types range from 
A to M. 

Our radial velocity monitoring program, using the CORALIE spectrograph attached to the 1.2\,m Swiss Euler Telescope 
at the ESO La Silla Observatory, is still ongoing but we have already found orbital parameters for 15 of our 21 program stars. Orbital periods
range from 100 up to more than 2000 days \citep{vanwinckel07}. 
The objects for which we do not yet have orbital parameters, have pulsational amplitudes which are too large for a
straightforward detection of the binary motion. The results of the radial velocity monitoring program will be subject of a forthcoming paper.

The orbital periods indicate that strong binary interaction must have occurred during
the late stages on the AGB. The binaries are now not in contact but the orbits are too 
small to accommodate a full-grown AGB star. The discs are all circumbinary since all orbits lie well within
the dust sublimation radius. The distribution of the mass-functions gives a range in minimal mass for the companion
between $1\,$M$_{\odot}$ and $2\,$M$_{\odot}$. The companion stars are likely to be unevolved main sequence stars \citep[e.g.][]{gielen07,vanwinckel07}.

These binary post-AGB stars are characterised by a depletion pattern
in their photospheres \citep{giridhar94,giridhar98,giridhar00,gonzalez97b,gonzalez97a, vanwinckel98, maas05}. 
This abundance pattern is the result of gas-dust separation followed by reaccretion
of the gas, which is poor in refractory elements. \citet{waters92} proposed that the most likely circumstance for this
process to occur is when the dust is trapped in a circumstellar disc.
Photospheric chemical studies of post-AGB candidates in the LMC have revealed that there also, depletion patterns are
common \citep{reyniers07}. The observations in the recent release of the SAGE database \citep{meixner06} showed that these depleted LMC sources
have infrared excesses similar to the Galactic binary post-AGB stars, and are therefore thought to be disc sources as well.

The disc formation itself is badly understood.
Possible formation scenarios of the discs include a wind capture scenario \citep[e.g.][]{mastrodemos99}, or a formation scenario
through non-conservative mass transfer in an interacting binary. In the first scenario, the AGB wind is captured by the companion. In the second
scenario, which is still not very well explored theoretically, the disc formation precedes the 
dust-grain formation and it is likely that the thermal history of the grains was very different from that in normal AGB winds. This could lead to very different
chemico-physical properties during formation. The sizes of the orbits suggest this scenario to be more likely.

We may witness the formation of a circumbinary disc by Roche-Lobe overflow in the 
evolved binary system SS\,Lep. The optical/IR interferometric observables can be best understood, assuming a Roche-lobe filling M star in a system with a circumsystem disc \citep{verhoelst07}.

Our program stars are therefore all likely binaries which are surrounded by a dusty stable disc. This disc seems to have an important impact on the objects
and in this paper we focus on the analysis of the IR spectra as probes for the disc physics.

\begin{table*}
\caption{The name, equatorial coordinates $\alpha$ and $\delta$(J2000), spectral type, effective
temperature T$_{eff}$, surface gravity $\log g$ and metallicity [Fe/H] of our sample stars.
For the model parameters we refer to \citet{deruyter06}. Also given is the orbital period \citep[see references in][]{deruyter06,gielen07}.
The total reddening $E(B-V)_{tot}$, the energy ratio $L_{IR}/L_*$ and the calculated distance,
assuming a luminosity of $L_*=5000\pm2000$\,L$_{\odot}$. Distances marked with * are upper limits due
to the aspect angle of the disc.}
\label{sterren}
\centering
% use packages: array
\begin{tabular}{llrrcccrrrc}
\hline \hline
N$^\circ$ & Name & $\alpha$ (J2000) & $\delta$ (J2000)  & T$_{eff}$ & $\log g$ & [Fe/H] & P$_{orbit}$ & $E(B-V)_{tot}$ & $L_{IR}/L_*$ & d\\ 
 & & (h m s) & ($^\circ$ ' '')  & (K) & (cgs) & & (days) & & (\%) & (kpc)  \\
\hline
1 & EP\,Lyr      & 19 18 17.5 & $+$27 50 38  & 7000 & 2.0 & -1.5   &               & 0.52$\pm$0.01 & 3$\pm$0    & 4.1$\pm$0.8      \\ 
2 & HD\,131356   & 14 57 00.7 & $-$68 50 23  & 6000 & 1.0 & -0.5   & 1490          & 0.20$\pm$0.01 & 50$\pm$2   & 3.0$\pm$0.6      \\ 
3 & HD\,213985   & 22 35 27.5 & $-$17 15 27  & 8250 & 1.5 & -1.0   & 259           & 0.27$\pm$0.01 & 24$\pm$1   & 3.1$\pm$0.6      \\ 
4 & HD\,52961    & 07 03 39.6 & $+$10 46 13  & 6000 & 0.5 & -4.8   & 1310          & 0.06$\pm$0.01 & 12$\pm$1   & 2.1$\pm$0.4      \\ 
5 & IRAS\,05208$-$2035  & 05 22 59.4 & $-$20 32 53  & 4000 & 0.5 & 0.0    & 236    & 0.00$\pm$0.00 & 38$\pm$2   & 3.9$\pm$0.8      \\ 
6 & IRAS\,09060$-$2807  & 09 08 10.1 & $-$28 19 10  & 6500 & 1.5 & -0.5   & 371    & 0.57$\pm$0.02 & 63$\pm$3   & 5.4$\pm$1.1      \\ 
7 & IRAS\,09144$-$4933  & 09 16 09.1 & $-$49 46 06  & 5750 & 0.5 & -0.5   & 1770   & 1.99$\pm$0.05 & 53$\pm$5   & 2.7$\pm$0.6      \\ 
8 & IRAS\,10174$-$5704  & 10 19 18.1 & $-$57 19 36  &      &     &        & 323    &  &  &                                         \\ 
9 & IRAS\,16230$-$3410  & 16 26 20.3 & $-$34 17 12  & 6250 & 1.0 & -0.5   &        & 0.56$\pm$0.02 & 60$\pm$3   & 6.1$\pm$1.2      \\ 
10 & IRAS\,17038$-$4815 & 17 07 36.3 & $-$48 19 08  & 4750 & 0.5 & -1.5   & 1381   & 0.22$\pm$0.02 & 69$\pm$5   & 4.5$\pm$1.0      \\ 
11 & IRAS\,17243$-$4348 & 17 27 56.1 & $-$43 50 48  & 6250 & 0.5 & 0.0    & 484    & 0.59$\pm$0.02 & 68$\pm$4   & 3.8$\pm$0.8      \\ 
12 & IRAS\,19125$+$0343 & 19 15 00.8 & $+$03 48 41  & 7750 & 1.0 & -0.5   & 517    & 1.08$\pm$0.02 & 52$\pm$3   & 1.8$\pm$0.4      \\ 
13 & IRAS\,19157$-$0247 & 19 18 22.5 & $-$02 42 09  & 7750 & 1.0 & 0.0    & 120.5  & 0.68$\pm$0.01 & 63$\pm$2   & 4.2$\pm$0.9      \\ 
14 & IRAS\,20056$+$1834 & 20 07 54.8 & $+$18 42 57  & 5850 & 0.7 & -0.4   &        & 0.51$\pm$0.02 & 905$\pm$42 & 10.9$\pm$2.3     \\ 
15 & RU\,Cen     & 12 09 23.7 & $-$45 25 35  & 6000 & 1.5 & -2.0   & 1489          & 0.55$\pm$0.01 & 13$\pm$1   & 2.3$\pm$0.5      \\ 
16 & SAO\,173329 & 07 16 08.3 & $-$23 27 02  & 7000 & 1.5 & -0.8   & 115.9         & 0.39$\pm$0.01 & 36$\pm$1   & 6.5$\pm$1.3      \\ 
17 & ST\,Pup     & 06 48 56.4 & $-$37 16 33  & 5750 & 0.5 & -1.5   & 410           & 0.00$\pm$0.00 & 55$\pm$1   & 5.7$\pm$1.2      \\ 
18 & SU\,Gem     & 06 14 00.8 & $+$27 42 12  & 5750 & 1.125 & -0.7 &               & 0.58$\pm$0.02 & 111$\pm$7  & 4.8$\pm$1.0*     \\ 
19 & SX\,Cen     & 12 21 12.6 & $-$49 12 41  & 6000 & 1.0 & -1.0   & 600           & 0.32$\pm$0.02 & 34$\pm$2   & 3.8$\pm$0.7      \\ 
20 & TW\,Cam     & 04 20 48.1 & $+$57 26 26  & 4800 & 0.0 & -0.5   &               & 0.40$\pm$0.02 & 42$\pm$3   & 3.2$\pm$0.6      \\ 
21 & UY\,CMa     & 06 18 16.4 & $-$17 02 35  & 5500 & 1.0 & 0.0    &               & 0.00$\pm$0.00 & 89$\pm$3   & 9.6$\pm$2.0*     \\
\hline                                                                             
\end{tabular}
\end{table*}

\section{Observations and data reduction}
\label{observations}

\subsection{SPITZER}

High- and low-resolution spectra for 21 post-AGB stars were obtained using the Infrared Spectrograph (IRS; \citealp{houck04}) aboard the SPITZER
Space Telescope \citep{werner04} in February 2005. The spectra were observed using combinations of the short-low (SL), short-high (SH) and
long-high (LH) modules. SL ($\lambda$=5.3-14.5\,$\mu$m) spectra have a resolving power of
R=$\lambda/\bigtriangleup\lambda \sim$ 100
, SH ($\lambda$=10.0-19.5\,$\mu$m) and LH
($\lambda$=19.3-37.0\,$\mu$m) spectra have a resolving power of $\sim$ 600.
Exposure times were chosen to achieve a S/N ratio of around 400 for the high-resolution modes,
which we complemented with short exposures in low-resolution mode with a S/N ratio around 100,
using the first generation of the exposure time calculator of the call for proposals.

The spectra were extracted from the SSC raw data pipeline version S13.2.0 products,
using the c2d Interactive Analysis reduction software package
\citep{kessler06,lahuis06}. This data
processing includes bad-pixel correction, extraction, defringing and order
matching. Individual orders are corrected for offsets, if necessary, by small scaling
corrections to match the bluer order.

\subsection{TIMMI2}

For stars where we lack the SPITZER IRS-SH observations we obtained additional ground-based N-band infrared
spectra with the Thermal Infrared Multi Mode Instrument 2 (TIMMI2, \citealp{reimann00,kaufl03}), mounted on the 3.6\,m telescope
at the ESO La Silla Observatory. The low-resolution ($ R \sim$ 160) N band grism
was used in combination with a 1.2\,arcsec slit, the pixel scale in the
spectroscopic mode of TIMMI2 is 0.45\,arcsec.
For the reduction of the spectra we used the method described in \citet{vanboekel05}.
We scaled the TIMMI2 spectra to the SPITZER spectra and found a very good agreement in spectral shape between the two data sets.

\section{Spectral energy distribution}
\label{sed}

For all sample stars spectral energy distributions were updated from the photometric data
as given by \citet{deruyter06}. The resulting SEDs are presented in Figs.~\ref{sed1} and \ref{sed2}.
The total extinction $E(B-V)_{tot}$ was determined by dereddening the observed photometry, using the average
extinction law of \citet{savage79}. The relation between $E(B-V)_{tot}$ and $A_V$ is given by $A_V=R_V \times E(B-V)$, with a 
typical value for $R_V=3.1$ \citep{savage79}. 
Minimising the difference between the dereddened observed optical fluxes and the
appropriate Kurucz model \citep{kurucz79} gives the total colour excess $E(B-V)_{tot}$ (Table~\ref{sterren}). Model parameters for our sample stars are given in Table~\ref{sterren} and are based on the analysis of high-resolution spectra as given by the literature (see \citet{deruyter06} for references). 
A considerable fraction of our sample stars show a photometric variability due to pulsation, so we use only
photometric maxima for the SED construction. 

The errors on the value for $E(B-V)_{tot}$ are calculated using a Monte-Carlo simulation
on the photometric data. We use an error of 0.05 for the photometric measurements in a Gaussian distribution.
Since we do not know the distances to the sources, we adopt a likely luminosity for evolved low-gravity objects of $L_*=5000\pm2000$\,L$_{\odot}$.
The distance to IRAS\,20056 is likely overestimated since visible light from this source probably reaches us only by scattering through a
nearly edge-on optically thick disc \citep{menzies88,gledhill01}.

We also corrected the infrared spectra for reddening by extending the average extinction law of \citet{savage79} 
with the theoretical extinction law of \citet{steenman89,steenman91} and by using the derived total extinction values. This is done under
the assumption that the extinction is fully due to interstellar extinction. Since the total extinction probably
consists of both an interstellar and a circumstellar component, the applied dereddening is thus a maximal correction.

\begin{figure}
\vspace{0cm}
\hspace{0cm}
\resizebox{8cm}{!}{ \includegraphics{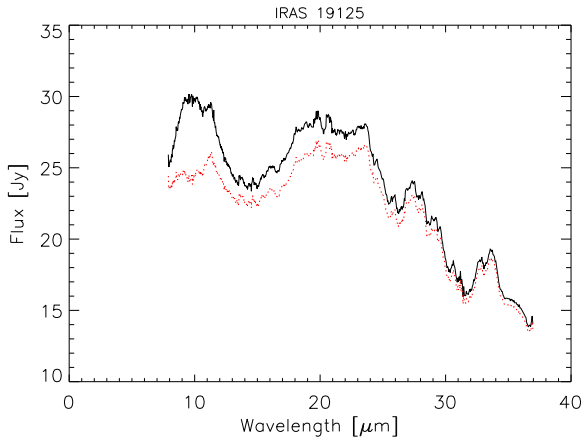}}
\caption{Comparison between the dereddened (black solid line) and reddened (red dotted line) of IRAS\,19125.
Dereddening mainly influences the spectral signature in the 10\,$\mu$m region.}
\label{plot_ontrood}%
\end{figure}

Dereddening the infrared spectra mainly influences the shape of the silicate feature around 9.8\,$\mu$m (see Fig.~\ref{plot_ontrood}). 
Since our total extinction values are on average quite small, the effect of deredding the infrared spectra is negligible for most stars.
For stars with a higher value for $E(B-V)_{tot}$, like IRAS\,09144 and IRAS\,19125, one has to be careful interpreting 
the shape of the silicate feature, since the partitioning between interstellar and circumstellar reddening is not known at this point.

We compute the energy ratio $L_{IR}/L_*$ to determine the amount of energy reprocessed by the CE (Table~\ref{sterren}).
80\% of our stars have a ratio $L_{IR}/L_*$ of 30\% and higher, showing that the absorption and re-radiation of stellar light
by the CE is extremely efficient.

\section{General overview}
\label{general}

Figures~\ref{fitting1} till \ref{fitting3} show the wide range in observed spectra. All spectra are characterised by many distinct spectral
structures in emission over a smooth continuum. There is quite some variety over the sample, with different continuum slopes, but for 
nearly all stars the spectrum is dominated by emission bands that can be attributed to silicate dust species.

In a few stars only, gas phase emission is detected in the form
of bandhead emission of $^{12}$CO$_2$ and $^{13}$CO$_2$. These stars are
HD\,52961, EP\,Lyr and IRAS\,10174.

Unique to our sample is the spectrum of EP\,Lyr in the 7-20\,$\mu$m
region. This spectral range is dominated by a strong
and narrow 11.3\,$\mu$m emission peak and a complex plateau with
narrow emission features in the 14-18\,$\mu$m region. The strongly
asymmetric band between 7\,$\mu$m and 10\,$\mu$m is real as well.
These features are observed in C-rich evolved stars and in the
carbon-rich component of the ISM and are associated with PAH
emission and the associated CH-out-of-plane and C-C-C bending
modes \citep[e.g.][]{vankerckhoven00,hony01,peeters02}.
Longward of 18\,$\mu$m, the spectrum is dominated by silicate emission.
EP\,Lyr, is the only object in our sample displaying a mixed chemistry.
It is remarkable that in our considerable sample of post-AGB stars, only one possible post-carbon star is found!

\subsection{Silicate dust features: Identification}
\label{silicatedust}

\begin{figure}
\vspace{0cm}
\hspace{0cm}
\includegraphics[height=9.3cm,width=9.5cm]{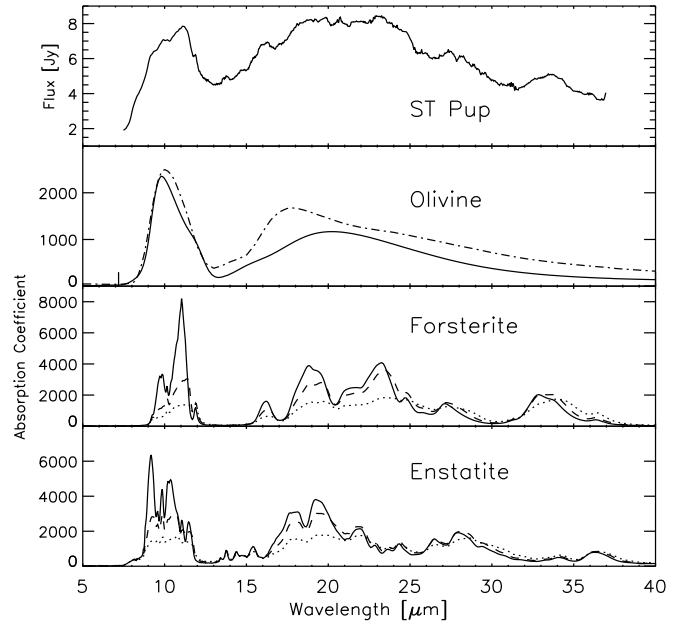}
\caption{The mass absorption coefficients [cm$^2$/g] of forsterite and enstatite in GRF approximation.
Grain sizes of 0.1\,$\mu$m (full line), 2.0\,$\mu$m (dashed line) and 4.0\,$\mu$m (dotted line) are plotted.
The third box shows olivine in 0.1\,$\mu$m GRF approximation. The Mg-rich member (Mg$_2$SiO$_4$) is plotted in full line,
the standard Mg-Fe member (MgFeSiO$_4$) is plotted in dot-dashed line.
As an example we also plotted the SPITZER spectrum of one of our sample stars, ST\,Pup.}
\label{plot_cryst}%
\end{figure}

Amorphous and crystalline silicates are the most commonly found dust species in the interstellar
and circumstellar environment \citep{molster02a,molster02b,molster02c,kemper04,min07}. To describe the glassy
and crystalline silicates with an olivine and pyroxene stoichiometry we will use the commonly used term
``amorphous and crystalline olivine and pyroxene''. 
Amorphous olivine (Mg$_{2x}$Fe$_{2(1-x)}$SiO$_4$, where 0$\leq$x$\leq$1 denotes the magnesium content) has very prominent
broad features around 9.8\,$\mu$m and 18\,$\mu$m (see Fig.~\ref{plot_cryst}), which are easily detected in our spectra. These features (also called the
10\,$\mu$m and 20\,$\mu$m features) arise respectively from the Si-O
stretching mode and the O-Si-O bending mode. For large grains the
9.8\,$\mu$m feature gets broader and shifts to redder wavelengths. Amorphous
pyroxene (Mg$_{x}$Fe$_{1-x}$SiO$_3$) shows a 10\,$\mu$m feature similar to that of
amorphous olivine, but shifted towards shorter wavelengths. Also the shape
of the 20\,$\mu$m feature is slightly different.

The Mg-rich end members of crystalline olivine and pyroxene, forsterite (Mg$_2$SiO$_4$) and enstatite (MgSiO$_3$),
show strong but narrow features at distinct wavelengths around 11.3 - 16.2 - 19.7 - 23.7 - 28 - 33.6\,$\mu$m 
(see Fig.~\ref{plot_cryst}), making them easily identifiable in our spectra.
Forsterite condenses directly from the gas-phase at high temperatures
($\approx 1500$\,K) or it may form by thermal annealing of amorphous
silicates, diffusing the iron out of the lattice-structure. Enstatite can
form in the gas-phase from a reaction between forsterite and silica (SiO$_2$), or it
may also form by a similar thermal annealing process as forsterite \citep{bradley83,tielens97}.
Laboratory experiments have indicated that silica can be formed when amorphous silicates anneal to forsterite \citep[e.g.][]{fabian00}.
No evidence for the presence of silica, with strong features around 9 and 21\,$\mu$m, can be found in our spectra.
There is also no evidence for the presence of Fe-rich crystalline silicates, like fayalite (Fe$_2$SiO$_4$).

Nearly all stars show strong crystalline dust features, both at short and long wavelengths,
showing that both hot and cool crystallines seem to be abundant.

In our spectra, the amorphous silicate dust seems to peak at 20\,$\mu$m, rather than 18\,$\mu$m, as would
be expected from synthetic spectra of amorphous silicates. This could point to the presence of
Mg-rich amorphous dust, which also shows this shift to redder wavelengths (see Fig.~\ref{plot_cryst}).

\subsection{Mean spectra and complexes}
\label{mean}

Different complexes at 10, 14, 16, 19, 23 and 33\,$\mu$m
can be identified in our spectra. In order to study the systematics between these complexes in our spectra we compare them to a
mean spectrum in that region. The mean complex spectra were obtained by adding continuum subtracted spectra of sample sources
with clear spectral structures, using a weighing factor proportional to the S/N in that spectral region.
The continuum was determined by linearly interpolating between the beginning and end of the studied region.
The 10\,$\mu$m complex is discussed below, other complexes are discussed in the appendix (Sect.~\ref{mean_appendix}).
The mean complex spectra are shown in Figs.~\ref{mean14} to ~\ref{mean33}. We also plot the mass absorption
coefficients of forsterite and enstatite in every complex.

\subsubsection{The 10\,$\mu$m complex (8-13\,$\mu$m)}

\begin{figure}
\vspace{0cm}
\hspace{0cm}
\resizebox{9.7cm}{!}{ \includegraphics{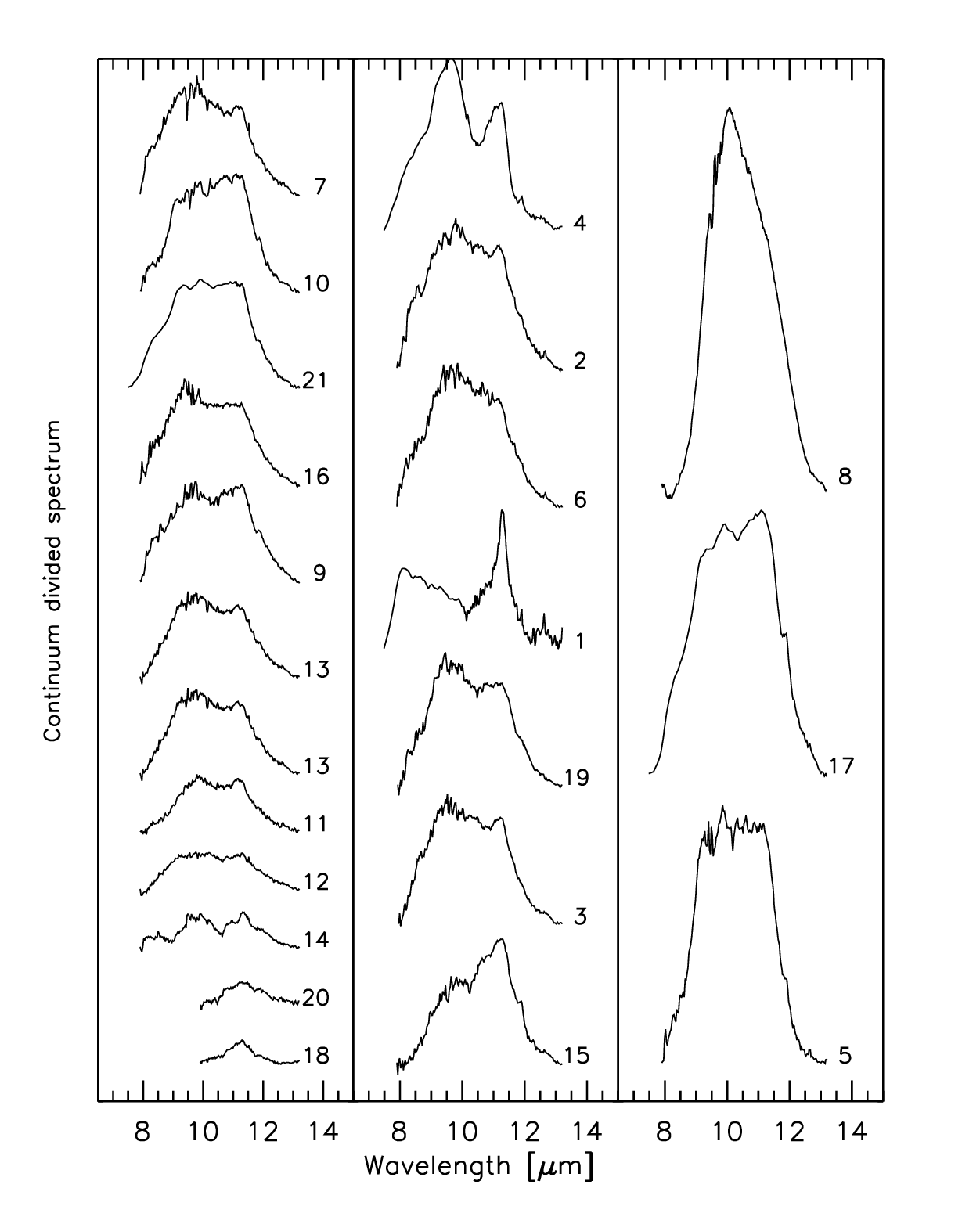}}
\caption{Continuum divided spectra in the 10\,$\mu$m region, ordered by peak value. The peak value
increases from left to right and from bottom to top. The numbering corresponds with the star numbers as given in Table~\ref{sterren}.
For SU\,Gem and TW\,Cam we lack flux points in the 7.5\,$\mu$m region to perform a good continuum determination.}
\label{peak_cont}%
\end{figure}

In Fig.~\ref{peak_cont} we plot the continuum divided spectra in the 10\,$\mu$m region, ordered by peak value.
We plot $1+F_{\nu,cs}/<F_{\nu,c}>$, where $F_{\nu,cs}$ is the continuum subtracted spectrum $(F_{\nu}-F_{\nu,c})$
and $<F_{\nu,c}>$ is the mean of the continuum. The continuum was determined by linearly interpolating between 7.5 and 13\,$\mu$m.
This method preserves the shape of the emission band and allows a good comparison between the profiles themselves.
The figure clearly illustrates the very wide variety of the spectral appearance of the warm silicates, and because of this large
variety a  mean spectrum was not constructed.
We do not find a relation between the strength and the shape of the silicate feature. 

\begin{figure}
\vspace{0cm}
\hspace{0cm}
\resizebox{9cm}{!}{ \includegraphics{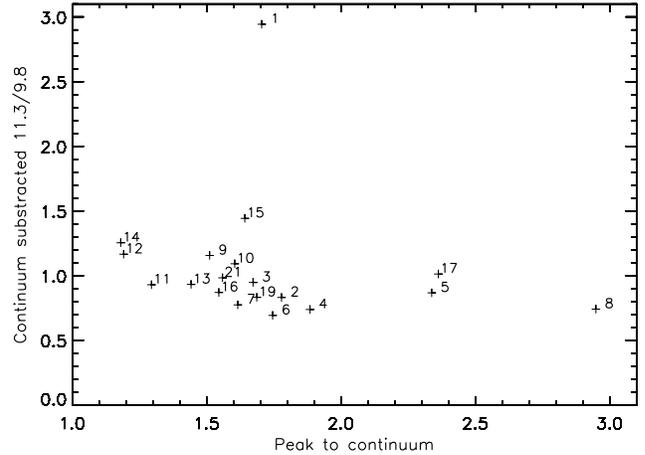}}
\caption{Continuum subtracted flux at 11.3 and 9.8\,$\mu$m versus the peak to continuum ratio of the 10\,$\mu$m silicate feature.
SU\,Gem and TW\,Cam are not plotted since for these stars we lack a good continuum determination.}
\label{peakcont_contsub}%
\end{figure}

In Fig.~\ref{peakcont_contsub} we show the continuum subtracted flux at 11.3 and 9.8\,$\mu$m versus the peak to continuum ratio of the 10\,$\mu$m silicate feature.
The 11.3/9.8\,$\mu$m ratio is a measure for the amount of processing that the dust has undergone, where the peak/continuum ratio is a measure
for the typical grain size. We find a high degree of crystallisation (11.3/9.8\,$\mu$m ratio) for all sources, but rather small peak to continuum values in general. EP\,Lyr is a clear outlier with a very strong 11.3/9.8\,$\mu$m ratio, due to a very sharp peak at 11.3\,$\mu$m which is due to PAH emission.
No clear correlation, and thus no evolutionary trend, is observed between the shape and the
strength of the 10\,$\mu$m feature in our sample stars, however.
This is in contrast with the clear correlation that is seen in young stellar objects \citep{vanboekel03,vanboekel05}, where stars with a strong feature have
low 11.3/9.8\,$\mu$m ratios and stars with a weaker silicate feature show higher 11.3/9.8\,$\mu$m ratios.
Sources that display a strong emission feature show rather unprocessed silicate band, similar to the ISM profile,
with little evidence for crystalline grains. Sources with a weak 10\,$\mu$m profile show a broader and flatter silicate feature,
dominated by larger grains.

\section{Analysis: Profile fitting}
\label{profilefit}

To fit the observed emission features with synthetic spectra, mass absorption coefficients for
different dust species need to be calculated. The conversion from laboratory measured optical constants of dust to mass
absorption coefficients is not straightforward and is largely dependent on the adopted size, shape, structure and chemical composition of
the dust \citep{min03,min05a}. These different dust approximations result in very different predicted emission
feature profiles, with a clear division between the synthetic spectra of homogeneous and
irregular grains. If one would assume homogeneous spherical particles, one could use Mie theory \citep{aden51,toon81}. However, cosmic dust grains are not perfect spheres, so we have to reside to other methods.
Examples of irregular dust approximations are 
CDE (continuous distribution of ellipsoids, \citealp{bohren83}),
GRF (Gaussian random field particles, \citealp[e.g.][]{grynko03,shkuratov05}) and DHS (distribution of hollow
spheres, \citealp{min03,min05a}) grain shapes. The DHS shaped particles are further characterised by
the fraction of the total volume occupied by the central vacuum inclusion, $f$,
over the range $0<f<f_{max}$. The value of $f_{max}$ reflects the degree of
irregularity of the particles \citep{min03,min05a}.
Cross sections in CDE are computed under the assumption that the grains
are in the Rayleigh limit (that the grains are much smaller than the
wavelength of radiation, thus smaller than 0.1\,$\mu$m), and do not allow the study of grain growth.
The spectra for the GRF particles were computed using the Discrete Dipole Approximation \citep{draine88}.
More details on the particle shapes and the computations of the spectra can be found in \citet{min07} and Min 2008 (in preparation).
Refractory indices for the materials used are taken from \citet{servoin73,dorschner95,henning96,jaeger98}.

Not only grain shape, but also grain size, has a profound influence on the mass absorption coefficients.
Since a circumstellar disc is the perfect environment for grain growth to occur, we also study
the effect of different grain sizes on the observed emission profiles. 
Different grain sizes produce emission features at different central wavelengths and with different profiles,
as shown in Figs~\ref{plot_cryst}. With increasing grain size,
emission features will become weaker and eventually disappear, leaving mainly a contribution to the thermal continuum.
This effect is already considerable for grain radii $a > 2$\,$\mu$m.
Previous studies \citep{bouwman01,honda04} in the 10\,$\mu$m region have shown that the variety of emission features can be described using two typical grain sizes:
generally 0.1\,$\mu$m to describe the grains with radii $< 1$\,$\mu$m, and 1.5-2\,$\mu$m for larger grains.

\begin{figure}
\vspace{0cm}
\hspace{0cm}
\rotatebox{90}{\resizebox{6.5cm}{!}{ \includegraphics{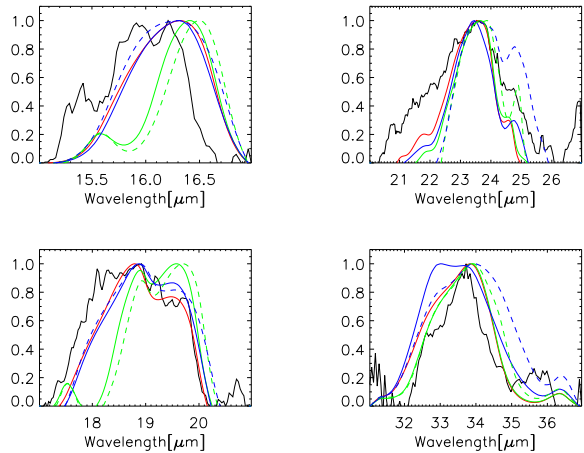}}}
\caption{Normalised and continuum subtracted emission features
of the mean of our sample stars, together with mass absorption coefficients for different
forsterite shape distributions. CDE grains are plotted in red, GRF in blue and DHS in green.
0.1\,$\mu$m grains are plotted in full lines, 2\,$\mu$m grains in dashed.}
\label{profiles1}%
\end{figure}

Comparing the observed crystalline emission features with calculated synthetic spectra (Fig.~\ref{profiles1}), we find
that irregular grains are needed to explain the dust profiles. When we look at the prominent forsterite 
features at 16-19-23.7-33.6\,$\mu$m, we find a good fit is achieved using grains in CDE, GRF or DHS (with $f=1.0$).
None of the tested dust shapes proves an accurate fit to the, in some stars very strong, forsterite 16\,$\mu$m feature, which seems to be shifted
to bluer wavelengths in the observed spectra. This was also seen in the disc spectra of \citet{molster02a}.
The best fit to this feature is achieved using dust in GRF approximation.
The fit of the 23.7\,$\mu$m feature is strongly improved when using large grains in DHS.
The GRF approximation does not result in a good fit to the 33.6\,$\mu$m feature, which is the purest forsterite feature,
since at other wavelengths the features may be blended with e.g. enstatite emission features. Both CDE and DHS reproduce the 
shape of this feature, but since the CDE approximation does not allow us to study the effect of large grains we 
use GRF and DHS approximations in our fitting routine.

A similar approach to determine the best enstatite or amorphous silicate dust approximations is not straightforward.
Enstatite emission features are mostly blended with forsterite features, but in the 14\,$\mu$m region there are some small
unblended features. These are ideal features to trace the enstatite fraction of the grains.
The central wavelengths of these peaks are at 13.8, 14.4 and 15.4\,$\mu$m, with the 15.4\,$\mu$m feature being the most prominent.
The sample sources that show clear emission in this region all show the 13.8\,$\mu$m feature and in lesser degree the 15.4\,$\mu$m feature.
The 14.4\,$\mu$m feature is absent but a rather strong unidentified feature appears around 14.8\,$\mu$m.
In order to keep the fitting homogeneous we will also use GRF and DHS dust approximations to describe the amorphous and crystalline pyroxene dust species.

\section{Full spectral fitting}
\label{fullfit}

\subsection{Method}

For an exact spectral modelling, full 2D radiative transfer 
in a realistic disc model should be studied. Such models are not yet
available so as a first approximation we assume 
the emission features to be formed in the thin surface layer of the disc.
Assuming the flux originates from an optically thin region, which
is motivated by the fact we see the dust features in emission, we can make linear 
combinations of the absorption profiles to calculate the model spectrum.

This approach is a first approximation but allows for a general study of trends in dust shapes, grain sizes and processing
in these discs. 
Disc models like the one of \citet{dullemond04}, which we used in our pilot study to fit
the SED of RU\,Cen and AC\,Her, do not allow for a detailed spectral modelling.   
The code computes the temperature structure and density of the disc. The
vertical scale height of the disc is computed by an iteration process,
demanding vertical hydrostatic equilibrium. Other important processes,
like dust settling and turbulent mixing, probably also occur in these discs but are
not included. The code also does not include an independent
dust species and grain size distribution throughout the disc, making a detailed mineralogy
study impossible. 

Our model emission profiles are then given by
$$F_\lambda \sim (\sum_i \alpha_i\kappa_i)\times(\sum_j \beta_j B_\lambda(T_j))$$
where $\kappa_i$
is the mass absorption coefficient of dust component $i$ and $\alpha_i$
gives the fraction of that dust component, $B_\lambda(T_j)$ denotes the
Planck function at temperature $T_j$ and $\beta_j$ the fraction of dust in
that given temperature. The temperature of dust grains will depend on grain size and grain shape, as well as the
distance to the central star and the integrated line-of-sight opacity from the stellar surface to the grains. This can only be calculated using
full radiative transfer, so we assume all grains to have the same dust temperatures, irrespective of size and shape.

In this modelling the continuum is given as a sum of Planck functions and we take it
as another dust component with a constant mass absorption coefficient.
To keep the number of free parameters in our fitting routine reasonable, we allow only two 
different dust temperatures and continuum temperatures (between 100 and 1000\,K), with different ratios. 
A fit with three temperatures was also tested and proved only a minor improvement.
The total number of fit parameters in the model is thus 15, with contributions of four silicate species, where we use two grain sizes, 
the continuum mass absorption coefficient, two dust and continuum temperatures which each their relative fractions.
 
To fit the spectra we minimise the reduced $\chi^2$ of the full SPITZER spectrum, extended with
the IRAS 60\,$\mu$m flux point to constrain the continuum at larger wavelengths, given by
$$\chi^2 = \frac{1}{N-M}\sum^N_{i=1}\left|\frac{F^{model}(\lambda_i)-F^{observed}(\lambda_i)}{\sigma_i}\right|^2,$$
with $N$ the number of wavelength points $\lambda_i$, $M$ the number of fit parameters and $\sigma_i$ the absolute error on the
observed flux at wavelength $\lambda_i$. The errors $\sigma_i$ represent the statistical noise on the spectra.
They are generated so that the errors are proportional to the square-root of the flux and they are scaled in such a way
that the best fit has a reduced $\chi^2$ of approximately 1.

The errors on the fit parameters are calculated using a Monte-Carlo simulation. We randomly add Gaussian noise with
a distribution of width $\sigma_i$ at each wavelength point. This generates 100 synthetic spectra, all consistent with our data,
on which we perform the same fitting procedure. This results in slightly different fit parameters for which we calculate
the mean, our best-fit value, and the standard deviation. The uncertainty on the mass absorption coefficients are not taken 
into account in the $\chi^2$-minimalisation. 

We find that the mass absorption coefficient of small and large amorphous grains
are quite similar, this degeneracy could introduce a large error on the derived fractions for small-large amorphous grains.

\begin{figure*}
\resizebox{8cm}{!}{\includegraphics{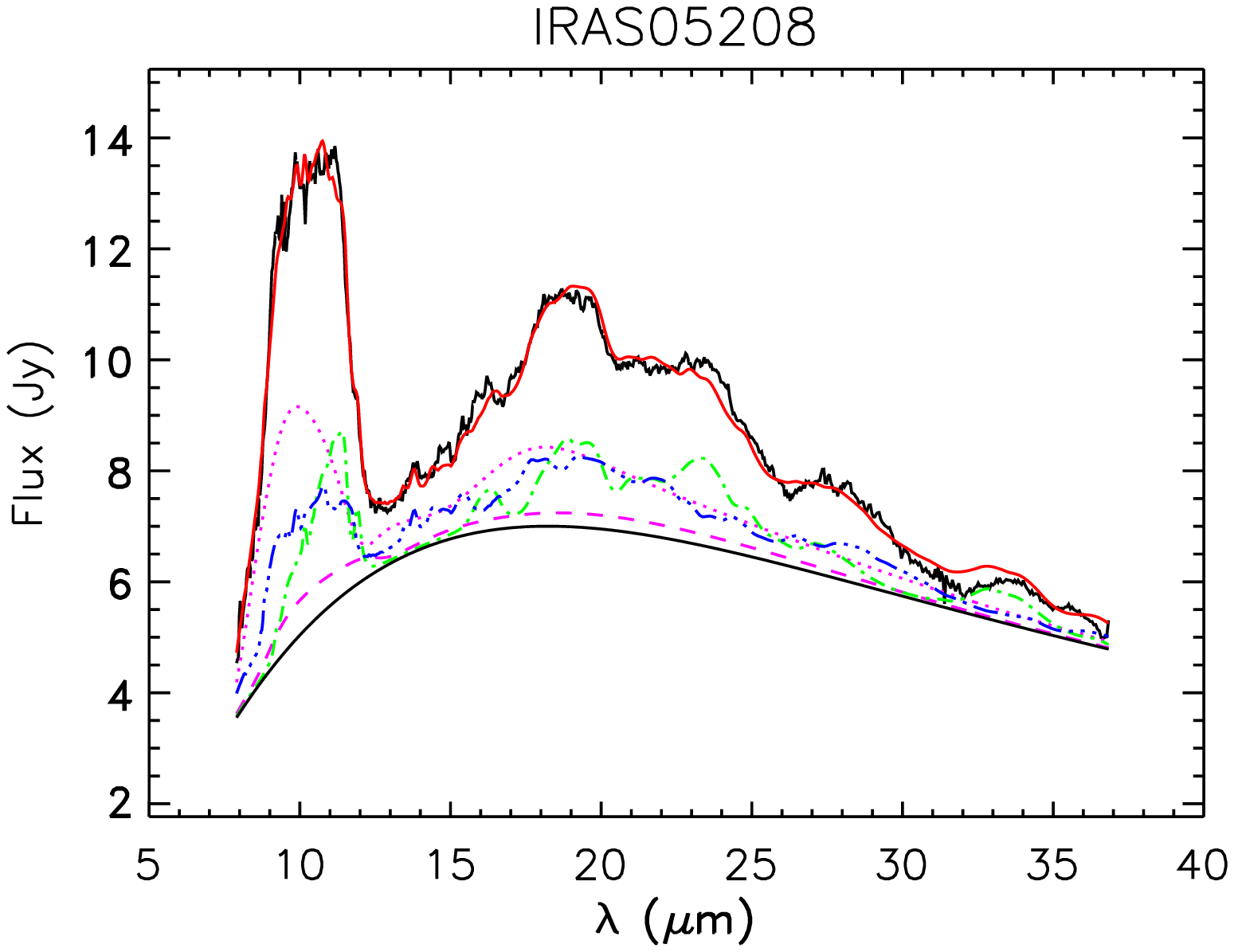}}
\resizebox{8cm}{!}{\includegraphics{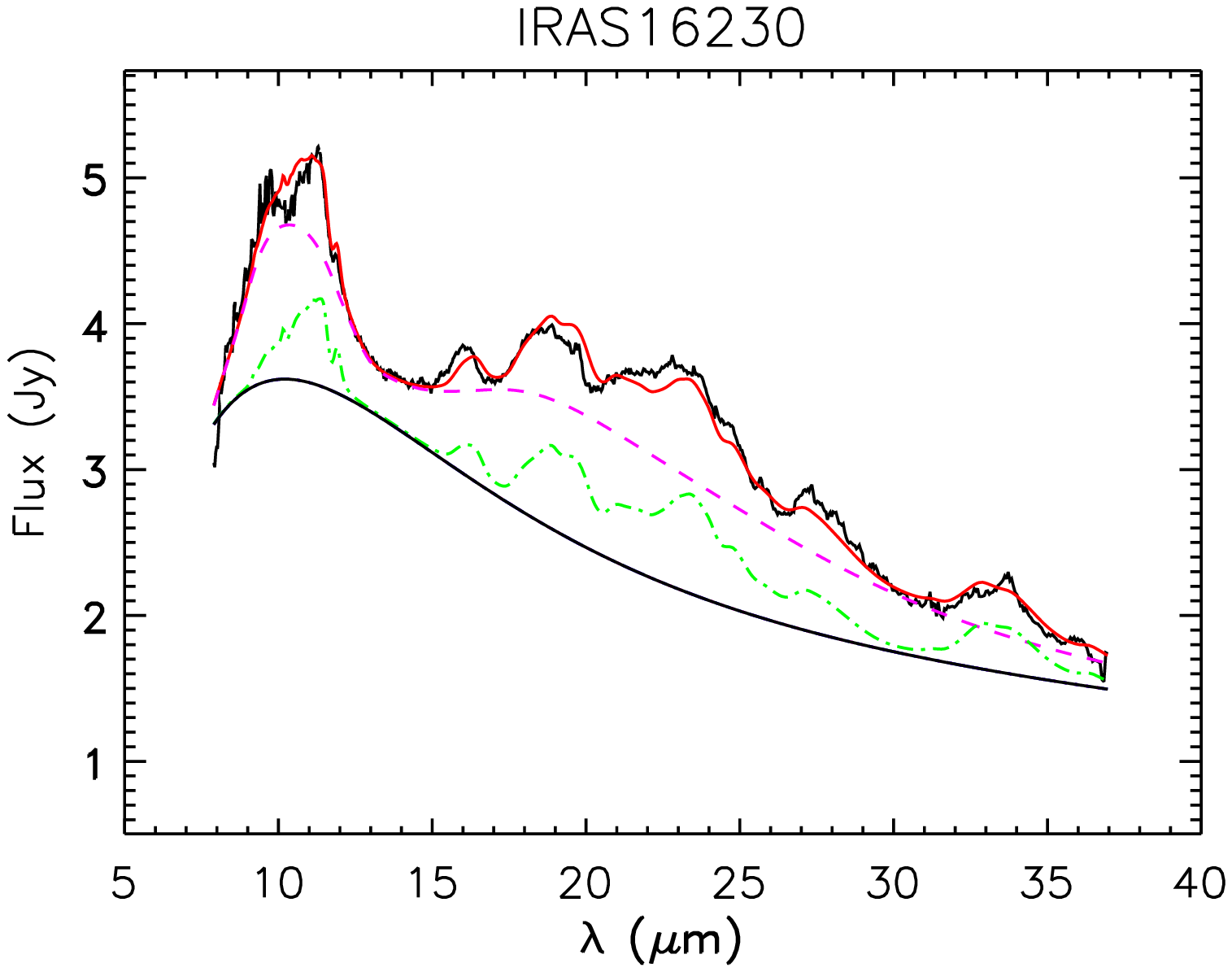}}
\end{figure*}
\begin{figure*}
\resizebox{8cm}{!}{\includegraphics{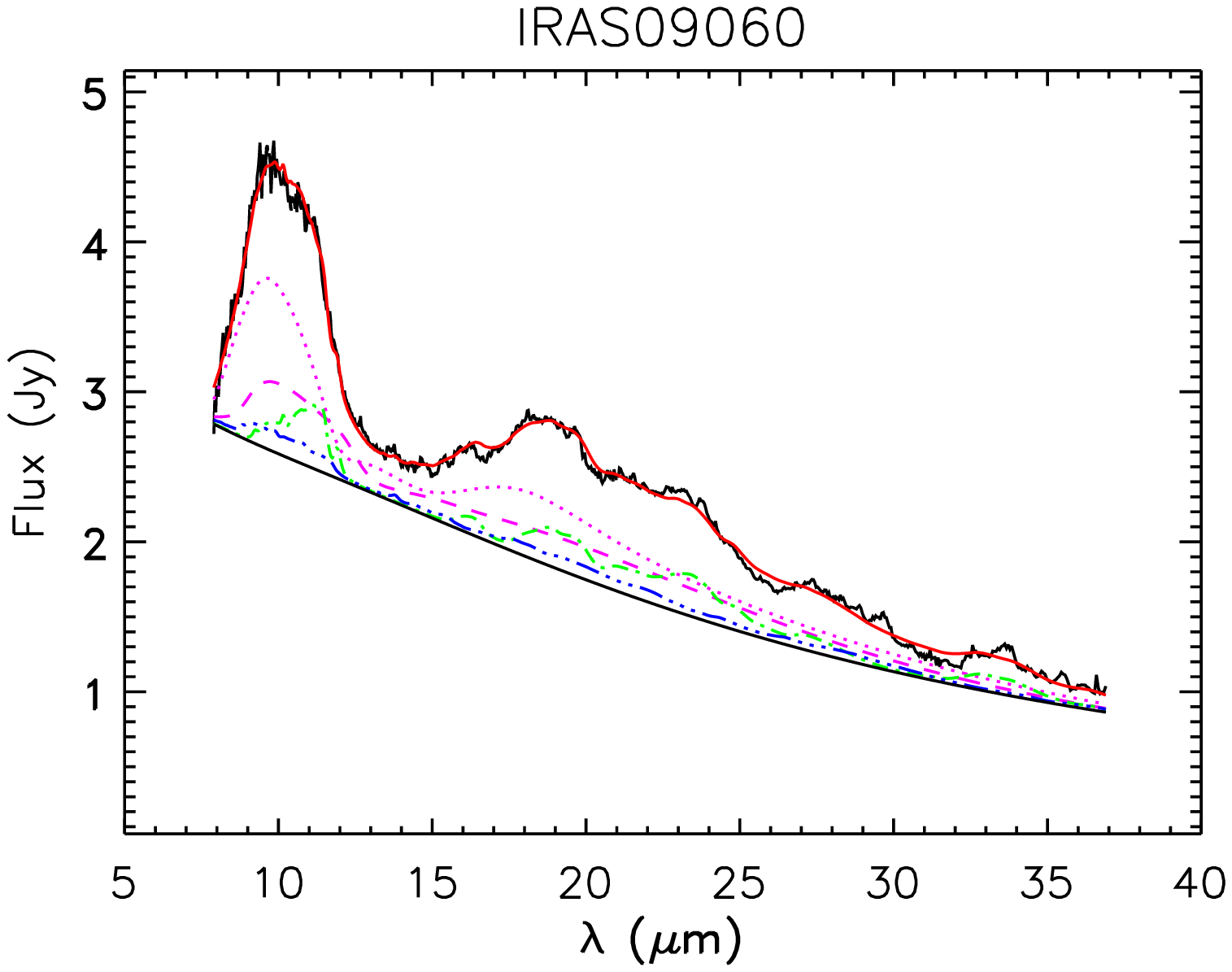}}
\resizebox{8cm}{!}{\includegraphics{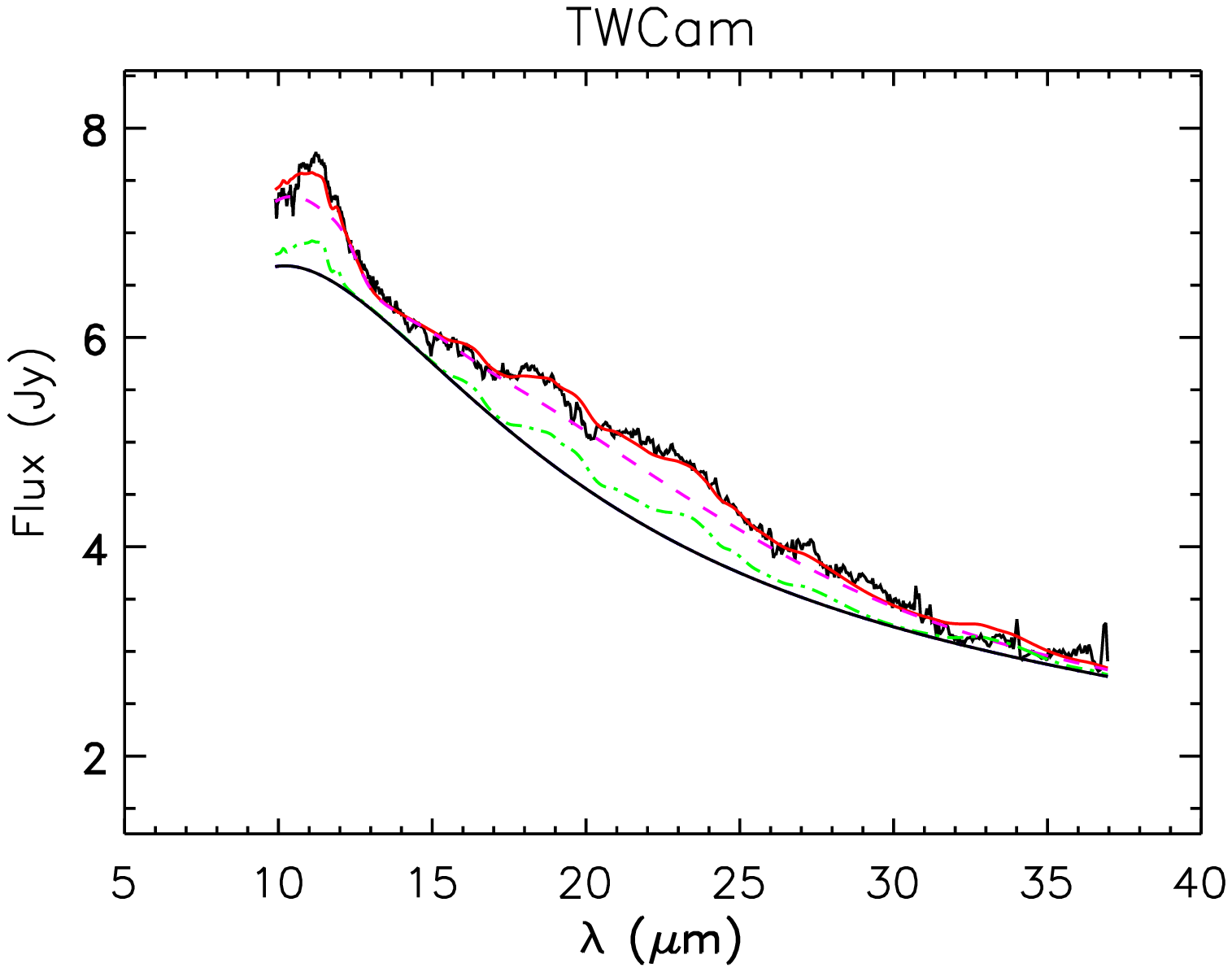}}
\caption{
Best model fits for some of our sample stars, showing the contribution of the different dust species.
Best fits for all sample stars are given in Figs.~\ref{fitting1}-\ref{fitting3}.
The observed spectrum (black curve) is plotted together with the best model fit (red curve) and the continuum (black solid line).
Forsterite is plotted in dash-dot lines (green) and enstatite in dash-dot-dotted lines (blue).
Small amorphous grains (2.0\,$\mu$m) are plotted as dotted lines (magenta) and large amorphous grains (4.0\,$\mu$m) as dashed lines (magenta).}
\label{fit_example}
\end{figure*}

\subsection{Results}

We tested both the DHS and the GRF dust approximations in the fitting routine, where the GRF approximation proved a 
far stronger match. In order to test for the presence of 
Fe-poor amorphous dust, as postulated in Sect.~\ref{silicatedust}, we perform the fitting
both with pure Mg-rich amorphous silicates ($x=1$, see Sect.~\ref{silicatedust}) 
and with the more standard Mg-Fe amorphous silicate dust ($x=0.5$). 

\begin{figure}
\vspace{0cm}
\hspace{0cm}
\resizebox{8cm}{!}{ \includegraphics{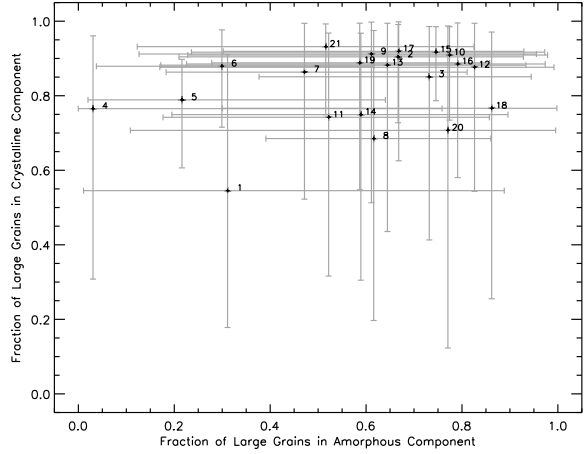}}
\caption{The mass fraction of large grains in the amorphous component versus the mass fraction of large
grains in the crystalline component, using the fitting with grain sizes of 0.1\,$\mu$m and 2.0\,$\mu$m. 
Crystalline grains are almost completely made up from large 2.0\,$\mu$m grains.}
\label{largelarge_oud}%
\end{figure}

When using 0.1\,$\mu$m and 2.0\,$\mu$m grain sizes, we find that nearly 80\% of the dust resides in the
2.0\,$\mu$m grains. The amount of grain growth in crystalline material versus amorphous material is plotted in Fig.~\ref{largelarge_oud}.
Almost all crystalline grains seem to be large, with a fraction of large grains in crystalline component higher than 0.7.
For the amorphous grains most sources have fractions between 0.3 and 0.9. No correlation between grain growth in crystalline
and amorphous grains is seen. An efficient removal of the smallest grains must have occurred in these discs 
or grain growth was the dominant factor to reduce the number of small grains at grain formation. 

Because of this lack of small grains, as a next step we performed a fitting using grain sizes of 2.0\,$\mu$m and 4.0\,$\mu$m.
This resulted in a better fit for 17/21 stars, giving slightly better values for $\chi^2$. For EP\,Lyr and HD\,52961 the quality of the fit
decreased considerably. This was to be expected, since the strong narrow crystalline features in these stars already indicated the presence of small grains.

In general the fit with the $x=1.0$ Mg-rich amorphous
silicates proved the best, be it only a minor improvement. The observed trends in the fit parameters and 
stellar characteristics do not change significantly between the fits when using $x=1.0$ or $x=0.5$ amorphous dust.
Of our sample stars, 6 stars show a better fit using the $x=0.5$ amorphous silicate dust.

The best model spectra, overplotted with the observed spectra are shown in Fig.~\ref{fit_example} and Figs.~\ref{fitting1}-\ref{fitting3}.
The resulting values of the best fit parameters are given in Table~\ref{fitresults1} and Table~\ref{fitresults2}.
The quality of our fitting is generally very good, 70\% of stars have a  $\chi^2 < 5$.

Some trends can be observed (e.g. Fig.~\ref{fit_example} and Figs.~\ref{fitting1}-\ref{fitting3}). 
As already explained in Sect.~\ref{profilefit}, the shape of the 33.6\,$\mu$m feature
is not well reproduced, although the strength is well modelled. Also the 29\,$\mu$m feature seems to be more prominent in 
the observed spectra, than in the model fits. The 19\,$\mu$m feature is slightly overestimated in
the model spectra, while the neighbouring 23\,$\mu$m feature is slightly underestimated. This discrepancy could be due to
a data reduction effect, since in this region there can be a bad overlap between the short and long SPITZER-IRS high-resolution bands.

The clear outliers in our modelling are EP\,Lyr and HD\,52961 with very high values of $\chi^2$. These stars have unusual features in their 
observed spectra, including strong narrow features and CO$_2$ emission lines. A detailed study of these outliers will be given in 
a follow-up article.

Nearly all sources show the presence of both hot and cool dust, both in dust temperature as in continuum temperature.
As seen in Table~\ref{fitresults1}, dust temperatures can differ strongly from continuum temperatures.

\subsection{Correlations}    

To gain insight in the dust formation process and evolution in the circumstellar environment
of our sample stars we look for trends in the derived fit parameters and correlations with
dust and stellar parameters.

\subsubsection{Mineralogy correlations}

\begin{figure}
\vspace{0cm}
\hspace{0cm}
\resizebox{8cm}{!}{ \includegraphics{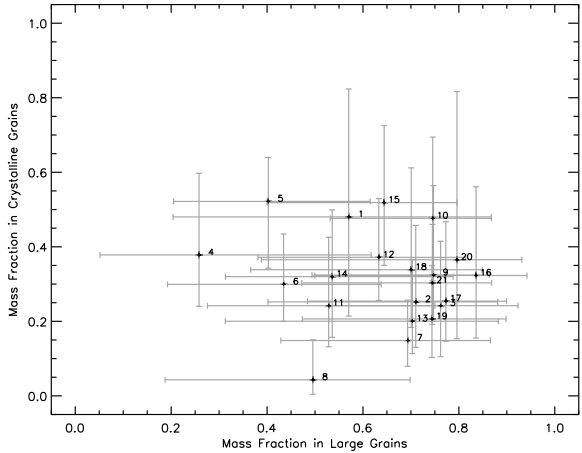}}
\caption{The mass fraction in large grains plotted against the mass fraction in crystalline grains, as derived
from our best fit parameters. A high degree of crystallinity is found.}
\label{crystlarge}%
\end{figure}

In Fig.~\ref{crystlarge} the mass fraction in large grains is plotted
against the mass fraction in crystalline grains. Mass fractions are calculated as fractions of the total dust mass, excluding
the dust responsible for the continuum. Note that small grains now indicate the 2.0\,$\mu$m grain sizes and large grains the
4.0\,$\mu$m grains.

85\% of our sources show a mass fraction in large grains
above 0.5. None of the sources have a mass fraction in large grains below 0.25.
Most sources have a mass fraction in crystalline grains between 0.1 and 0.6, strongly centred around 0.3.
There is one clear outlier: IRAS\,10174 with a degree of crystallinity of 0.05. This source shows the more standard
ISM profile, dominated by amorphous grains. 
No clear correlation can be found.

The fraction of large grains in the crystalline component has values between 0.25 and 0.7, while the fraction of large
grains in the amorphous component has values ranging from 0.3 till 0.95. Grain growth appears to be more efficient
in the amorphous dust component. No correlation between grain growth in crystalline and amorphous grains is seen however.

The fraction of enstatite grains in the crystalline component lies between 0.1 and 0.5, showing forsterite to be the dominant crystalline species.
No correlation between the enstatite fraction and crystallinity can be found.

\begin{figure}
\vspace{0cm}
\hspace{0cm}
\resizebox{8cm}{!}{ \includegraphics{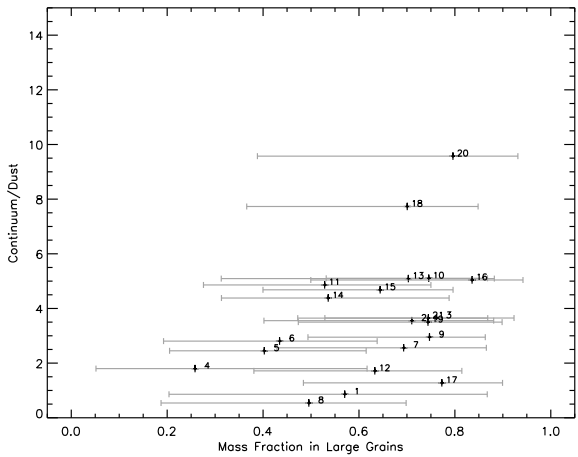}}
\caption{The continuum to dust ratio of the observed spectra plotted against the mass fraction on large grains.}
\label{ratiolarge}%
\end{figure}

In Fig.~\ref{ratiolarge} we plot the continuum to dust luminosity ratio of our observed spectra against the mass fraction in large and
crystalline grains.
A correlation can be found between the continuum/dust ratio and the mass fraction in large grains.
Sources with a high fraction of large grains show a high value for the continuum/dust ratio. This could be expected since larger grains
show less prominent emission features and have a larger continuum contribution. A high value for the continuum/dust ratio also indicates the
presence of even larger grains in the disc. This indicates that the abundance of these larger grains correlate with the abundance of the 4\,$\mu$m grains,
and thus that the size distribution continues beyond 4\,$\mu$m grains.	

\subsubsection{Central star correlations}

By comparing the derived fit parameters with stellar characteristics, like $T_{eff}$, $L_{IR}/L_*$ and the orbital period,
we can investigate possible evolutionary trends in our sample. However, no such correlation between fit parameters and stellar parameters are found.
This was also seen in the SEDs (Sect.~\ref{sed}), where nearly all stars show a similar SED, irrespective of the central star.
Our results do not show an obvious evolutionary trend in the mineralogy of our sample stars.
   
\section{Discussion}
\label{discussion}

The mineralogy of our sample stars show that the dust is purely O-rich.
Amorphous and crystalline silicate dust species prevail and no
features of a C-rich component are found, except the PAH emission
feature seen in EP\,Lyr. 

This is remarkable
since some of our sample stars are thought to have initial masses which would
make them evolve to carbon stars on the AGB on single-star evolutionary tracks.
The lack of third dredge-up is also seen in the two objects with clear CO$_2$ gas emission lines.
EP\,Lyr has a $^{12}$C$/^{13}$C ratio of $\approx 9$ \citep{gonzalez97a}, illustrating that $^{12}$C is not enriched during the preceding AGB evolution.
The internal chemical evolution of our sources seems to have been cut short by binary interaction processes.

EP\,Lyr is the only star in our sample which also has a mixed chemistry, and strongly resembles HD\,44179. 
Like HD\,44179, EP\,Lyr also is strongly depleted \citep{gonzalez97a}. 
PAH emission features dominate the spectrum till 20\,$\mu$m, at larger wavelengths crystalline silicates start to dominate. 
The observed broad and asymmetric emission feature at 8.2\,$\mu$m is very similar to the ``class C'' objects as discussed in \citet{peeters02}.
The famous ``Egg Nebula'' also shows this ``class C'' 8.2\,$\mu$m emission feature.
EP\,Lyr, together with HD\,52961, will be studied in detail in a future contribution.

Some binary post-AGB stars with oxygen-rich discs are known that did evolve to carbon stars, like HD\,44179,
which has a large carbon-rich resolved nebula.
These stars show infrared spectra which are indicative for mixed chemistries, with features of both oxygen-rich and carbon-rich species.
The most likely scenerio for this is that the formation of the O-rich disc antedated the C-rich transition of the central star.
Whether other of our sample stars will also undergo this evolution is still unknown.

Our full spectral fitting indicates a high degree of dust grain processing. 
The dust seems to consist of considerably large grains, with grain sizes larger than 2\,$\mu$m. 
An efficient removal of small grains must have occurred in the discs.
The dust shape is highly irregular, showing that Mie theory is not applicable for the dust in these discs.
The spectra of nearly all stars show a high degree of crystallinity, where Mg-rich end members of olivine and pyroxene silicates dominate. 
The dust condensation sequence of dust in winds of oxygen-rich AGB stars predicts the formation of aluminium- or calcium- rich dust grains, like
corundum (Al$_2$O$_3$), spinel (MgAl$_2$O$_4$) or anorthite (CaAl$_2$Si$_2$O$_8$) \citep{tielens97,cami02}.
There is however no evidence for the presence of these dust species.

Most features are well reproduced and only the 14.7 and 32.5\,$\mu$m features remain unidentified (Sect.~\ref{mean_appendix}). 
Diopside (CaMgSiO$_3$) could be a possible 
candidate (which has a feature around 14.7\,$\mu$m and a weak feature around 32.1\,$\mu$m). However, for diopside only mass absorption coefficients 
for very small grains are available, so we cannot include it in the fitting procedure in a homogeneous way.

The dust is highly magnesium-rich, leaving a large fraction of iron unaccounted for.
Previous studies \citep{gielen07} suggest the iron may be locked in the form of metallic iron. 
Photospheric depletion in iron, which we detect in our sample stars \citep{maas02,vanwinckel98},
can be understood when the iron is locked up in the circumstellar dust \citep{waters92}.
The lack of iron in the detected silicates is therefore surprising. If both the crystalline and amorphous silicates are devoid of
iron, this could mean that iron is stored in metallic iron or iron-oxide
\citep{sofia06}. Metallic iron has no distinct features but still makes a
significant contribution in opacity, especially at shorter wavelengths, making it very hard to detect directly.

In our fitting method, for both dust temperatures equal silicate fractions are used.
This seems to fit features indicative of both hot and cool crystalline dust, meaning we do not have
a strong radial gradient in crystallinity throughout the disc. 
The inner regions of the discs, with temperatures above the annealing temperature, are expected to be fully crystalline.
The presence of a considerable amount of cool crystalline grains implies thus that we have strong turbulent mixing in the disc or that
the crystalline grains were already abundant at disc formation.
It is interesting to note that the very young circumbinary disc in the evolved binary SS\,Lep \citep{verhoelst07}, 
which may be formed in a similar process as the discs around our post-AGB sources, is
dominated by small and large amorphous grains. The crystalline component is very small \citep{schutz05}.
This would indicate that the crystallisation process happens during disc
evolution and is not already present at formation.

The evolution of these discs is still unknown. Our analysis shows no 
clear correlation between dust parameters and any fundamental parameter (T$_{eff}$, orbital period) 
of the central star. In a recent study of \citet{chesneau07}, a compact dusty disc was discovered in ``the Ant'',
a well studied bipolar planetary nebula. Interferometric MIDI observations provided evidence for a flat, nearly edge-on disc,
primarily composed of amorphous silicates. This is in contrast with the high crystallinity observed in the discs around binary
post-AGB stars, suggesting that the disc in the Ant is relatively young. Whether there is a link between this disc and the post-AGB discs
remains unclear.

Some degeneracy is present in our spectral fits. We performed the spectral fitting both with pure Mg-rich amorphous silicates ($x=1$) 
and with the more standard Mg-Fe amorphous silicate dust ($x=0.5$). Both models often have similar $\chi^2$ values and the best model
strongly varies from star to star. Some stars also show equally well-fitting models, when using 0.1-2.0\,$\mu$m or 2.0-4.0\,$\mu$m grain sizes.

\subsection{Comparison with young stellar objects}

The mineralogy of the observed spectra shows a striking resemblance to the infrared spectra of 
young stellar objects, like Herbig Ae/Be stars \citep{lisse07,vanboekel05} or T\,Tauri stars \citep{watson07}, and primitive comets such as Hale-Bopp \citep{lisse07,bouwman03,min05b}.
There also, amorphous silicates and Mg-rich crystalline features prevail.  
The dusty disc is the relic of the star formation process so both silicates together with a carbon-rich
component in the form of PAH emission are often detected, depending on the disc geometry \citep{acke04}.

We compare our findings with the ones discussed in \citet{vanboekel05}. They present spectroscopic
observations of a large sample of Herbig Ae/Be stars in the 10\,$\mu$m region. Similar studies on young
stellar objects have also been done by e.g. \citet{bouwman01}.
One has to be careful comparing studies, since our study includes a far larger wavelength range,
and an exact comparison is thus not possible.

The degree of crystallinity found in the Herbig Ae/Be stars is clearly smaller than for
our sample of post-AGB binaries. The Herbig stars all show a degree of crystallinity below 0.35,
whereas our stars have significantly higher values. \citet{vanboekel05} use both small (0.1\,$\mu$m) and large (1.5\,$\mu$m)
grains to fit the observed emission features and find that most sources have a mass fraction in large grains 
of more than 80\%.

For both young and evolved objects a substantial removal of the smallest grains has occurred,
but seems to be more efficient in the discs around the evolved stars. A similar physical process might
be responsible for the observed grain size distribution in both cases, like aggregation of small grains 
or the removal of small grains by radiation pressure.
Since several processes occur in the formation of the disc it could also be that silicate grains
in the post-AGB disc are formed at large grain sizes. These large grains could be broken up by 
grain interactions in the disc, producing a small fraction of small grains.

\citet{vanboekel05} found a clear correlation between the crystallinity and the dominance of enstatite
or forsterite in these grains. For sources with a high degree of crystallinity most crystalline grains appear to be in
the form of enstatite, while for sources with a low crystallinity, forsterite seems to be the dominant species.
For all our sample stars forsterite is the dominant species, despite the high crystallinity factors.
This difference could relate to the initial dust species when the disc is formed. For discs around YSO, which are formed from the ISM, the abundant dust species
is likely amorphous olivine. Forsterite is expected to be the dominant species formed by thermal annealing, while enstatite
is expected to be the dominant species formed by chemical equilibrium processes in most of the inner disc.
Our discs seem to prefer the formation of forsterite, but the initial dust species is not known.
The innermost disc regions are hot enough to crystallise dust, but the very high degree of crystallinity seems to point
to the presence of another crystallisation processes, possibly at disc formation.

\section{Conclusions}
\label{conclusions}

We present high-resolution TIMMI2 and SPITZER infrared spectra of 21 binary post-AGB stars surrounded by
a stable Keplerian disc. We summarise our main conclusions:

\begin{itemize}

\item
Almost all discs display only O-rich spectral signatures. The noticeable exception is EP\,Lyr, which
shows a very similar spectrum as the central star of the Red Rectangle. PAH emission features dominate the spectrum
till 20\,$\mu$m, at larger wavelengths crystalline silicates start to dominate.

\item 
Our mineralogy study indicates the dominance
of Mg-rich amorphous and crystalline silicate dust in the disc. 
The high crystallinity and the large fraction of large grains, as deduced from our full spectral fitting, show strong
dust grain processing in the discs. 

\item 
The temperature estimates from our fitting routine show that a significant fraction of crystalline grains must be cool. 
This shows that radial mixing is efficient is these discs or indicate a different thermal history at grain formation.

\item 
Trend analysis of our fitting parameters show no clear correlation with stellar characteristics. 
For the moment it is not clear if and how the observed diversity in observed spectra relates to specific structural elements
of the disc, the star and/or the orbits or whether we witness directly an evolutionary change between different sources.

\end{itemize}

To further improve our understanding of these circumbinary discs, as a next step, we will combine our photometric and spectroscopic data with interferometric measurements.
Comparing spatial information from the MIDI and AMBER interferometric instruments with a realistic disc model \citep{dullemond04},
constrained by photometric and spectroscopic data, will allow us, not only to study the mineralogy, but also the structure of the discs.
So far, interferometric measurements for five of our sample stars have been obtained \citep{deroo06,deroo07c}.

Probing the dust processing in the discs around evolved objects proves to be an excellent complement to study physics in planet-forming
young discs.

\begin{acknowledgements}
The authors want to acknowledge: the 1.2\,m Mercator staff
as well as the observers from the Instituut voor Sterrenkunde who contributed
to the monitoring observations using the Mercator telescope.
CG acknowledges support of the Fund for Scientific Research of Flanders
(FWO) under the grant G.0178.02. and G.0470.07.
We also thank Fred Lahuis for his assistance with the SPITZER data reduction.
\end{acknowledgements}

\bibliographystyle{aa}
\bibliography{0053bib.bib}

\onecolumn
\begin{appendix}
\section{Tables}

\begin{table}[h]
\caption{ Best fit parameters deduced from our full spectral fitting.
Listed the $\chi^2$, dust and continuum temperatures and their relative fractions.}
\label{fitresults1}
\centering
\begin{tabular}{llrllllll}
\hline \hline
N$^\circ$ & Name & $\chi^2$ &  $T_{dust1}$ & $T_{dust2}$ & Fraction & $T_{cont1}$ & $T_{cont2}$ & Fraction \\
       &      &          &     (K)     & (K)         & $T_{dust1}$- $T_{dust2}$    & (K)         & (K)         & $T_{cont1}$-$T_{cont2}$   \\
\hline
  1 &EP\,Lyr  & 56.1 &$ 116_{  16}^{  97}$ &$ 237_{  37}^{ 277}$ &$ 0.80_{ 0.30}^{ 0.10}- 0.20_{ 0.10}^{ 0.30}$ &$ 195_{ 100}^{  45}$ &$ 921_{ 342}^{  80}$ &$ 0.98_{ 0.02}^{ 0.01}- 0.02_{ 0.01}^{ 0.02}$\\
  2 &HD\,131356  & 3.5 &$ 226_{  46}^{ 269}$ &$ 928_{ 187}^{  73}$ &$ 0.80_{ 0.50}^{ 0.10}- 0.20_{ 0.10}^{ 0.50}$ &$ 195_{ 104}^{  12}$ &$ 534_{  50}^{  95}$ &$ 0.92_{ 0.02}^{ 0.03}- 0.08_{ 0.03}^{ 0.02}$\\
  3 &HD\,213985  & 4.1 &$ 185_{  87}^{  45}$ &$ 926_{ 333}^{  74}$ &$ 0.80_{ 0.40}^{ 0.10}- 0.20_{ 0.10}^{ 0.40}$ &$ 187_{  91}^{  13}$ &$ 838_{ 162}^{  73}$ &$ 0.98_{ 0.00}^{ 0.01}- 0.02_{ 0.01}^{ 0.00}$\\
  4 &HD\,52961  & 72.2 &$ 199_{ 140}^{  10}$ &$ 845_{ 106}^{ 121}$ &$ 0.90_{ 0.70}^{ 0.00}- 0.10_{ 0.00}^{ 0.70}$ &$ 109_{   9}^{ 497}$ &$ 998_{ 139}^{   2}$ &$ 0.99_{ 0.03}^{ 0.00}- 0.01_{ 0.00}^{ 0.03}$\\
  5 &IRAS\,05208  & 4.5 &$ 291_{ 101}^{ 176}$ &$ 904_{ 124}^{  97}$ &$ 0.80_{ 0.30}^{ 0.10}- 0.20_{ 0.10}^{ 0.30}$ &$ 177_{  79}^{  30}$ &$ 374_{  76}^{  26}$ &$ 0.84_{ 0.03}^{ 0.02}- 0.16_{ 0.02}^{ 0.03}$\\
  6 &IRAS\,09060  & 3.6 &$ 206_{   6}^{ 176}$ &$ 755_{ 163}^{ 140}$ &$ 0.90_{ 0.30}^{ 0.00}- 0.10_{ 0.00}^{ 0.30}$ &$ 225_{ 122}^{  86}$ &$ 805_{ 198}^{ 162}$ &$ 0.93_{ 0.02}^{ 0.02}- 0.07_{ 0.02}^{ 0.02}$\\
  7 &IRAS\,09144  & 6.1 &$ 212_{  24}^{  91}$ &$ 535_{  78}^{ 137}$ &$ 0.90_{ 0.20}^{ 0.00}- 0.10_{ 0.00}^{ 0.20}$ &$ 196_{ 105}^{  16}$ &$ 767_{  89}^{  89}$ &$ 0.94_{ 0.02}^{ 0.02}- 0.06_{ 0.02}^{ 0.02}$\\
  8 &IRAS\,10174  & 13.9 &$ 234_{ 125}^{  72}$ &$ 378_{  79}^{  47}$ &$ 0.70_{ 0.40}^{ 0.20}- 0.30_{ 0.20}^{ 0.40}$ &$ 125_{  25}^{  76}$ &$ 322_{  22}^{  95}$ &$ 0.96_{ 0.04}^{ 0.02}- 0.04_{ 0.02}^{ 0.04}$\\
  9 &IRAS\,16230  & 4.9 &$ 210_{  22}^{ 143}$ &$ 507_{  74}^{ 182}$ &$ 0.90_{ 0.20}^{ 0.00}- 0.10_{ 0.00}^{ 0.20}$ &$ 151_{  51}^{ 154}$ &$ 592_{  93}^{ 286}$ &$ 0.94_{ 0.03}^{ 0.01}- 0.06_{ 0.01}^{ 0.03}$\\
 10 &IRAS\,17038  & 2.9 &$ 250_{  69}^{ 140}$ &$ 854_{ 192}^{ 120}$ &$ 0.80_{ 0.20}^{ 0.10}- 0.20_{ 0.10}^{ 0.20}$ &$ 198_{ 139}^{   2}$ &$ 563_{  85}^{  84}$ &$ 0.96_{ 0.02}^{ 0.01}- 0.04_{ 0.01}^{ 0.02}$\\
 11 &IRAS\,17243  & 2.3 &$ 210_{  38}^{ 110}$ &$ 441_{  79}^{  89}$ &$ 0.80_{ 0.50}^{ 0.10}- 0.20_{ 0.10}^{ 0.50}$ &$ 202_{  15}^{ 113}$ &$ 633_{  39}^{ 168}$ &$ 0.91_{ 0.01}^{ 0.03}- 0.09_{ 0.03}^{ 0.01}$\\
 12 &IRAS\,19125  & 3.9 &$ 102_{   2}^{ 139}$ &$ 204_{   4}^{ 277}$ &$ 0.90_{ 0.10}^{ 0.00}- 0.10_{ 0.00}^{ 0.10}$ &$ 321_{ 138}^{ 169}$ &$ 702_{  84}^{ 126}$ &$ 0.84_{ 0.04}^{ 0.04}- 0.16_{ 0.04}^{ 0.04}$\\
 13 &IRAS\,19157  & 5.5 &$ 199_{ 111}^{  26}$ &$ 609_{ 159}^{ 176}$ &$ 0.90_{ 0.40}^{ 0.00}- 0.10_{ 0.00}^{ 0.40}$ &$ 186_{  89}^{  14}$ &$ 689_{ 137}^{  85}$ &$ 0.96_{ 0.02}^{ 0.01}- 0.04_{ 0.01}^{ 0.02}$\\
 14 &IRAS\,20056  & 3.8 &$ 118_{  18}^{  84}$ &$ 219_{  19}^{  83}$ &$ 0.60_{ 0.60}^{ 0.20}- 0.40_{ 0.20}^{ 0.60}$ &$ 238_{  51}^{  87}$ &$ 658_{  72}^{ 119}$ &$ 0.88_{ 0.04}^{ 0.01}- 0.12_{ 0.01}^{ 0.04}$\\
 15 &RU\,Cen  & 3.4 &$ 233_{  67}^{  87}$ &$ 517_{ 115}^{ 200}$ &$ 0.80_{ 0.40}^{ 0.10}- 0.20_{ 0.10}^{ 0.40}$ &$ 199_{   0}^{   1}$ &$ 545_{ 125}^{  77}$ &$ 0.98_{ 0.02}^{ 0.01}- 0.02_{ 0.01}^{ 0.02}$\\
 16 &SAO\,173329  & 3.1 &$ 205_{  36}^{ 160}$ &$ 764_{ 133}^{ 179}$ &$ 0.80_{ 0.50}^{ 0.10}- 0.20_{ 0.10}^{ 0.50}$ &$ 195_{ 106}^{   5}$ &$ 581_{ 100}^{  23}$ &$ 0.93_{ 0.02}^{ 0.01}- 0.07_{ 0.01}^{ 0.02}$\\
 17 &ST\,Pup  & 8.4 &$ 207_{  13}^{  99}$ &$ 480_{  86}^{  86}$ &$ 0.80_{ 0.20}^{ 0.10}- 0.20_{ 0.10}^{ 0.20}$ &$ 207_{  13}^{  99}$ &$ 524_{  57}^{ 189}$ &$ 0.95_{ 0.03}^{ 0.02}- 0.05_{ 0.02}^{ 0.03}$\\
 18 &SU\,Gem  & 1.8 &$ 241_{  98}^{ 336}$ &$ 629_{ 184}^{ 284}$ &$ 0.80_{ 0.30}^{ 0.10}- 0.20_{ 0.10}^{ 0.30}$ &$ 177_{  79}^{  23}$ &$ 763_{ 192}^{ 137}$ &$ 0.95_{ 0.02}^{ 0.02}- 0.05_{ 0.02}^{ 0.02}$\\
 19 &SX\,Cen  & 4.3 &$ 225_{  44}^{ 202}$ &$ 918_{ 183}^{  83}$ &$ 0.80_{ 0.30}^{ 0.10}- 0.20_{ 0.10}^{ 0.30}$ &$ 191_{  96}^{   9}$ &$ 659_{  80}^{  95}$ &$ 0.94_{ 0.02}^{ 0.02}- 0.06_{ 0.02}^{ 0.02}$\\
 20 &TW\,Cam  & 2.3 &$ 216_{  82}^{ 123}$ &$ 390_{ 171}^{ 107}$ &$ 0.70_{ 0.40}^{ 0.20}- 0.30_{ 0.20}^{ 0.40}$ &$ 131_{  31}^{ 321}$ &$ 546_{  46}^{ 205}$ &$ 0.95_{ 0.02}^{ 0.01}- 0.05_{ 0.01}^{ 0.02}$\\
 21 &UY\,CMa  & 2.9 &$ 205_{  12}^{ 104}$ &$ 761_{ 113}^{ 131}$ &$ 0.90_{ 0.00}^{ 0.00}- 0.10_{ 0.00}^{ 0.00}$ &$ 203_{   3}^{ 119}$ &$ 511_{  11}^{ 126}$ &$ 0.84_{ 0.02}^{ 0.03}- 0.16_{ 0.03}^{ 0.02}$\\

\hline
\end{tabular}
\end{table}
   
\begin{table}[h]
\caption{Best fit parameters deduced from our full spectral fitting. The abundances of small (2.0\,$\mu$m) and large (4.0\,$\mu$m) grains of the various
dust species are given as fractions of the total mass, excluding the dust responsible for the continuum emission.
The last column gives the continuum flux contribution, listed as a percentage of the total integrated flux over the 
full wavelength range.}
\label{fitresults2}
\centering
\begin{tabular}{llccccc}
\hline \hline   
 N$^\circ$ & Name & Olivine & Pyroxene & Forsterite & Enstatite & Continuum\\
           &      & Small  -  Large & Small  -  Large &  Small  -   Large & Small  -   Large &\\
\hline
  1    &EP\,Lyr    &$ 0.17_{ 0.17}^{ 7.59}    -   5.61_{ 5.61}^{46.82}$    &$19.22_{17.10}^{34.23}    -  26.98_{24.42}^{40.96}$    &$20.42_{14.25}^{32.78}    -   7.80_{ 7.46}^{27.63}$    &$ 3.17_{ 3.13}^{26.21}    -  16.63_{14.58}^{29.16}$    &$58.26_{ 8.63}^{ 8.69}$\\
  2    &HD\,131356    &$ 1.58_{ 1.57}^{25.29}    -  29.75_{21.09}^{15.77}$    &$12.99_{11.34}^{18.61}    -  30.47_{25.35}^{21.47}$    &$13.93_{ 7.29}^{10.91}    -   3.63_{ 3.57}^{18.73}$    &$ 0.49_{ 0.49}^{ 5.85}    -   7.17_{ 6.82}^{17.49}$    &$77.78_{ 2.72}^{ 3.10}$\\
  3    &HD\,213985    &$ 1.51_{ 1.52}^{21.49}    -  28.81_{21.16}^{14.62}$    &$14.12_{12.08}^{15.94}    -  31.37_{22.90}^{27.40}$    &$ 7.64_{ 5.58}^{10.20}    -   8.50_{ 7.60}^{17.33}$    &$ 0.55_{ 0.56}^{ 6.53}    -   7.50_{ 6.31}^{10.07}$    &$75.05_{ 5.53}^{ 3.51}$\\
  4    &HD\,52961    &$ 0.21_{ 0.21}^{15.75}    -   0.59_{ 0.59}^{45.99}$    &$54.40_{30.88}^{16.42}    -   7.00_{ 7.04}^{55.05}$    &$17.33_{12.99}^{15.22}    -  11.53_{10.64}^{22.97}$    &$ 2.26_{ 2.26}^{21.87}    -   6.69_{ 6.49}^{32.31}$    &$64.91_{ 6.83}^{ 4.75}$\\
  5    &IRAS\,05208    &$ 1.91_{ 1.90}^{16.88}    -   7.39_{ 6.59}^{12.81}$    &$29.09_{13.86}^{11.19}    -   9.42_{ 8.97}^{25.21}$    &$23.04_{10.05}^{ 8.08}    -   2.70_{ 2.69}^{17.47}$    &$ 5.70_{ 5.25}^{10.35}    -  20.75_{15.60}^{13.58}$    &$67.64_{ 4.27}^{ 2.74}$\\
  6    &IRAS\,09060    &$ 7.20_{ 7.04}^{24.66}    -  19.67_{17.20}^{16.68}$    &$34.08_{18.50}^{13.73}    -   9.12_{ 9.02}^{32.11}$    &$14.03_{ 6.33}^{ 7.88}    -   5.10_{ 4.97}^{15.99}$    &$ 1.25_{ 1.24}^{10.01}    -   9.55_{ 7.17}^{10.30}$    &$71.99_{ 3.46}^{ 2.60}$\\
  7    &IRAS\,09144    &$ 1.05_{ 1.06}^{23.75}    -  32.86_{19.81}^{15.70}$    &$20.98_{14.76}^{21.45}    -  30.25_{23.38}^{27.10}$    &$ 8.39_{ 5.42}^{ 8.05}    -   2.96_{ 2.89}^{10.25}$    &$ 0.26_{ 0.26}^{ 6.45}    -   3.23_{ 3.18}^{10.74}$    &$72.99_{ 3.25}^{ 3.48}$\\
  8    &IRAS\,10174    &$21.58_{18.72}^{27.48}    -  28.94_{24.31}^{20.10}$    &$27.36_{15.16}^{15.22}    -  17.81_{15.55}^{21.52}$    &$ 0.85_{ 0.84}^{ 6.88}    -   1.70_{ 1.66}^{ 9.53}$    &$ 0.68_{ 0.68}^{ 8.72}    -   1.09_{ 1.09}^{ 7.09}$    &$35.70_{ 5.72}^{ 6.39}$\\
  9    &IRAS\,16230    &$ 3.13_{ 3.13}^{30.37}    -  43.75_{30.34}^{19.08}$    &$ 4.15_{ 3.96}^{12.04}    -  16.56_{15.24}^{18.26}$    &$18.02_{ 8.89}^{11.45}    -   9.20_{ 7.55}^{22.74}$    &$ 0.00_{ 0.00}^{ 0.00}    -   5.19_{ 4.95}^{20.91}$    &$77.27_{ 2.66}^{ 4.17}$\\
 10    &IRAS\,17038    &$ 0.79_{ 0.80}^{56.99}    -  22.98_{17.86}^{19.89}$    &$ 4.02_{ 3.95}^{13.05}    -  24.56_{17.78}^{18.20}$    &$19.33_{11.24}^{11.07}    -  10.25_{ 9.80}^{23.93}$    &$ 1.31_{ 1.31}^{ 8.97}    -  16.76_{12.88}^{18.14}$    &$81.62_{ 2.78}^{ 2.00}$\\
 11    &IRAS\,17243    &$ 8.62_{ 8.41}^{31.13}    -  31.99_{24.71}^{17.51}$    &$22.07_{13.70}^{16.69}    -  13.14_{12.05}^{28.77}$    &$16.36_{ 8.44}^{ 9.05}    -   1.27_{ 1.25}^{15.32}$    &$ 0.12_{ 0.12}^{ 4.41}    -   6.43_{ 5.36}^{13.45}$    &$83.09_{ 2.53}^{ 1.90}$\\
 12    &IRAS\,19125    &$10.68_{ 9.79}^{24.63}    -   9.13_{ 8.38}^{19.94}$    &$15.45_{11.32}^{21.11}    -  27.57_{21.85}^{25.18}$    &$ 9.04_{ 5.23}^{ 6.52}    -  10.35_{ 6.74}^{12.02}$    &$ 1.54_{ 1.53}^{11.44}    -  16.24_{ 9.74}^{11.04}$    &$72.70_{ 3.23}^{ 3.73}$\\
 13    &IRAS\,19157    &$ 9.14_{ 9.12}^{40.25}    -  48.99_{36.60}^{20.74}$    &$ 9.67_{ 7.79}^{13.47}    -  12.17_{11.69}^{25.87}$    &$10.71_{ 7.70}^{ 9.97}    -   5.13_{ 5.00}^{16.70}$    &$ 0.18_{ 0.18}^{ 8.99}    -   4.02_{ 3.78}^{14.16}$    &$80.17_{ 4.80}^{ 2.85}$\\
 14    &IRAS\,20056    &$ 6.12_{ 6.08}^{27.24}    -  20.84_{17.18}^{23.73}$    &$27.05_{19.61}^{20.02}    -  14.01_{13.25}^{35.34}$    &$11.59_{ 7.25}^{ 9.43}    -   5.04_{ 4.86}^{15.70}$    &$ 1.67_{ 1.67}^{10.66}    -  13.69_{11.05}^{15.29}$    &$81.36_{ 3.32}^{ 2.73}$\\
 15    &RU\,Cen    &$ 0.98_{ 0.98}^{16.00}    -  22.37_{17.52}^{16.23}$    &$ 5.74_{ 5.32}^{11.49}    -  19.03_{16.49}^{17.47}$    &$27.79_{11.46}^{14.30}    -  10.31_{ 8.72}^{21.44}$    &$ 1.07_{ 1.07}^{12.34}    -  12.70_{10.10}^{17.09}$    &$80.23_{ 3.34}^{ 2.89}$\\
 16    &SAO\,173329    &$ 1.07_{ 1.07}^{23.25}    -  41.52_{28.71}^{18.10}$    &$ 4.19_{ 4.15}^{20.45}    -  20.93_{16.77}^{18.36}$    &$ 9.10_{ 7.03}^{11.36}    -   7.62_{ 7.27}^{24.12}$    &$ 2.06_{ 2.04}^{25.12}    -  13.52_{10.40}^{21.78}$    &$83.30_{ 2.29}^{ 2.81}$\\
 17    &ST\,Pup    &$ 2.28_{ 2.29}^{32.70}    -  26.07_{17.27}^{15.63}$    &$ 9.33_{ 8.53}^{18.19}    -  36.91_{27.32}^{18.59}$    &$10.69_{ 6.94}^{ 6.40}    -   9.14_{ 8.28}^{19.19}$    &$ 0.43_{ 0.43}^{11.60}    -   5.15_{ 4.49}^{14.65}$    &$56.07_{ 4.14}^{ 4.32}$\\
 18    &SU\,Gem    &$ 3.52_{ 3.51}^{43.03}    -  45.56_{32.84}^{21.97}$    &$ 4.68_{ 4.46}^{20.74}    -  12.39_{11.23}^{21.56}$    &$21.23_{11.60}^{16.84}    -   5.50_{ 5.22}^{19.78}$    &$ 0.51_{ 0.51}^{ 9.92}    -   6.61_{ 5.97}^{17.57}$    &$86.85_{ 6.38}^{ 2.62}$\\
 19    &SX\,Cen    &$ 3.69_{ 3.70}^{31.56}    -  40.53_{23.30}^{16.94}$    &$12.07_{10.11}^{15.15}    -  23.07_{20.93}^{22.39}$    &$ 9.00_{ 5.24}^{ 7.35}    -   4.15_{ 4.08}^{16.22}$    &$ 0.84_{ 0.84}^{ 7.80}    -   6.64_{ 5.77}^{12.72}$    &$75.85_{ 3.51}^{ 3.10}$\\
 20    &TW\,Cam    &$ 2.26_{ 2.27}^{44.45}    -  56.52_{44.46}^{26.81}$    &$ 0.02_{ 0.02}^{ 0.00}    -   4.65_{ 4.63}^{38.76}$    &$17.29_{11.45}^{33.38}    -   8.64_{ 8.50}^{40.91}$    &$ 0.82_{ 0.82}^{13.00}    -   9.79_{ 8.79}^{35.32}$    &$90.61_{ 2.14}^{ 2.12}$\\
 21    &UY\,CMa    &$ 0.94_{ 0.94}^{30.95}    -  14.75_{11.31}^{14.58}$    &$ 8.63_{ 7.64}^{18.69}    -  45.36_{25.12}^{17.50}$    &$15.63_{ 8.53}^{ 9.36}    -   8.54_{ 7.81}^{16.88}$    &$ 0.33_{ 0.33}^{ 5.66}    -   5.82_{ 5.06}^{15.94}$    &$79.11_{ 1.97}^{ 2.10}$\\

\hline
\end{tabular}
\end{table}  

\section{Figures}

\begin{figure}[h]
\vspace{0cm}
\hspace{0cm}
\resizebox{6cm}{!}{\includegraphics{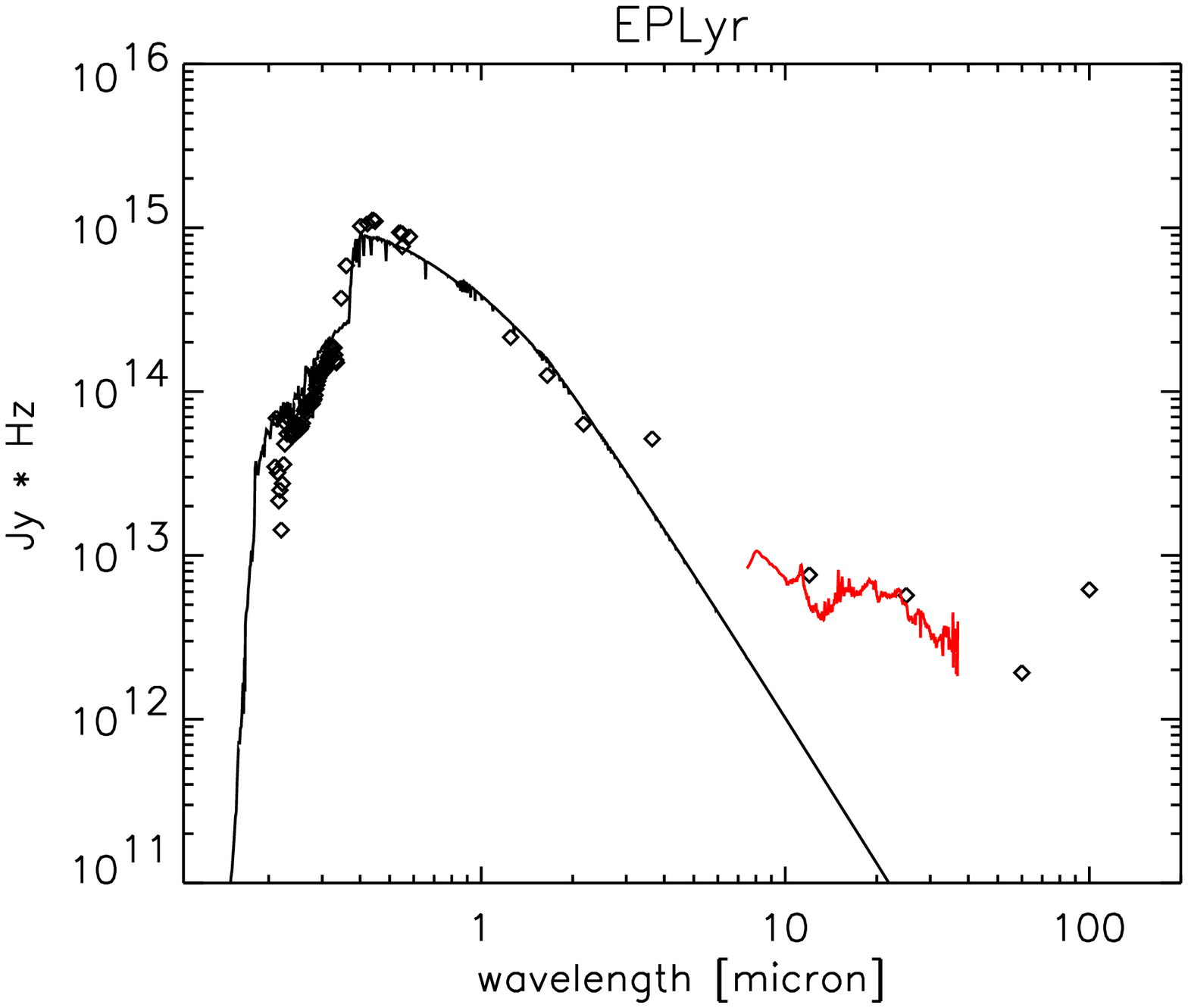}}
\resizebox{6cm}{!}{\includegraphics{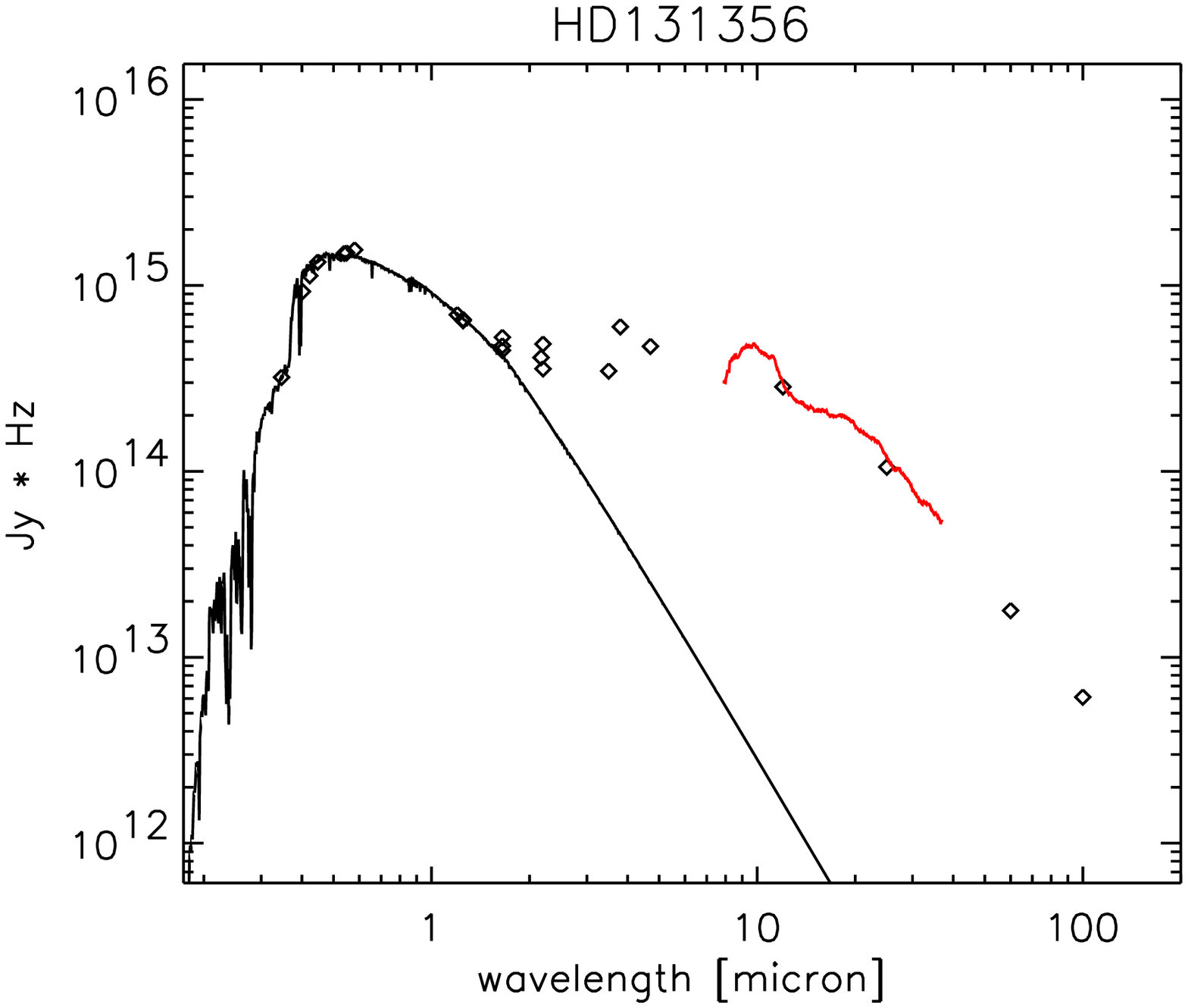}}
\resizebox{6cm}{!}{\includegraphics{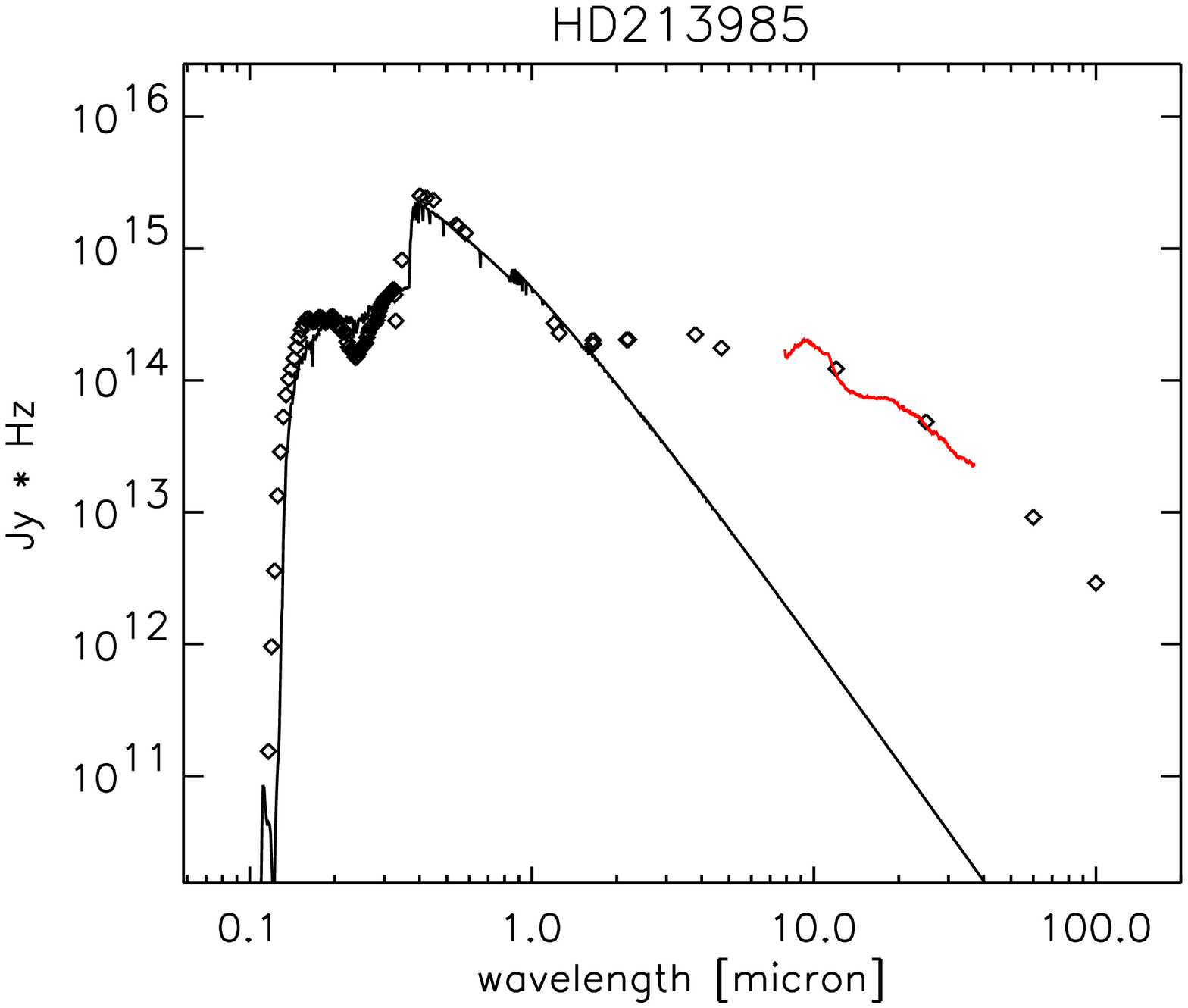}}
\resizebox{6cm}{!}{\includegraphics{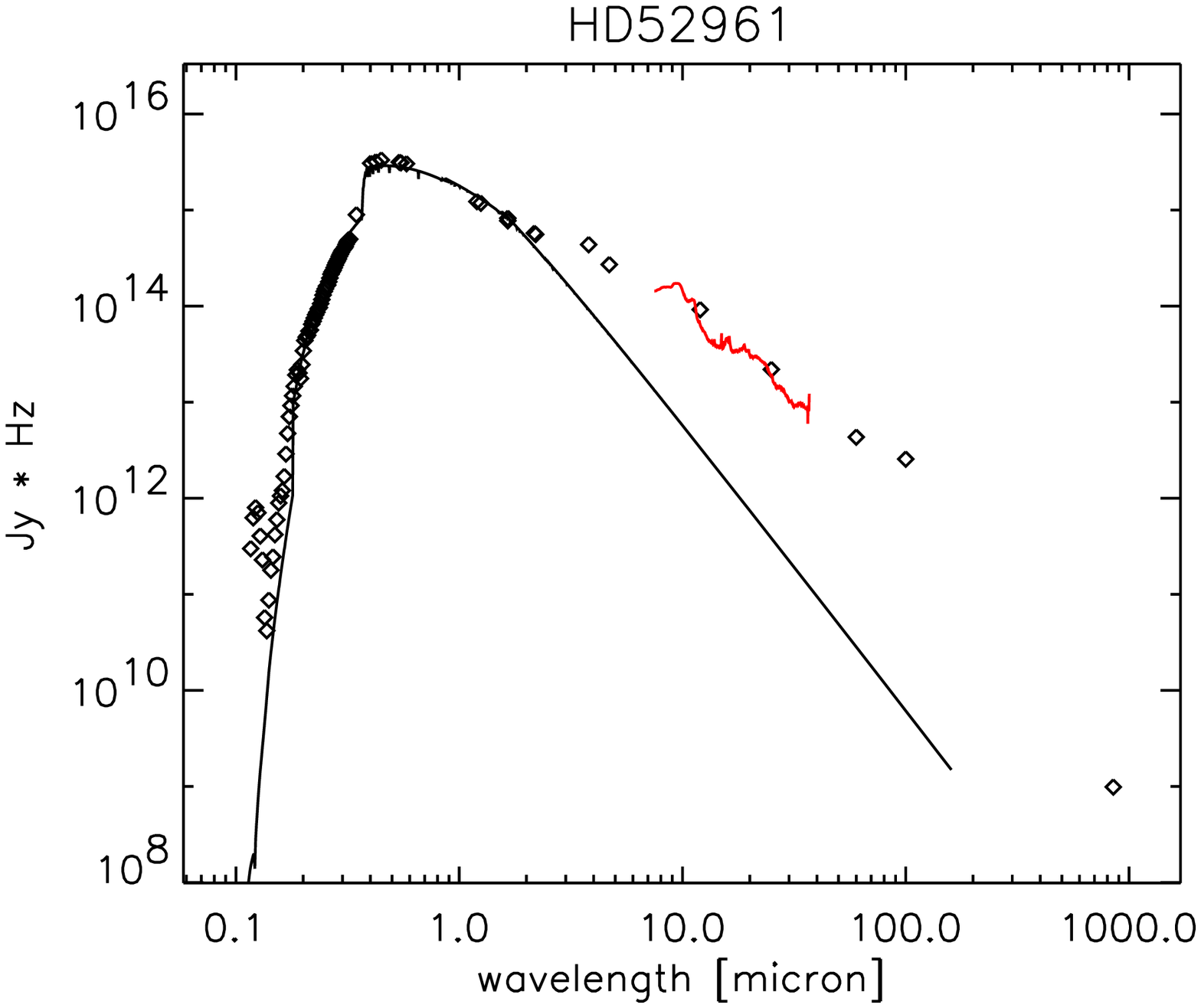}}
\resizebox{6cm}{!}{\includegraphics{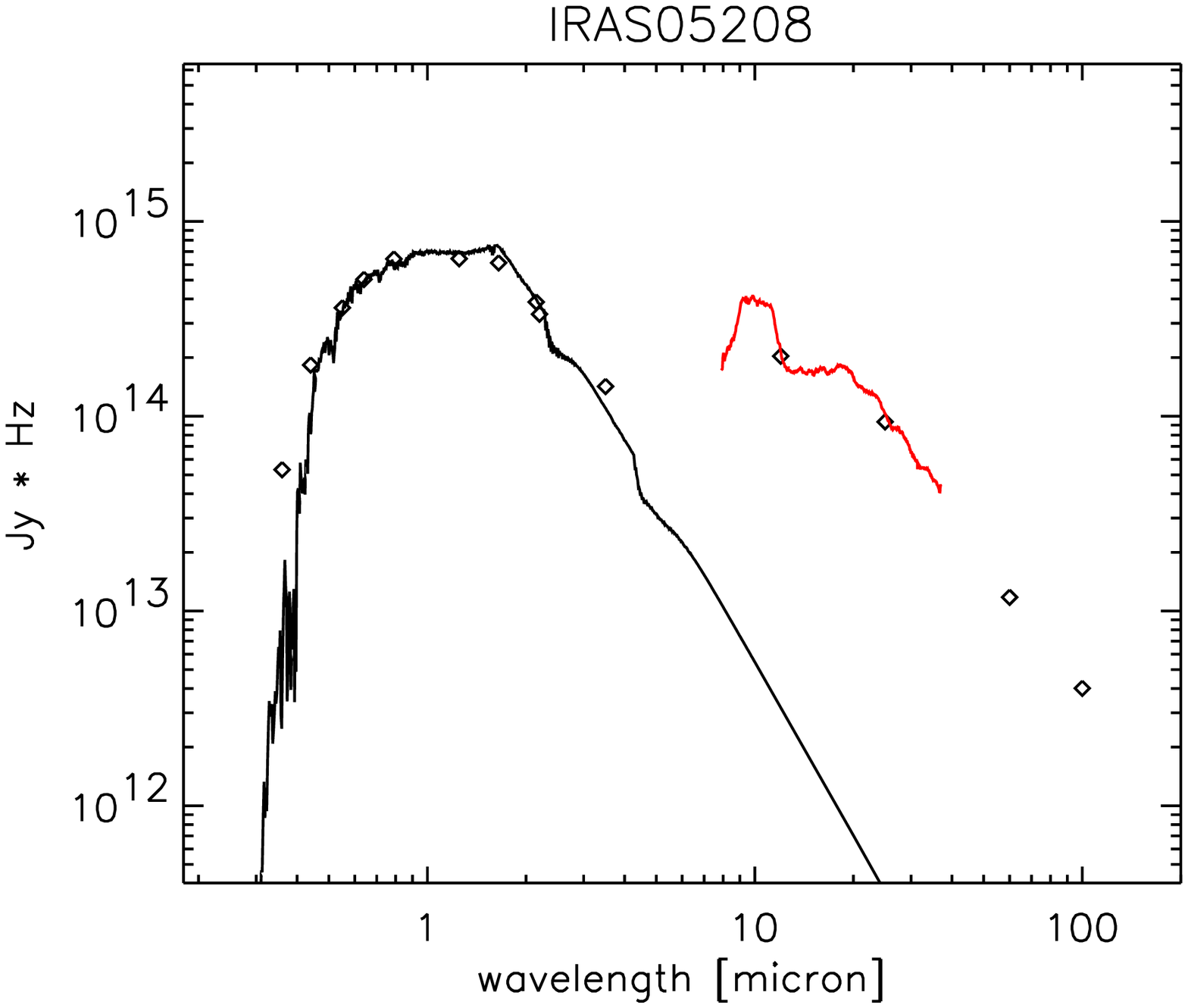}}
\resizebox{6cm}{!}{\includegraphics{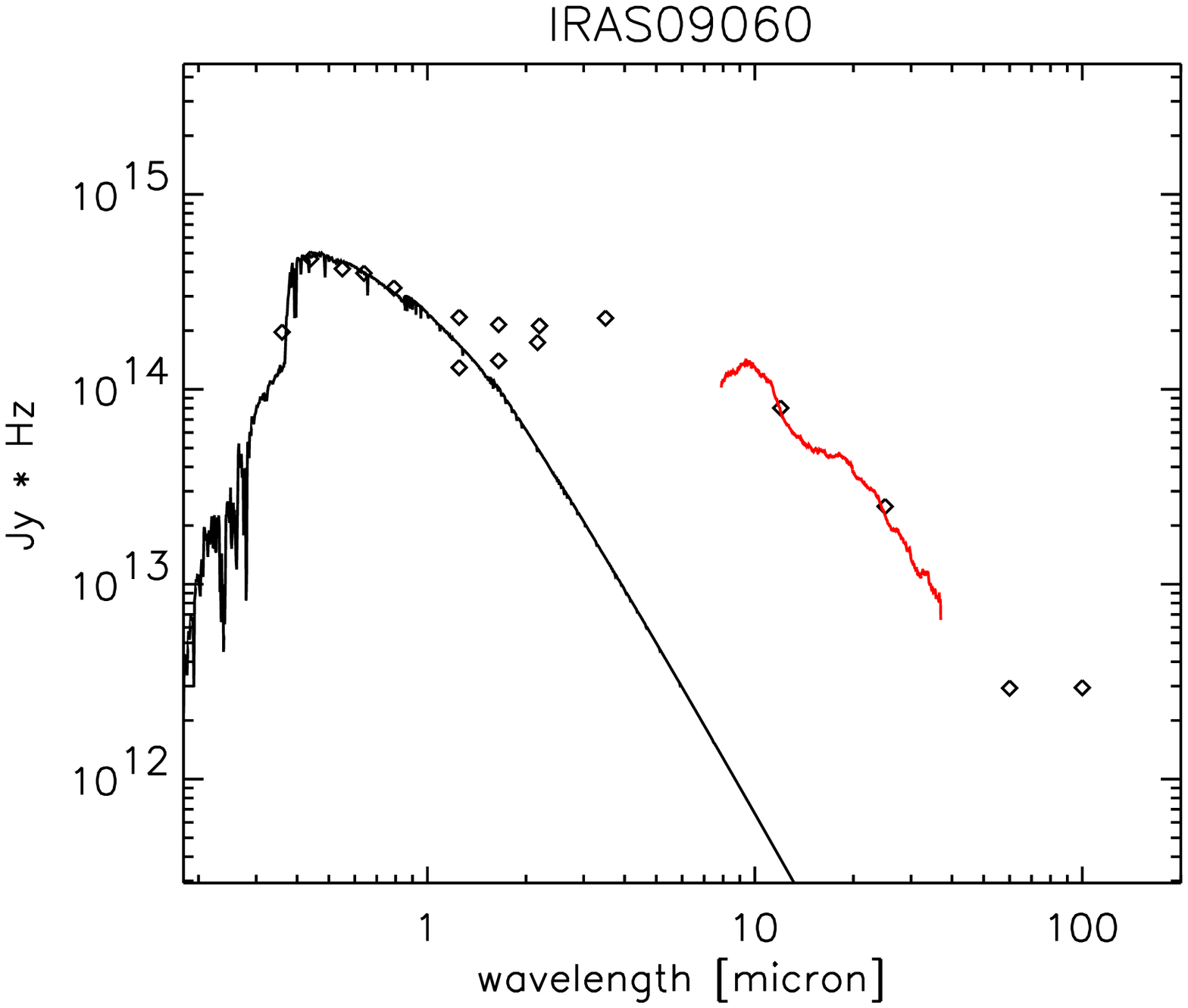}}
\resizebox{6cm}{!}{\includegraphics{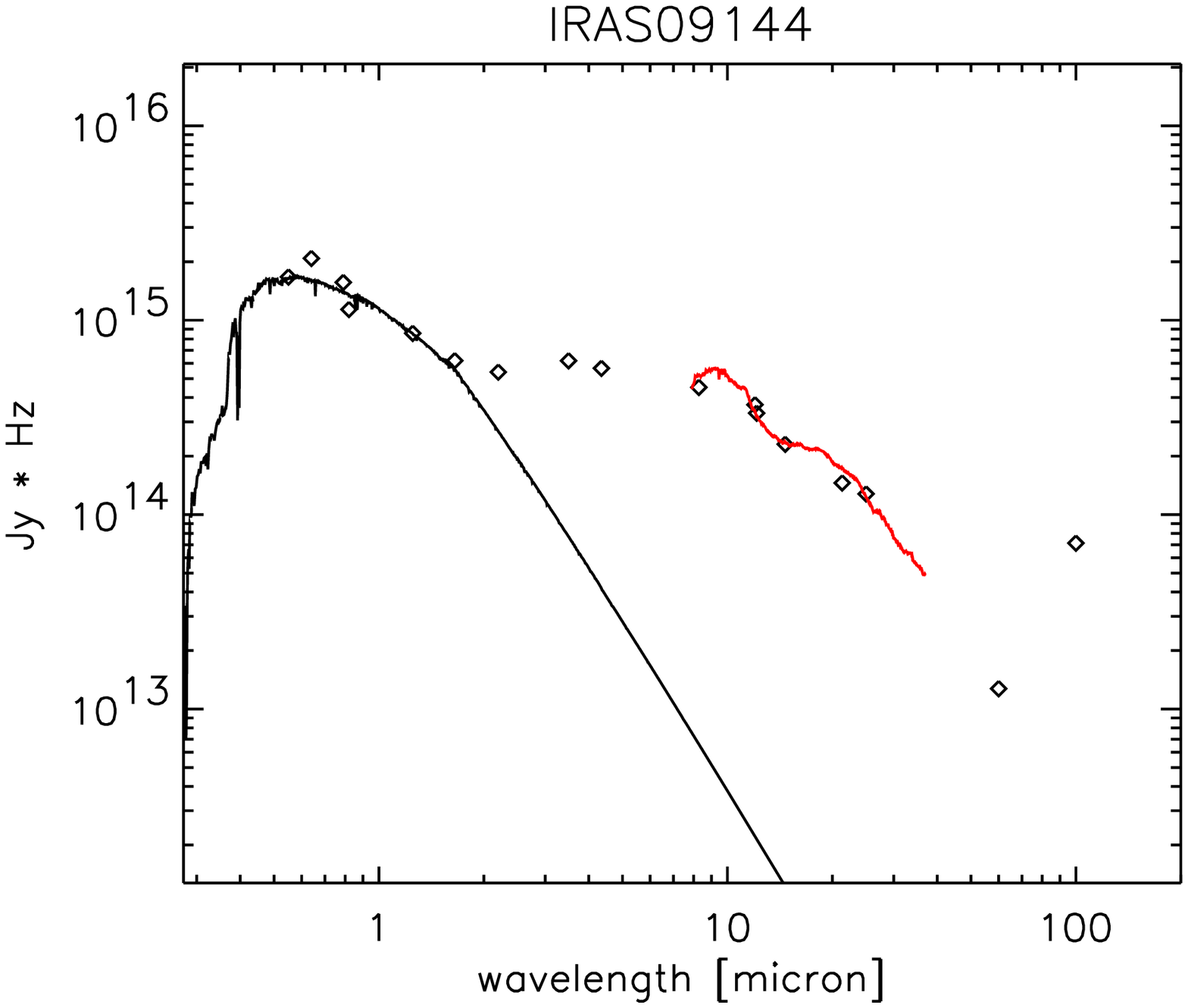}}
\resizebox{6cm}{!}{\includegraphics{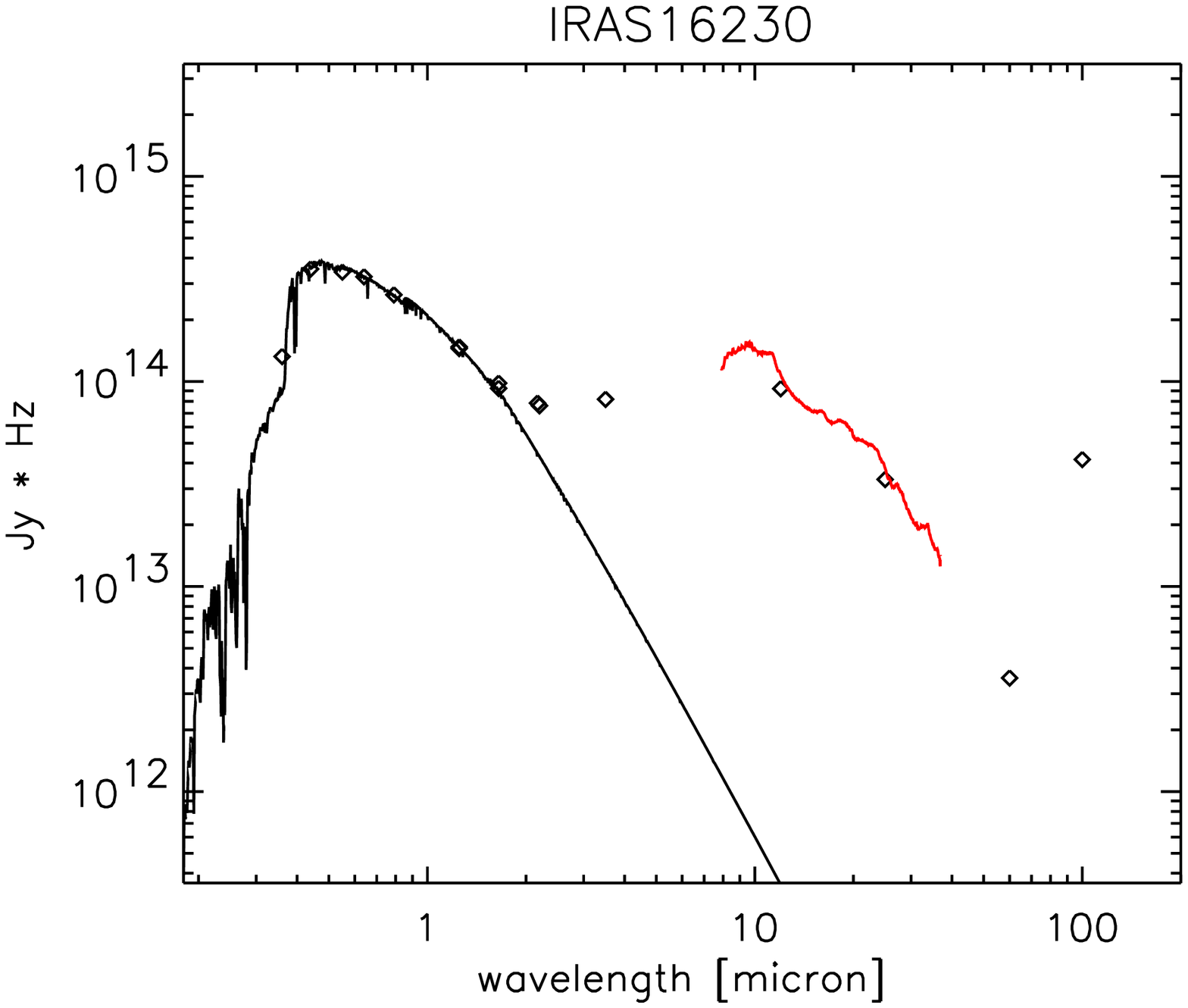}}
\resizebox{6cm}{!}{\includegraphics{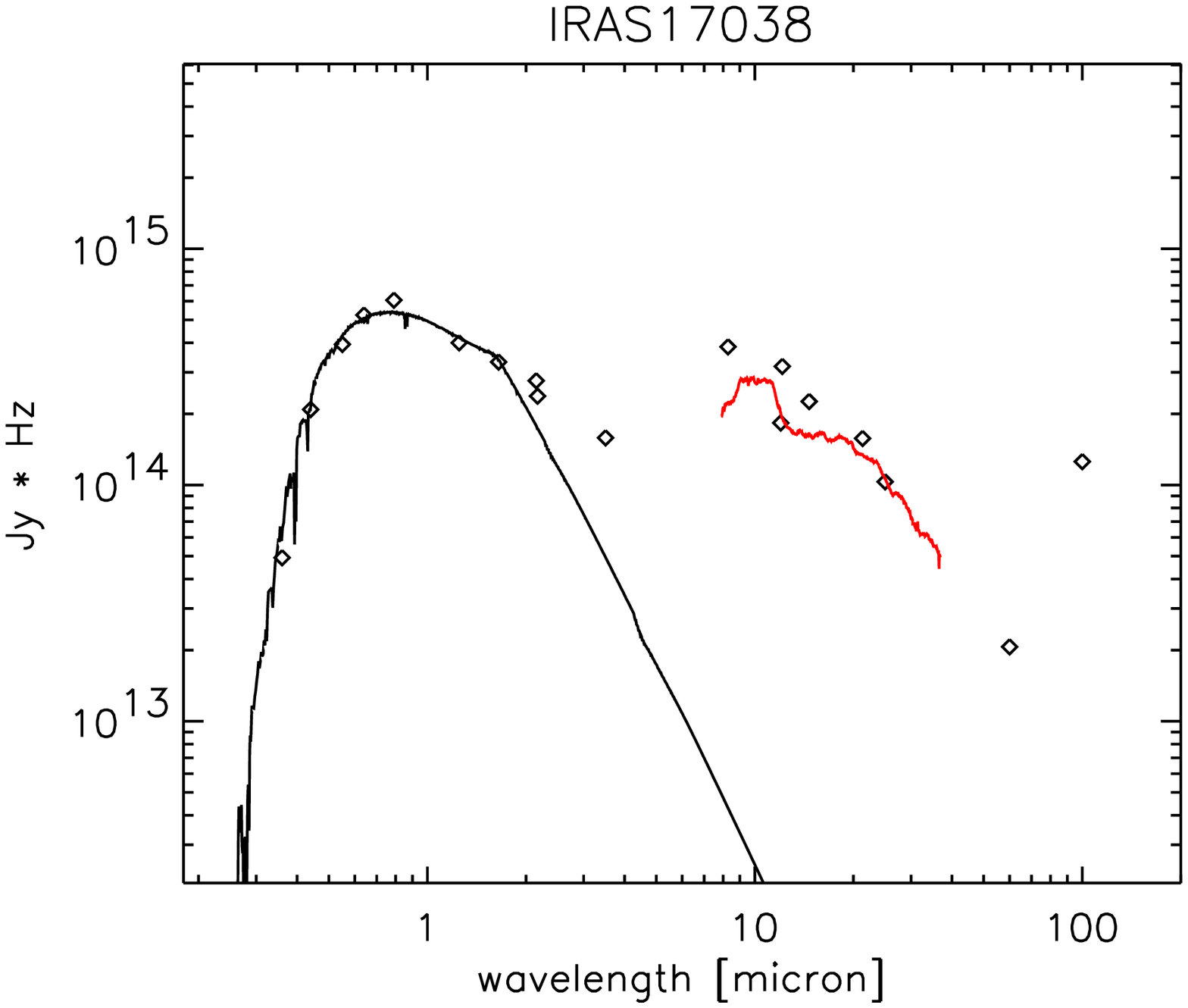}}
\resizebox{6cm}{!}{\includegraphics{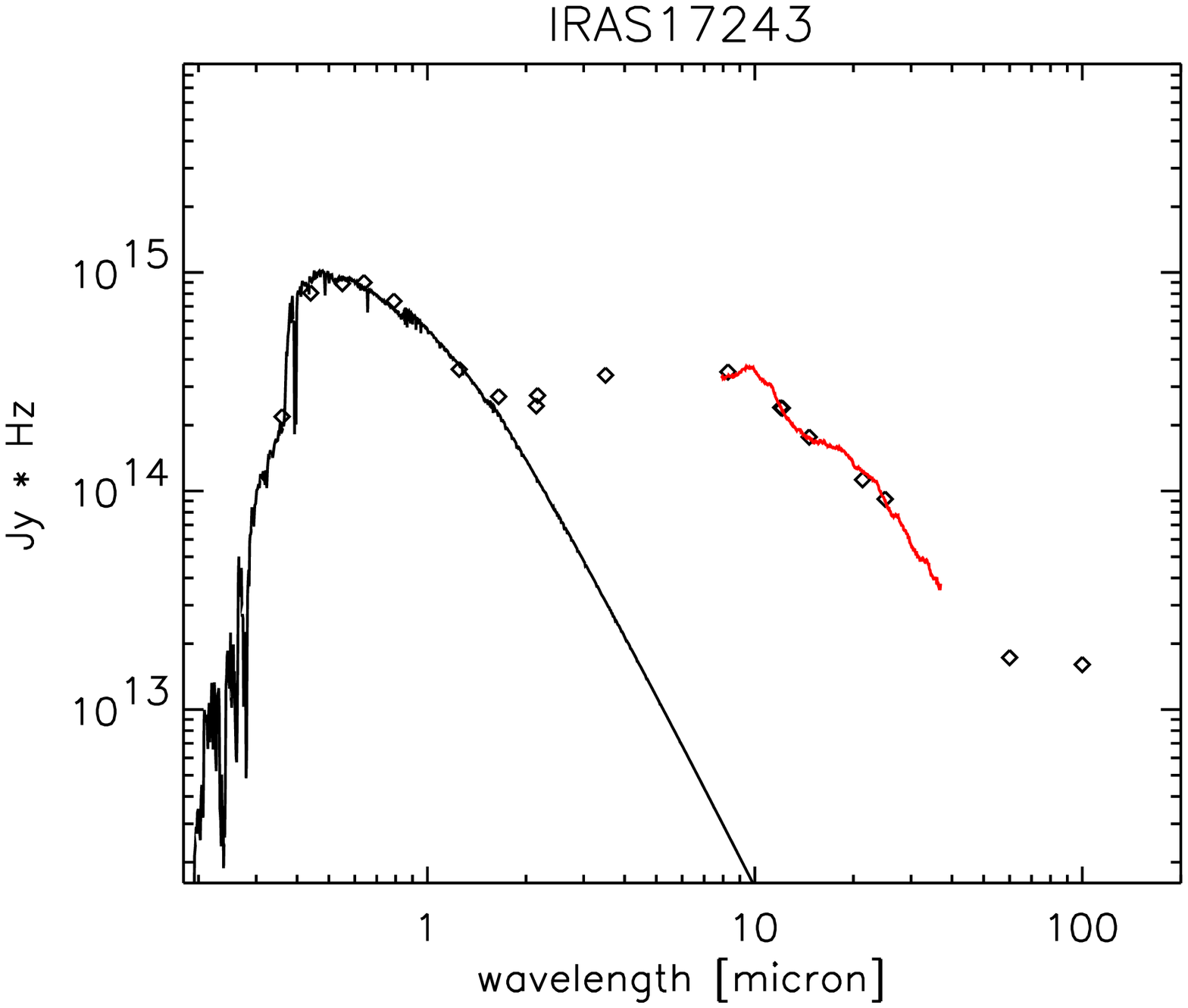}}
\resizebox{6cm}{!}{\includegraphics{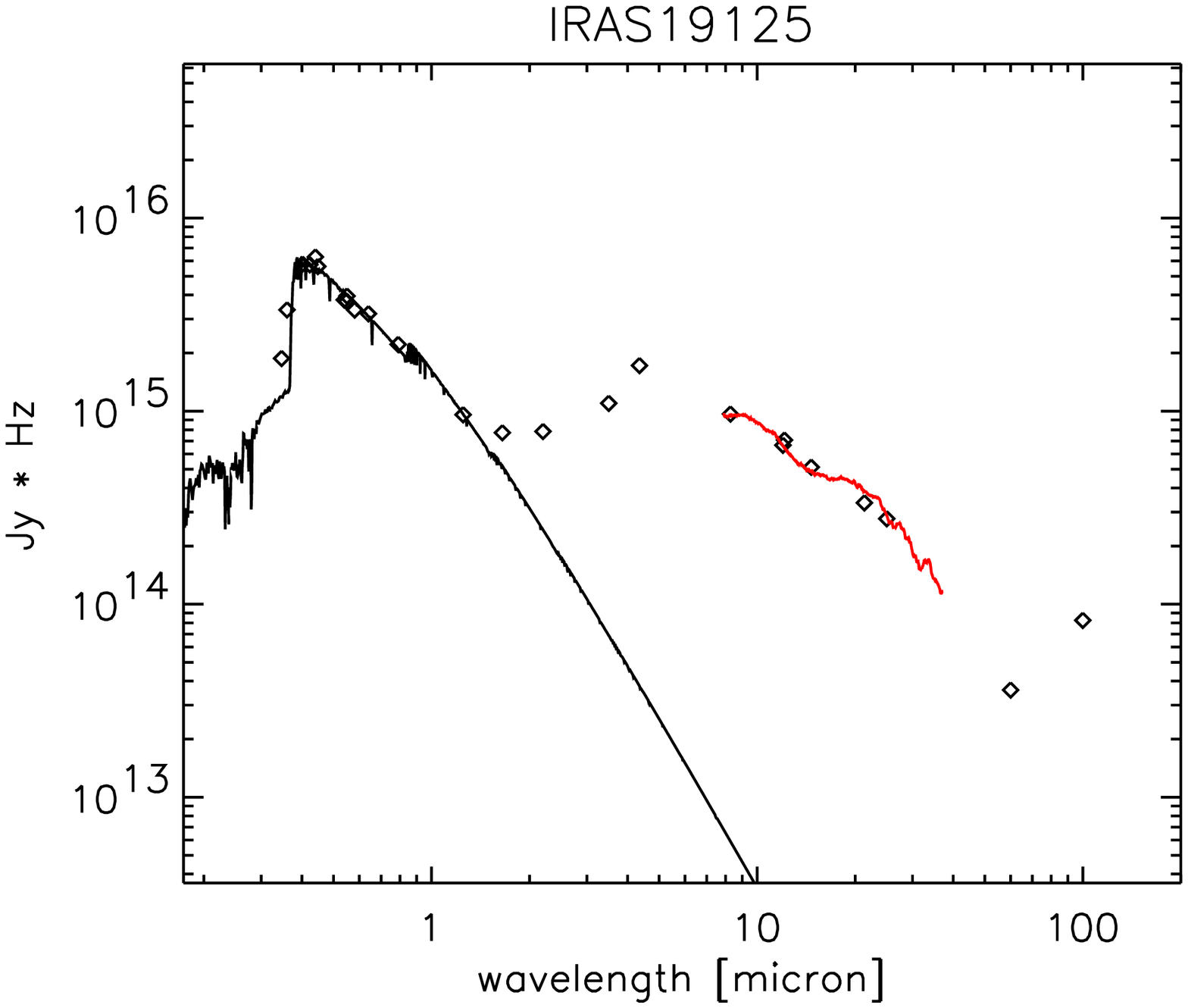}}
\resizebox{6cm}{!}{\includegraphics{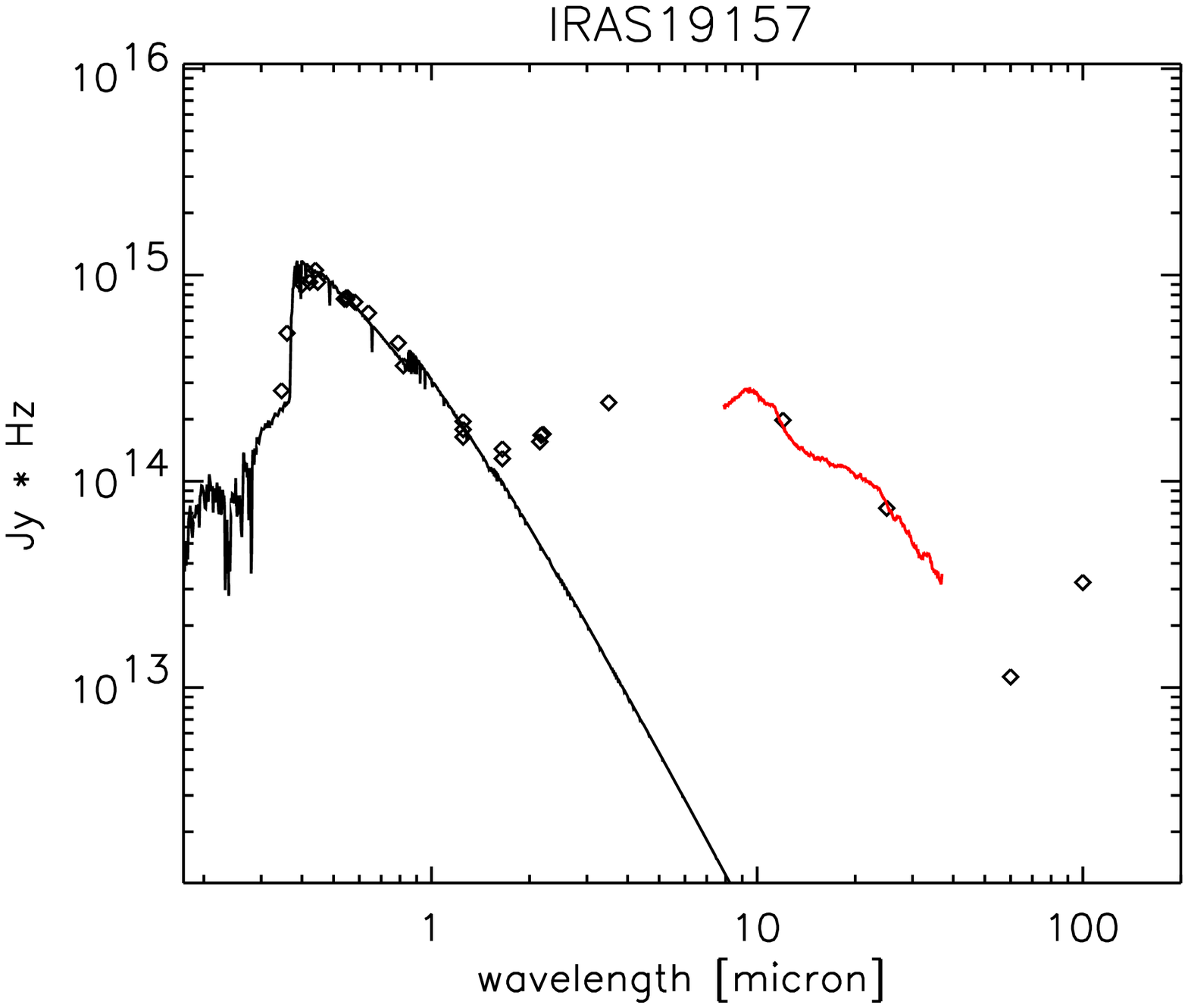}}
\caption{The SEDs of our sample stars. The dereddened fluxes (diamonds) are given together with the scaled
photospheric Kurucz model (solid black line). The corresponding TIMMI2 and SPITZER spectrum are overplotted in red.}
\label{sed1}
\end{figure}
\begin{figure}
\vspace{0cm}
\hspace{0cm}
\resizebox{6cm}{!}{\includegraphics{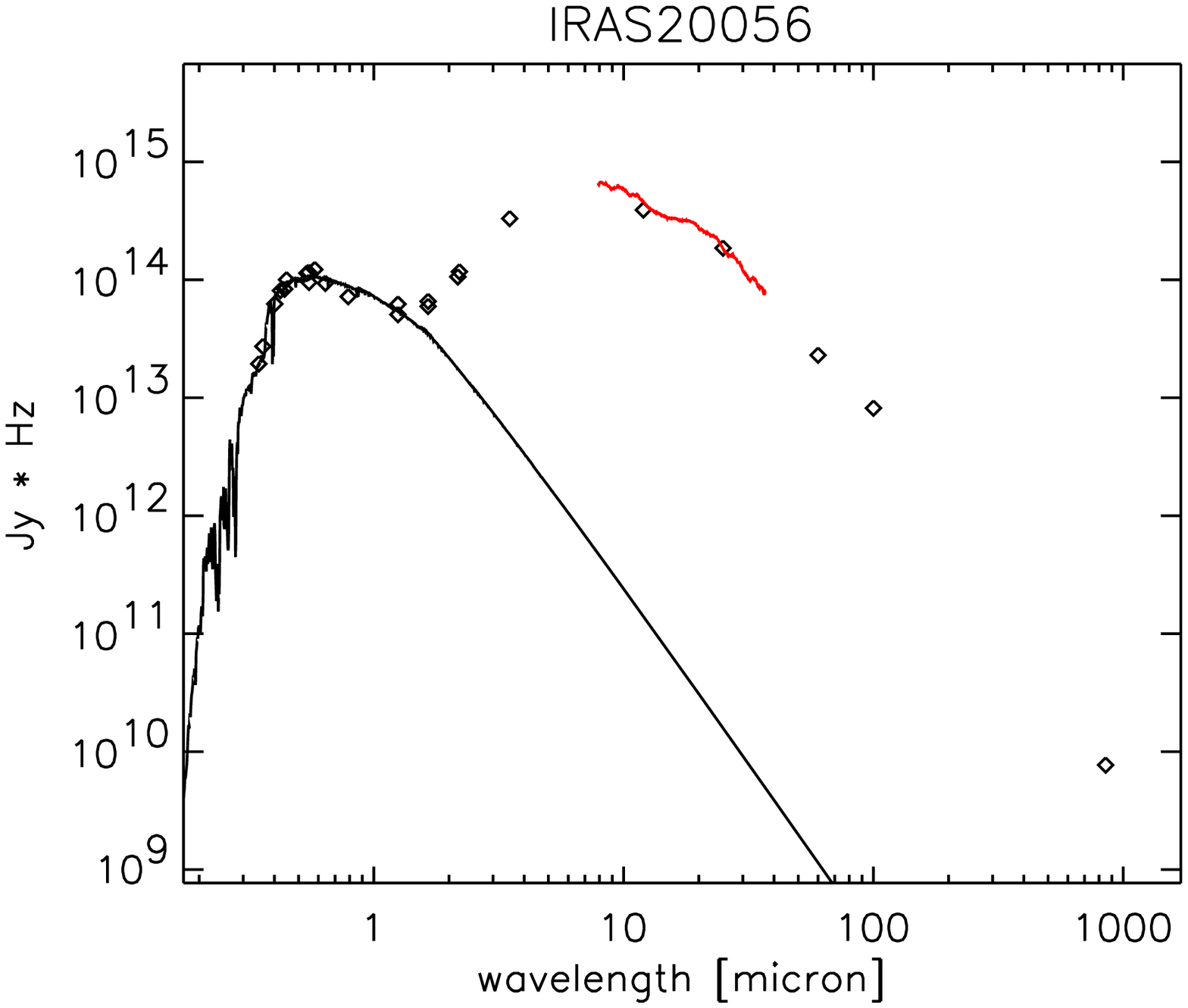}}
\resizebox{6cm}{!}{\includegraphics{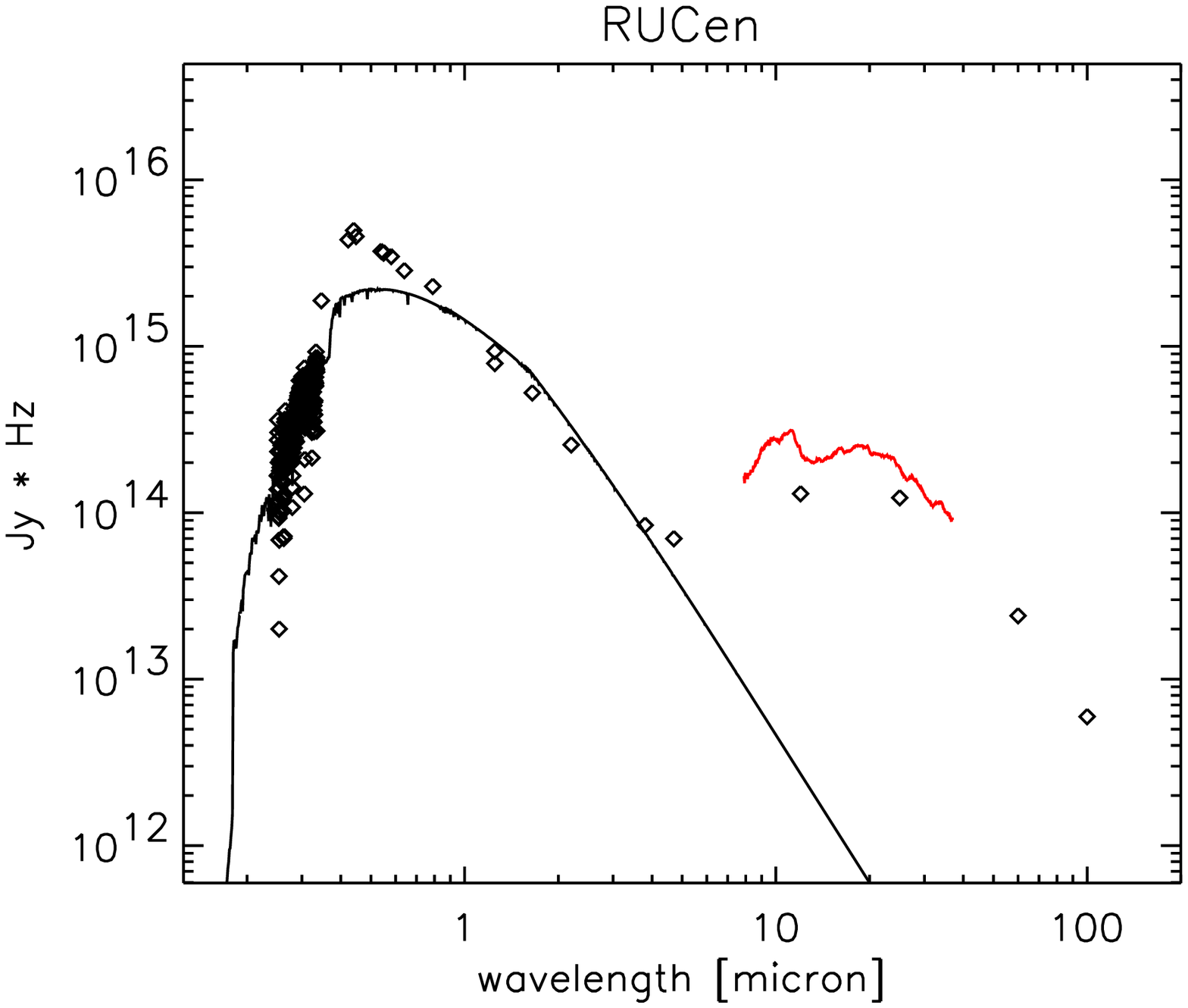}}
\resizebox{6cm}{!}{\includegraphics{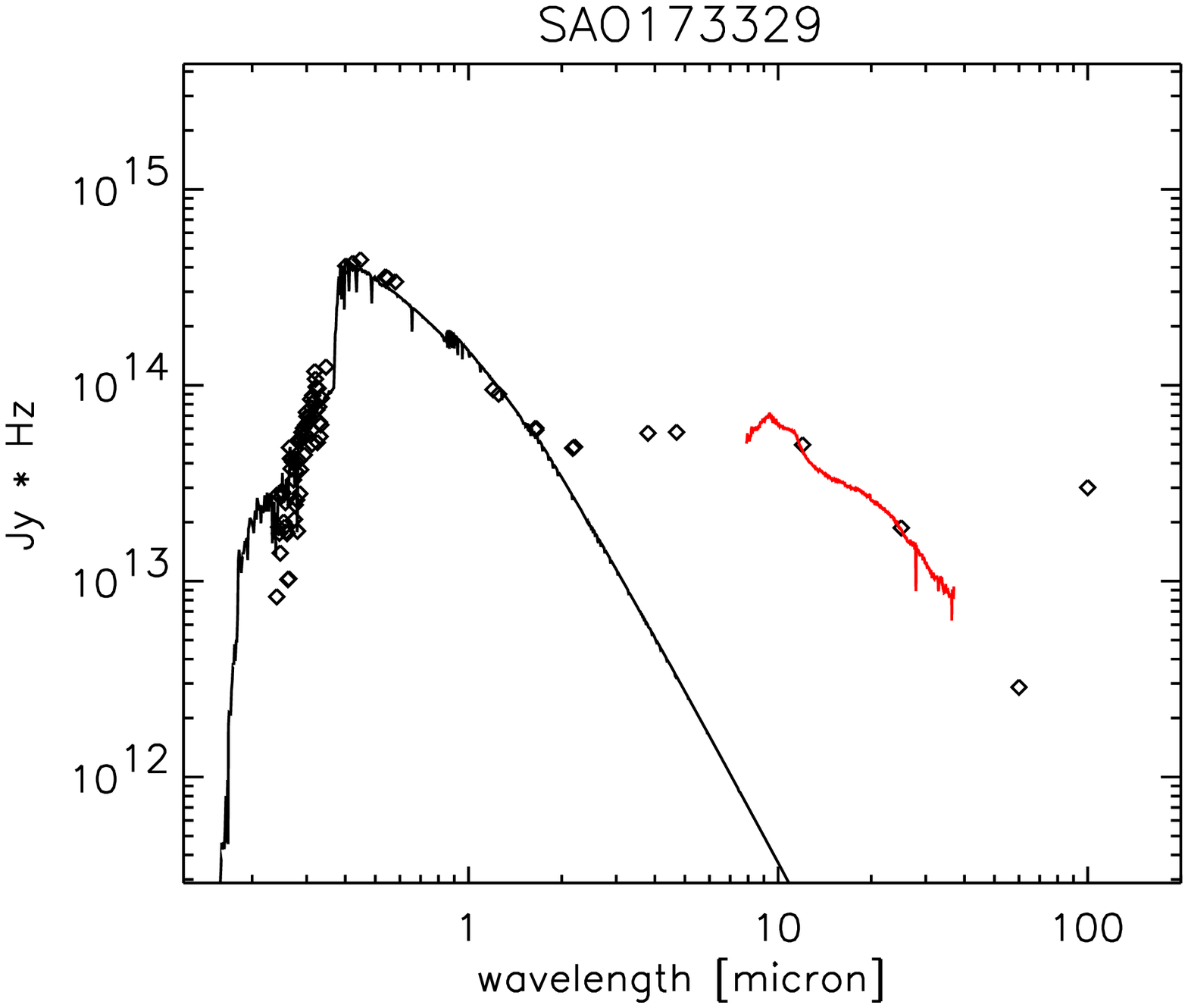}}
\resizebox{6cm}{!}{\includegraphics{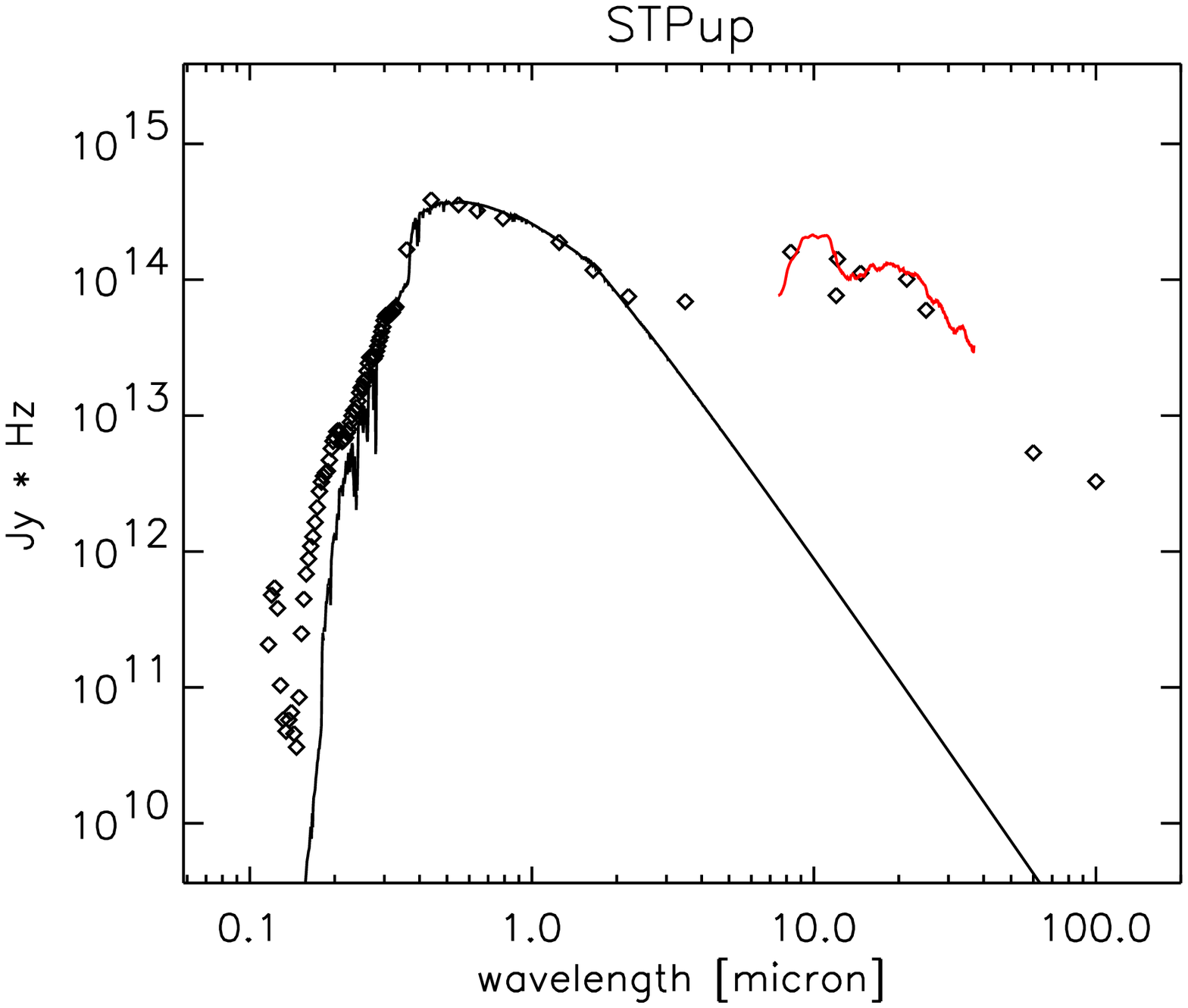}}
\resizebox{6cm}{!}{\includegraphics{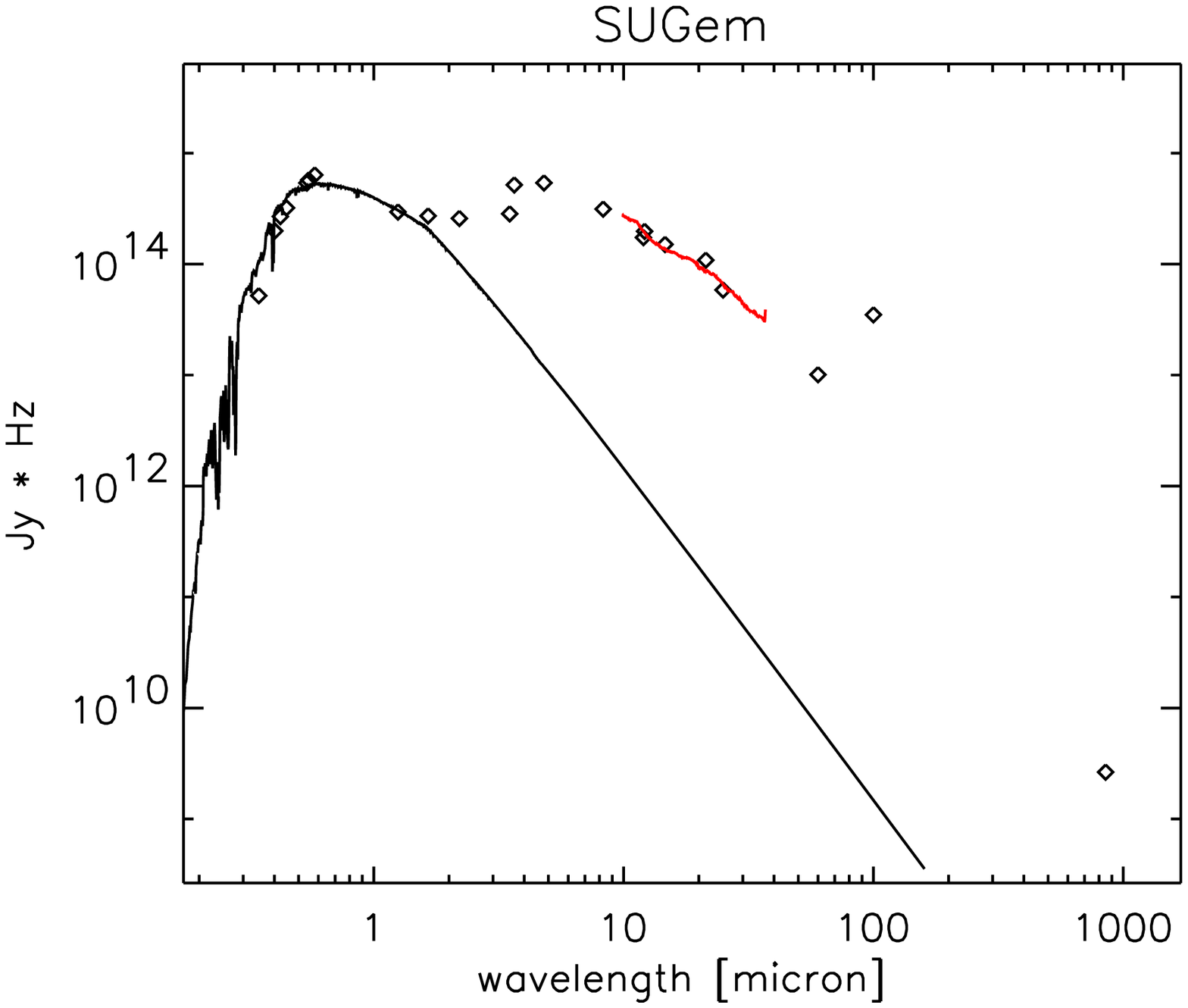}}
\resizebox{6cm}{!}{\includegraphics{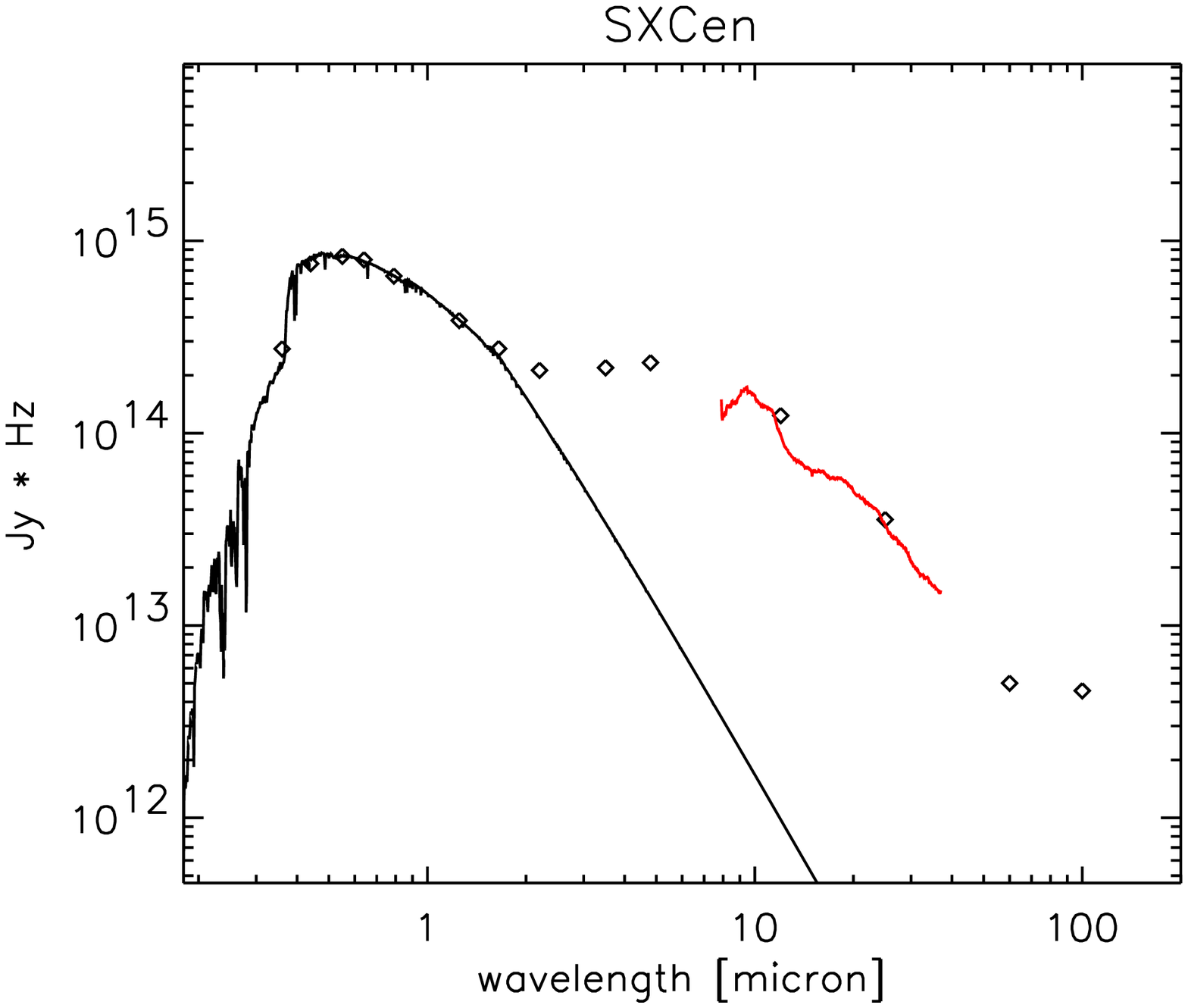}}
\resizebox{6cm}{!}{\includegraphics{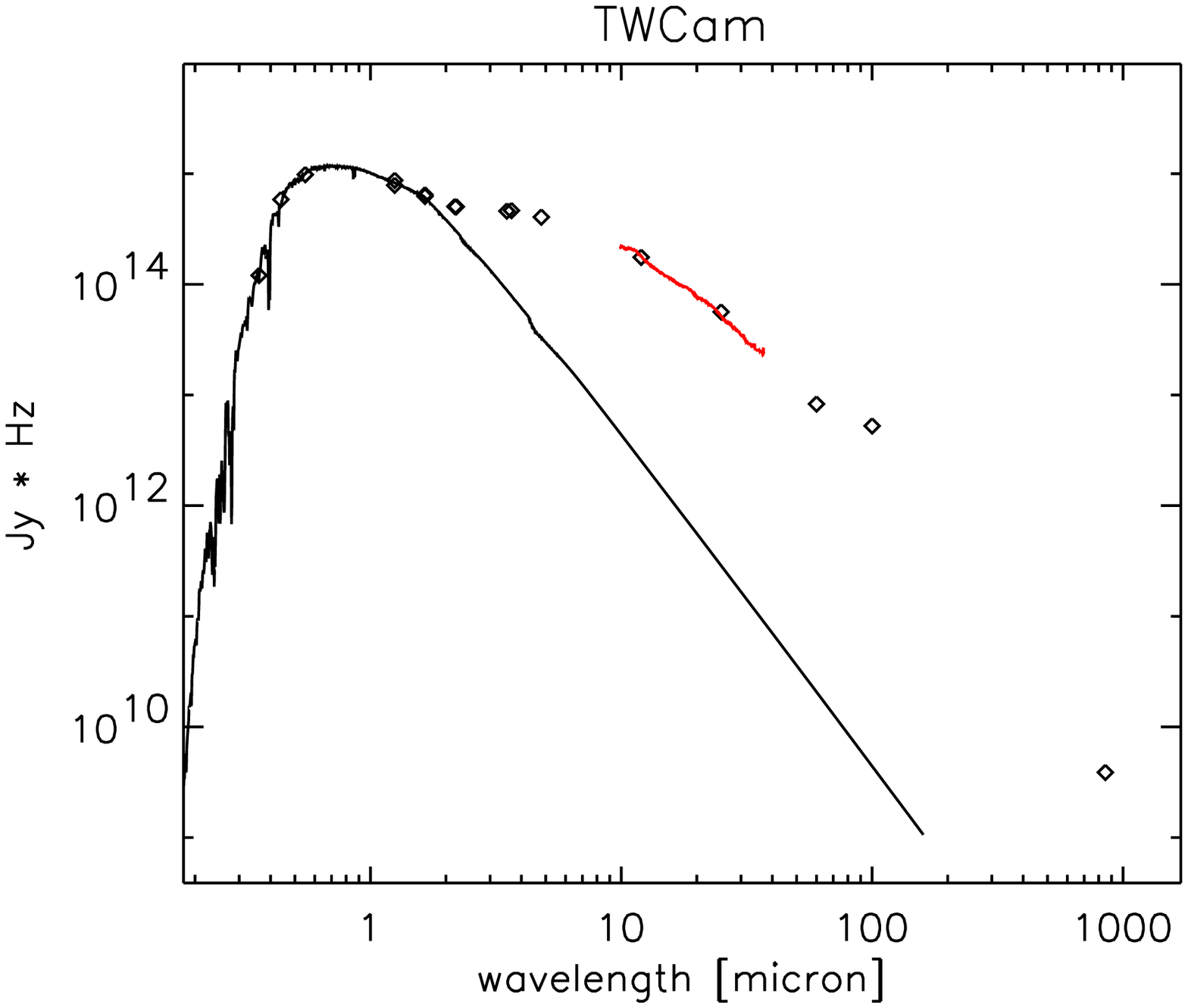}}
\resizebox{6cm}{!}{\includegraphics{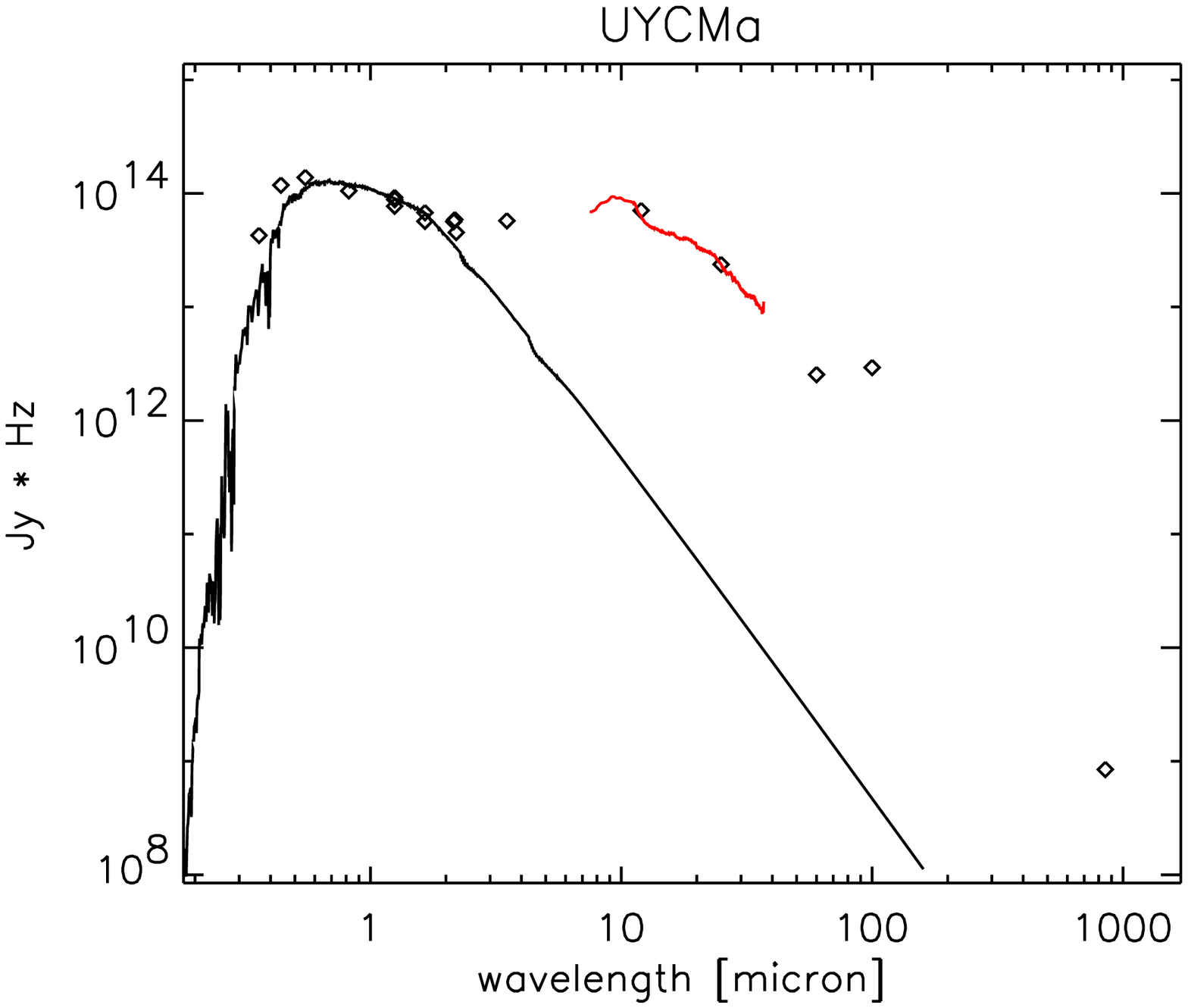}}
\caption{See previous caption.}
\label{sed2}
\end{figure}

\twocolumn
\subsection{Mean spectra and complexes}
\label{mean_appendix}

\subsection{The 14\,$\mu$m complex (13-15\,$\mu$m)}

\begin{figure}[ht]
\vspace{0cm}
\hspace{0cm}
\resizebox{9.5cm}{!}{\includegraphics{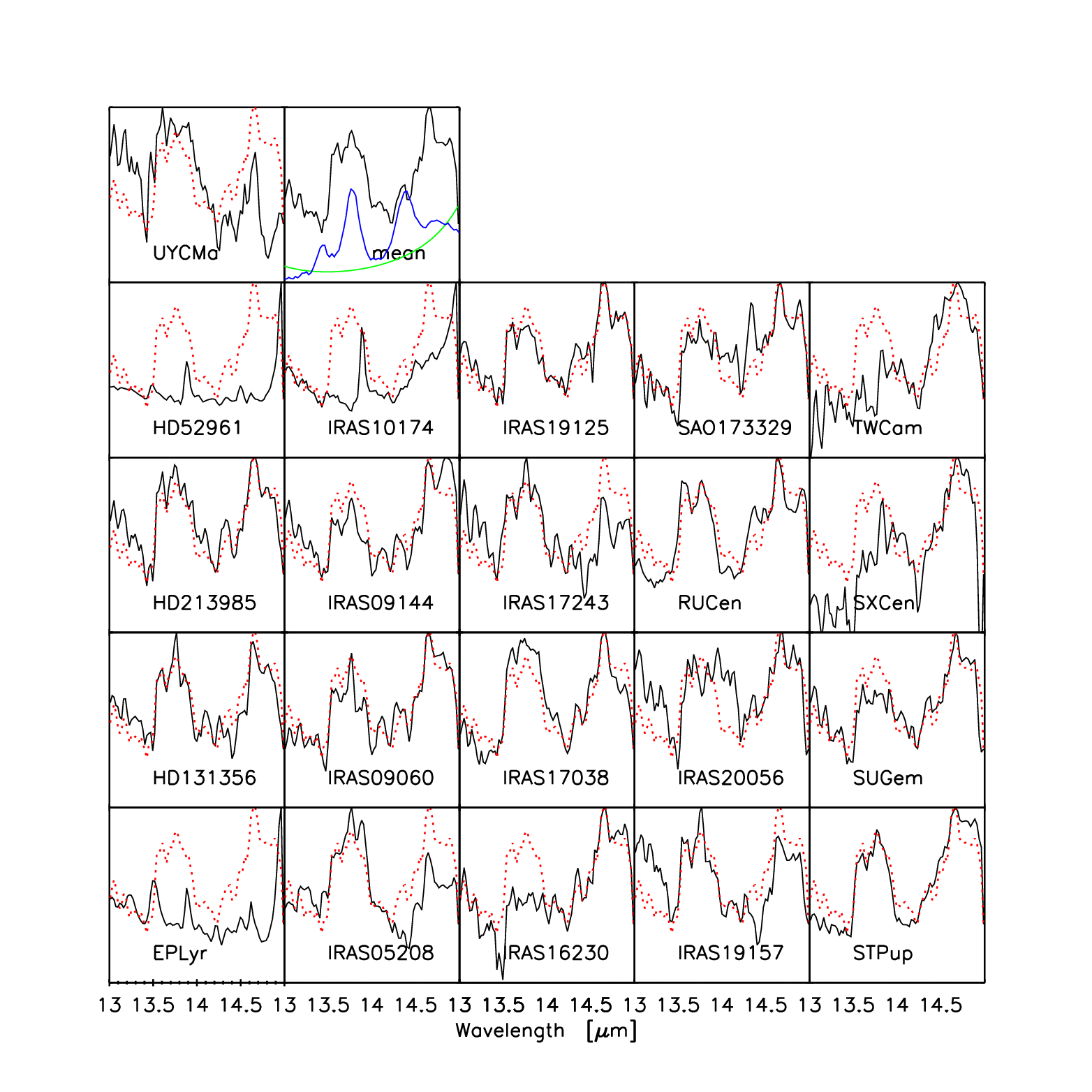}}
\caption{The 14\,$\mu$m complex, continuum subtracted and normalised. Overplotted in dotted line (red) the mean spectrum.
The mass absorption coefficients of forsterite and enstatite (in CDE approximation) are plotted in green and
blue.}
\label{mean14}
\end{figure}

In Fig.~\ref{mean14} the 14\,$\mu$m complexes are plotted.
The mean spectrum has two strong emission peaks, around 13.7 and 14.7\,$\mu$m.
The profile at 13.7\,$\mu$m seams to blend of two peaks, of which one can be
identified as enstatite. The small feature in the mean spectrum at 14.3\,$\mu$m
is also due to enstatite. The feature around 14.7\,$\mu$m remains unidentified. It
seems to be made up of a broader feature from 14.3\,$\mu$m till 15\,$\mu$m, with 
a strong narrow peak at 14.7\,$\mu$m.
In the sample stars, there is some variation in the ratio between the two strong peaks
at 13.7 and 14.7\,$\mu$m, meaning they are probably due to different dust species.

\subsection{The 16\,$\mu$m complex (15-17\,$\mu$m)}

\begin{figure}[ht]
\vspace{0cm}
\hspace{0cm}
\resizebox{9.5cm}{!}{\includegraphics{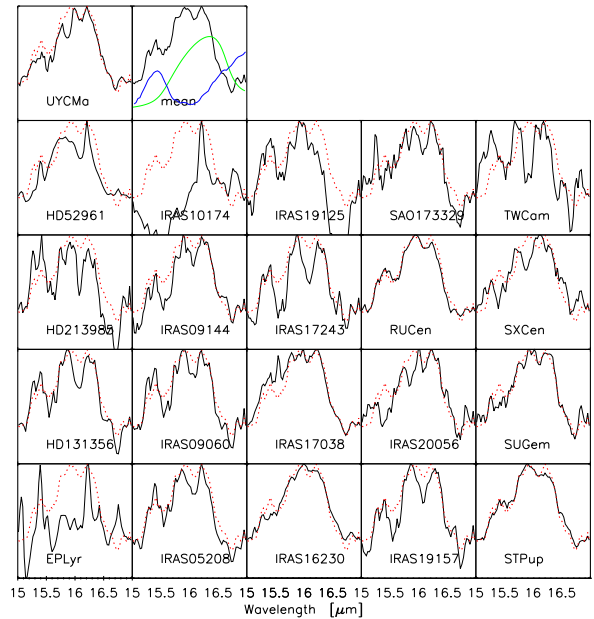}}
\caption{The 16\,$\mu$m complex, continuum subtracted and normalised. Overplotted in dotted line (red) the mean spectrum.
The mass absorption coefficients of forsterite and enstatite (in CDE approximation) are plotted in green and
blue.}
\label{mean16}
\end{figure}

In Fig.~\ref{mean16} the 16\,$\mu$m complexes are plotted.
The mean spectrum shows a clear broad feature around 16\,$\mu$m. The shape and strength
is very similar to the forsterite feature at 16.2\,$\mu$m, with a contribution of enstatite at 15.3\,$\mu$m.
On top of this broad band
more narrow features at 15.9\,$\mu$m and 16.2\,$\mu$m can be discriminated.
The small peaks at 15.4\,$\mu$m and 16.2\,$\mu$m are due
to CO$_2$ gas emission, which can only be clearly seen in EP\,Lyr, HD\,52961 and IRAS\,10174.

Most sample sources show a similar profile as the mean spectrum,
although the ratio between the two main contributors sometimes differs,
due to a different enstatite/forsterite ratio. 
Stars with a weak 16\,$\mu$m feature (like TW\,Cam) show a very clear separation between the three features,
with widths of about 0.3\,$\mu$m. This shows that the mean spectrum probably consist of a very broad forsterite feature
topped with more narrow features of another dust (or gas?) species.
HD\,52961 has a strong profile that is shifted bluewards in comparison to the mean spectrum,
with clear CO$_2$ emission lines.

\subsection{The 19\,$\mu$m complex (17-21\,$\mu$m)}

\begin{figure}[ht]
\vspace{0cm}
\hspace{0cm}
\resizebox{9.5cm}{!}{\includegraphics{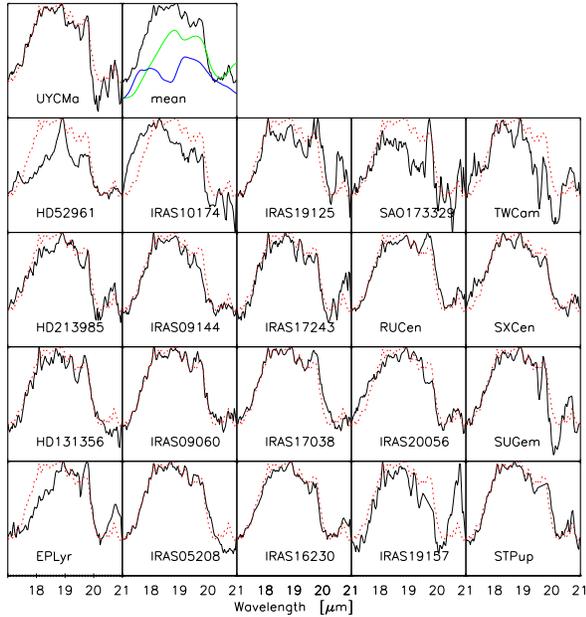}}
\caption{The 19\,$\mu$m complex, continuum subtracted and normalised. Overplotted in dotted line (red) the mean spectrum.
The mass absorption coefficients of forsterite and enstatite (in CDE approximation) are plotted in green and
blue.}
\label{mean19}
\end{figure}

In Fig.~\ref{mean19} the 19\,$\mu$m complexes are plotted.
The very broad profile is again a blend of forsterite and enstatite.
The sometimes very sharp feature around 19.7\,$\mu$m is a data reduction artefact.
Most stars have the same profile shape, except for HD\,52961, EP\,Lyr and IRAS\,10174.
The first two are clearly outliers in our sample stars, since they have very distinct spectra,
very different from other sample stars. Looking at the Spitzer spectrum of IRAS\,10174,
this star has almost no crystalline features and the observed complex is actually due to poor continuum
subtraction and normalisation.

\subsection{The 23\,$\mu$m complex (20-27\,$\mu$m)}

\begin{figure}[ht]
\vspace{0cm}
\hspace{0cm}
\resizebox{9.5cm}{!}{\includegraphics{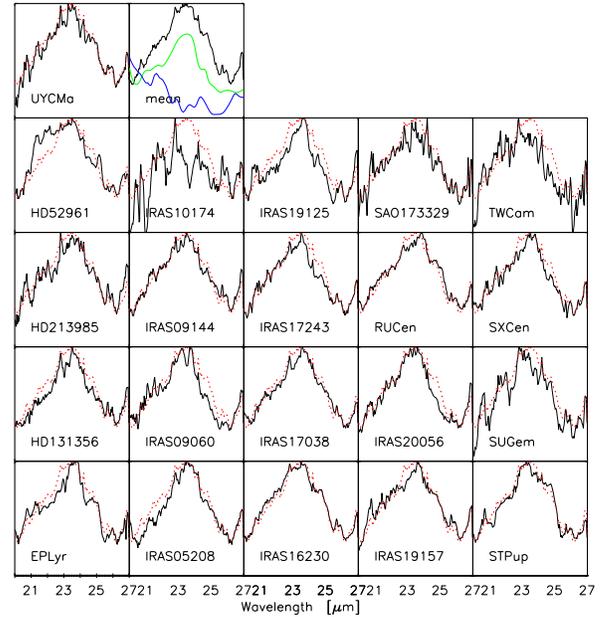}}
\caption{The 23\,$\mu$m complex, continuum subtracted and normalised. Overplotted in dotted line (red) the mean spectrum.
The mass absorption coefficients of forsterite and enstatite (in CDE approximation) are plotted in green and
blue.}
\label{mean23}
\end{figure}

In Fig.~\ref{mean23} the 23\,$\mu$m complexes are plotted.
The mean profile is dominated by forsterite and there is an extremely good
agreement between the mean spectrum and sample stars. Remarkable little variation
is detected in the whole sample.

\subsection{The 33\,$\mu$m complex (31-37\,$\mu$m)}

\begin{figure}
\vspace{0cm}
\hspace{0cm}
\resizebox{9.5cm}{!}{ \includegraphics{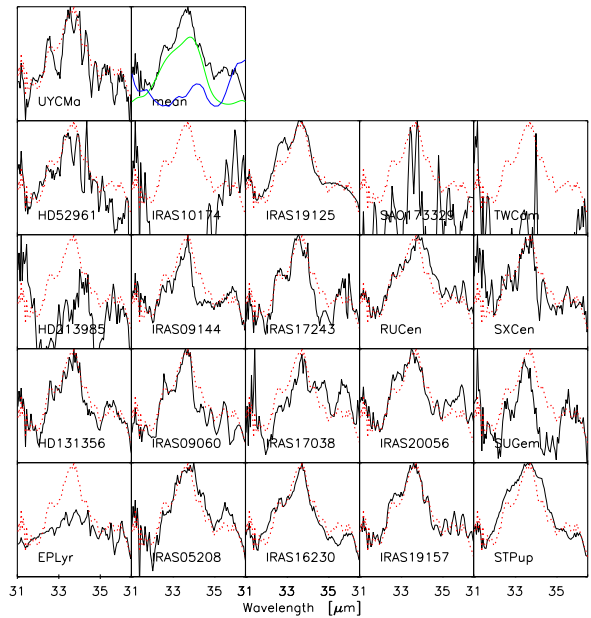}}
\caption{The 33\,$\mu$m complex, continuum subtracted and normalised. Overplotted in dotted line (red) the mean spectrum.
The mass absorption coefficients of forsterite and enstatite (in CDE approximation) are plotted in green and
blue.}
\label{mean33}%
\end{figure}

In Fig.~\ref{mean33} the 33\,$\mu$m complexes are plotted.
The observed profile is mainly due to forsterite at 33.6\,$\mu$m. At 32.5\,$\mu$m
another feature can be observed. The ratio 32.5/33.6\,$\mu$m varies slightly from source to 
source, indicating that a different dust species is responsible for the feature at 32.5\,$\mu$m.
In ST\,Pup, the feature is clearly broader than in other sample sources.
The bump at 35.7\,$\mu$m is not a reliable result since it sits at the end of the spectrum.  

\onecolumn
\begin{figure}
\vspace{0cm}
\hspace{0cm}
\resizebox{9cm}{!}{\includegraphics{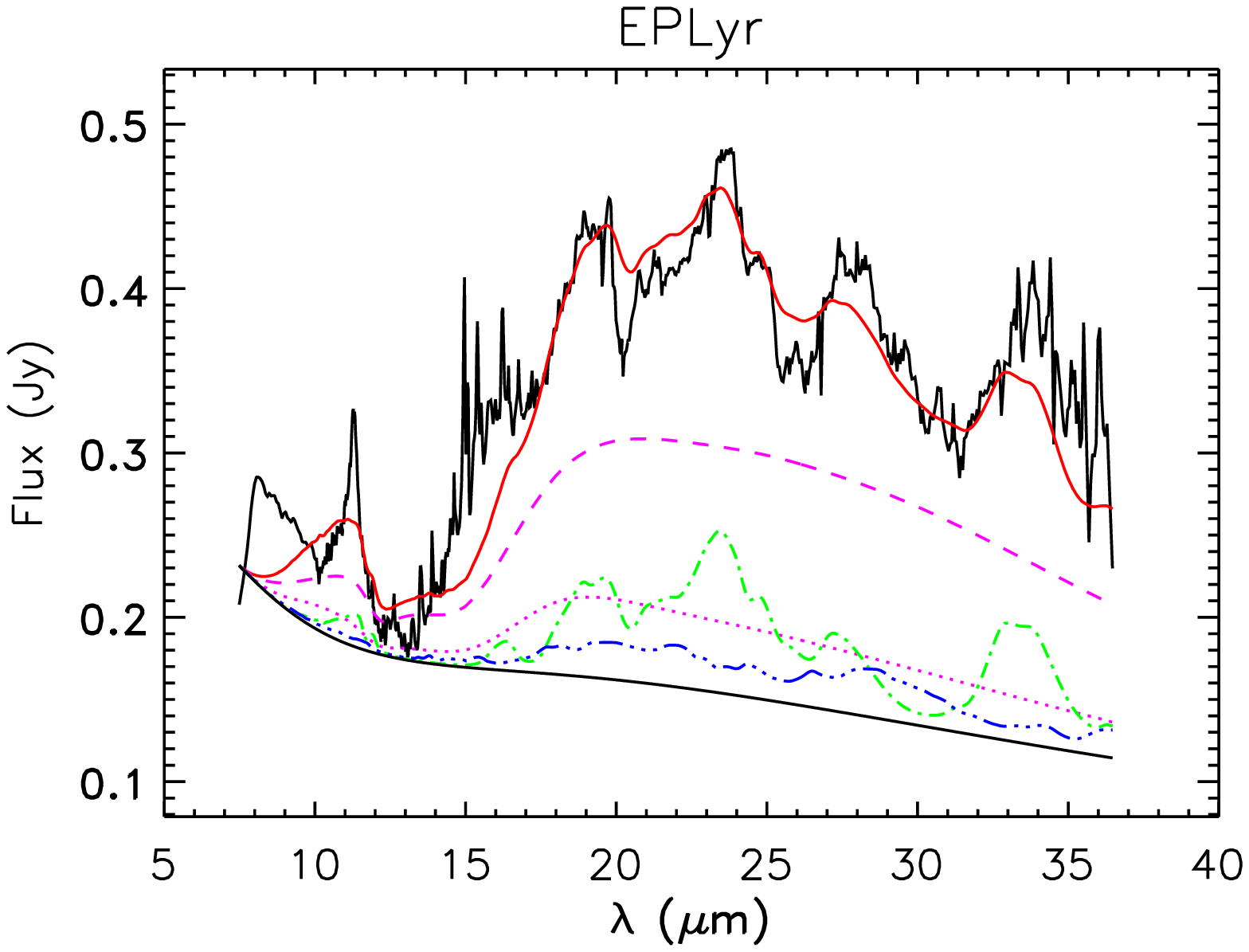}}
\resizebox{9cm}{!}{\includegraphics{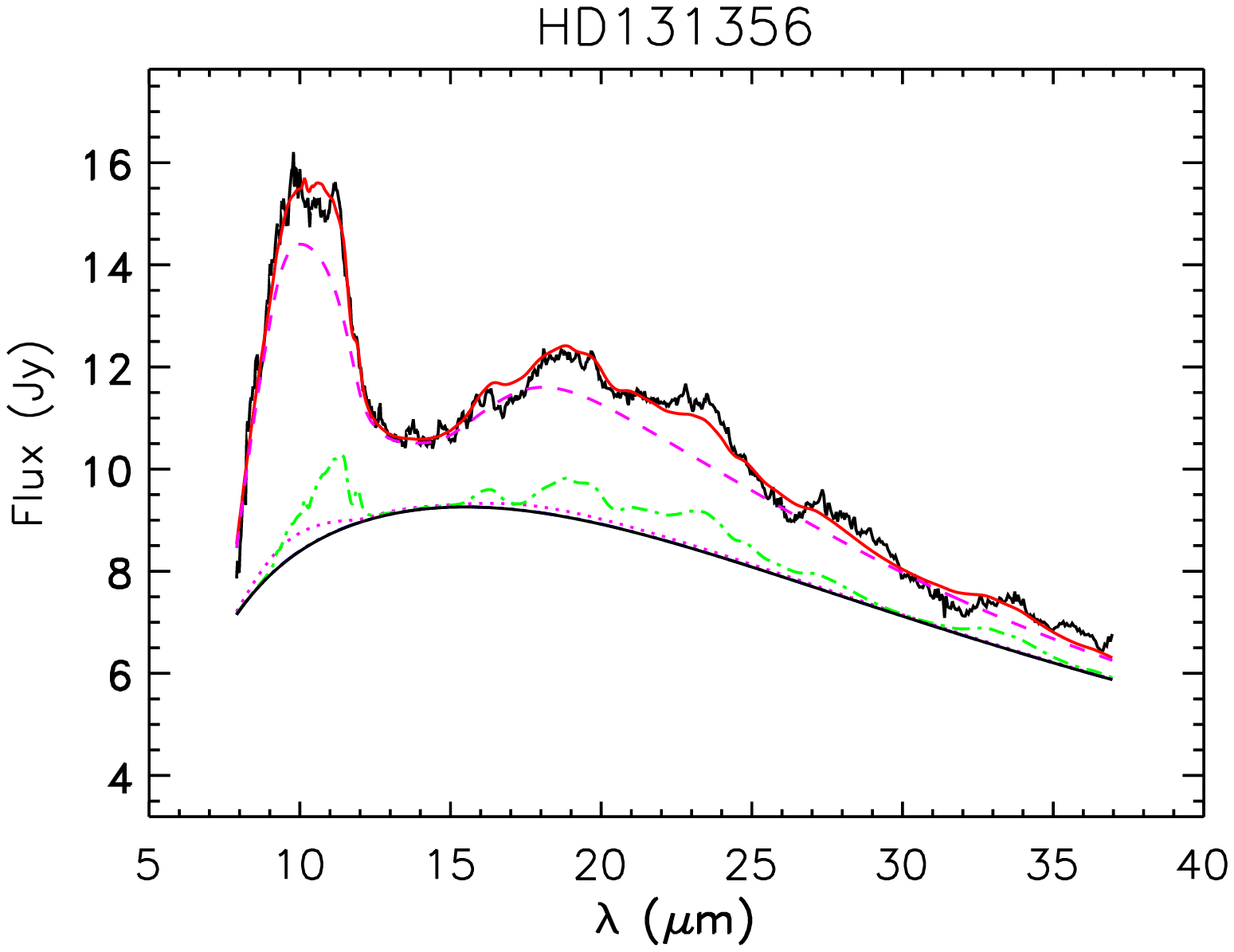}}
\resizebox{9cm}{!}{\includegraphics{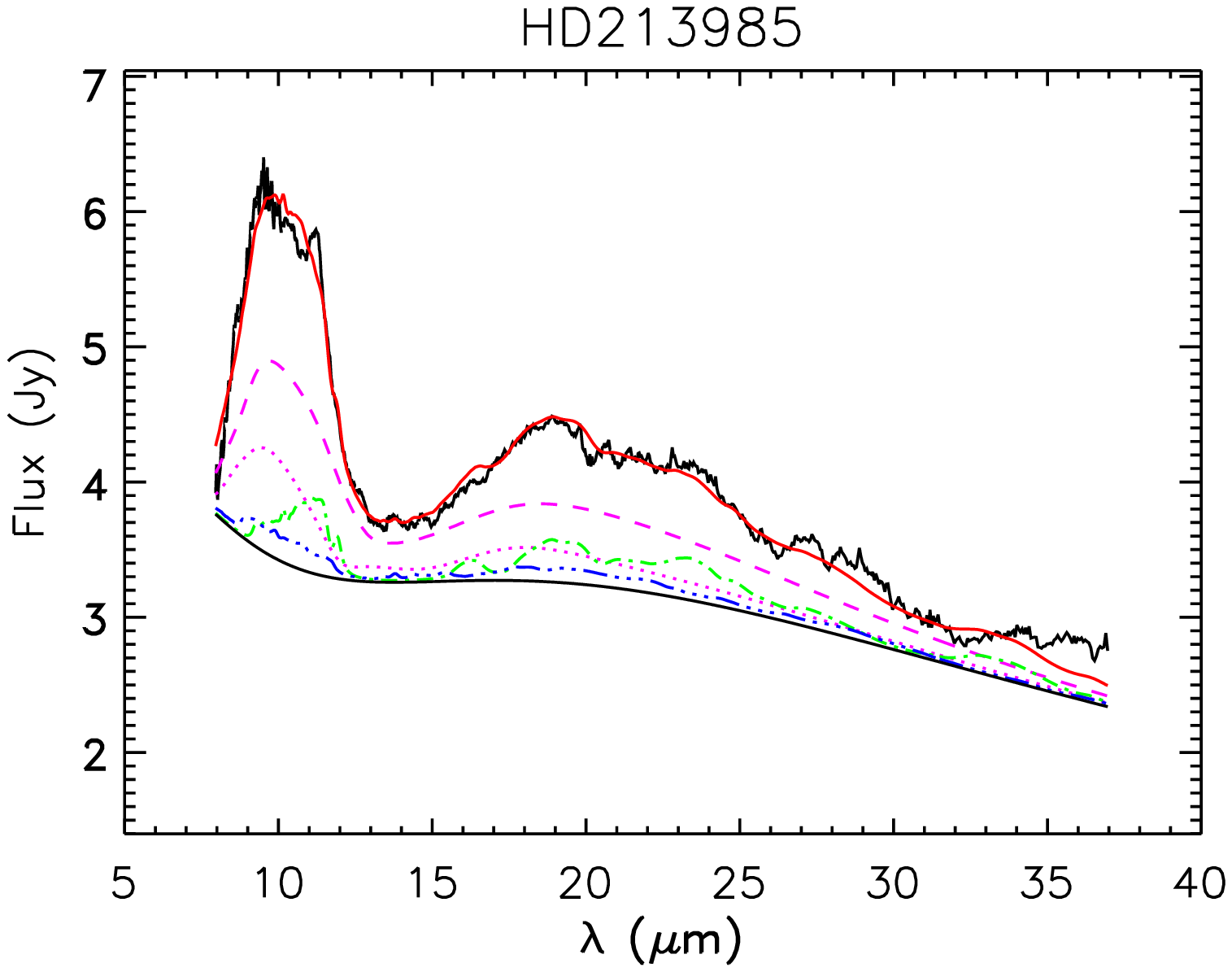}}
\resizebox{9cm}{!}{\includegraphics{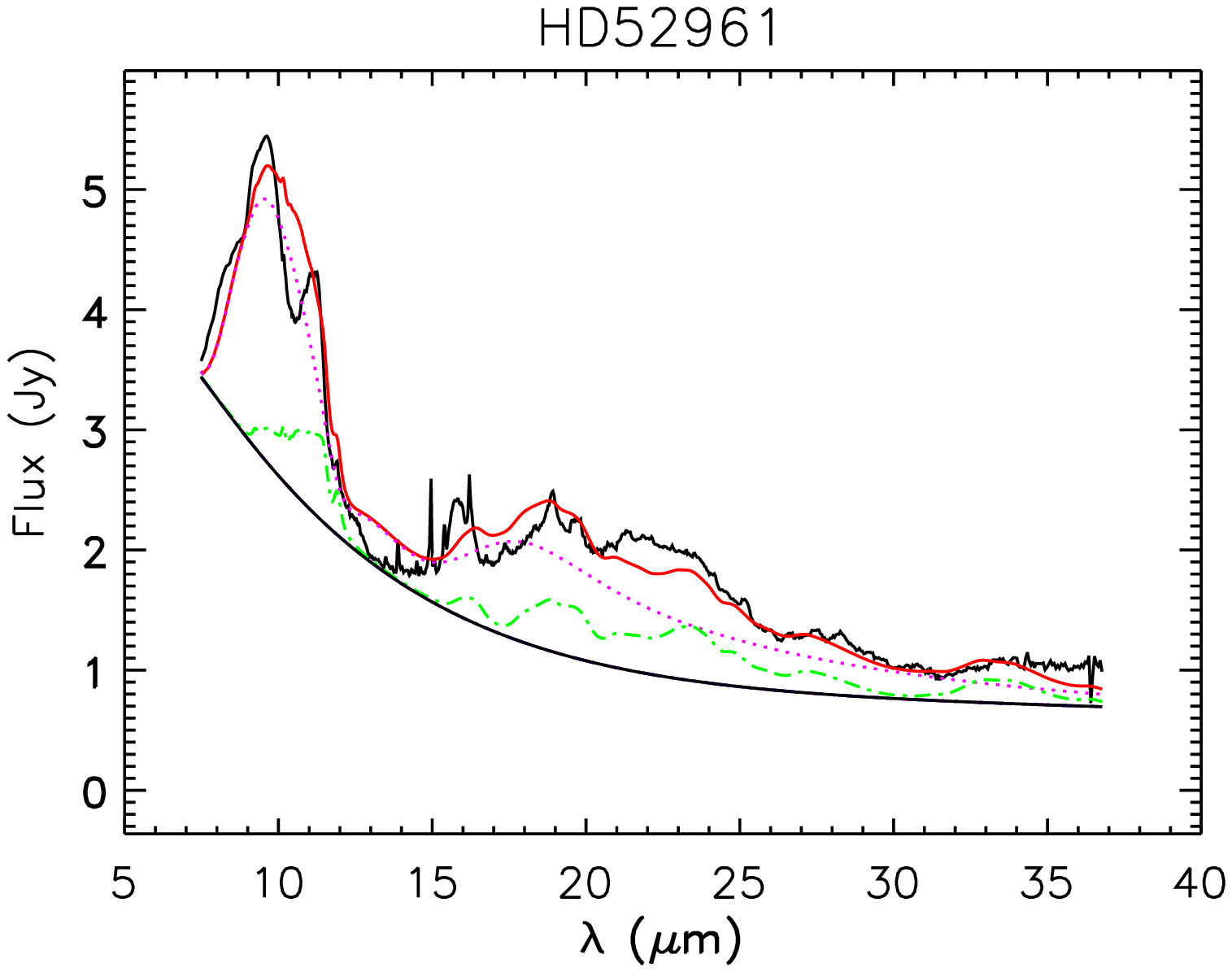}}
\resizebox{9cm}{!}{\includegraphics{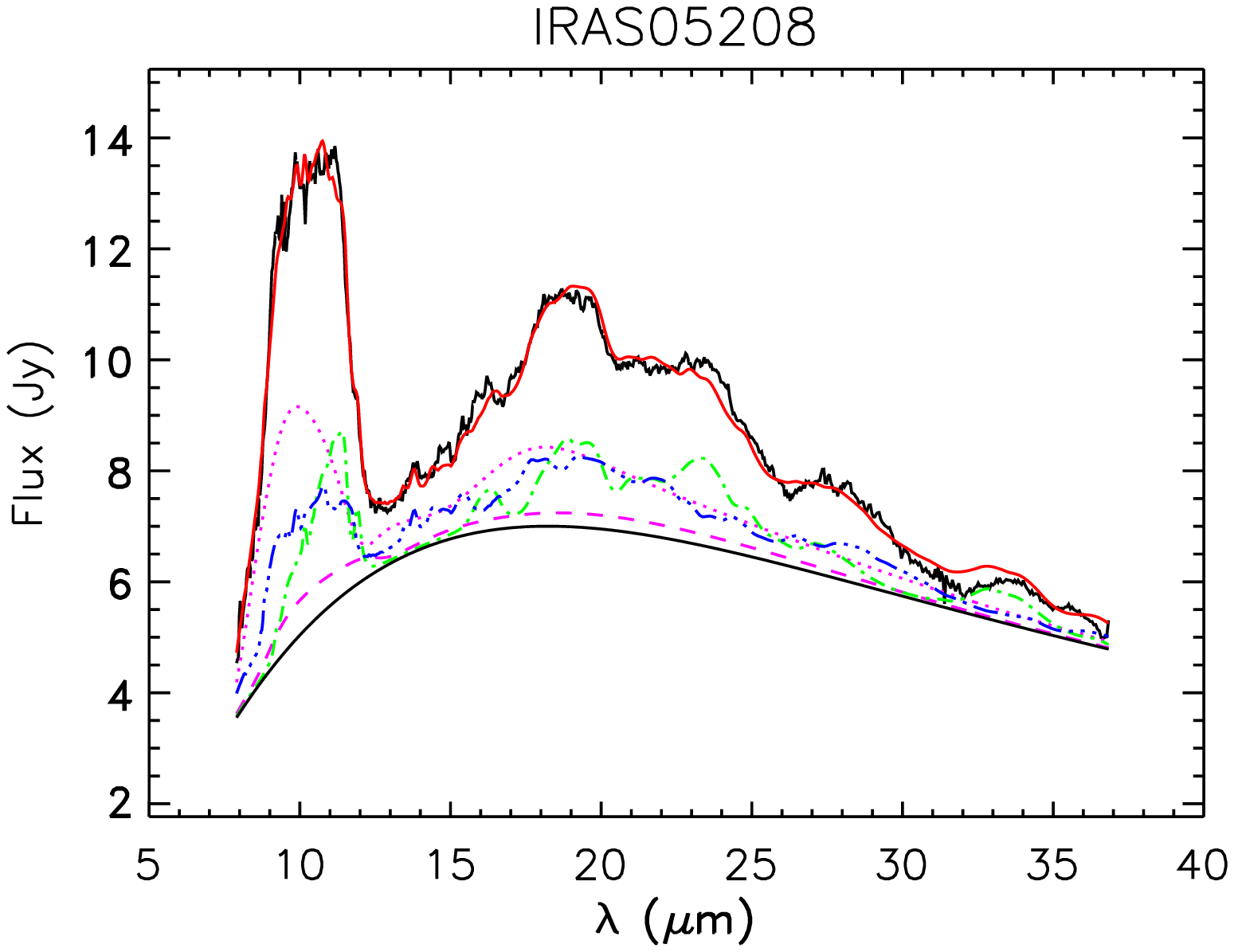}}
\resizebox{9cm}{!}{\includegraphics{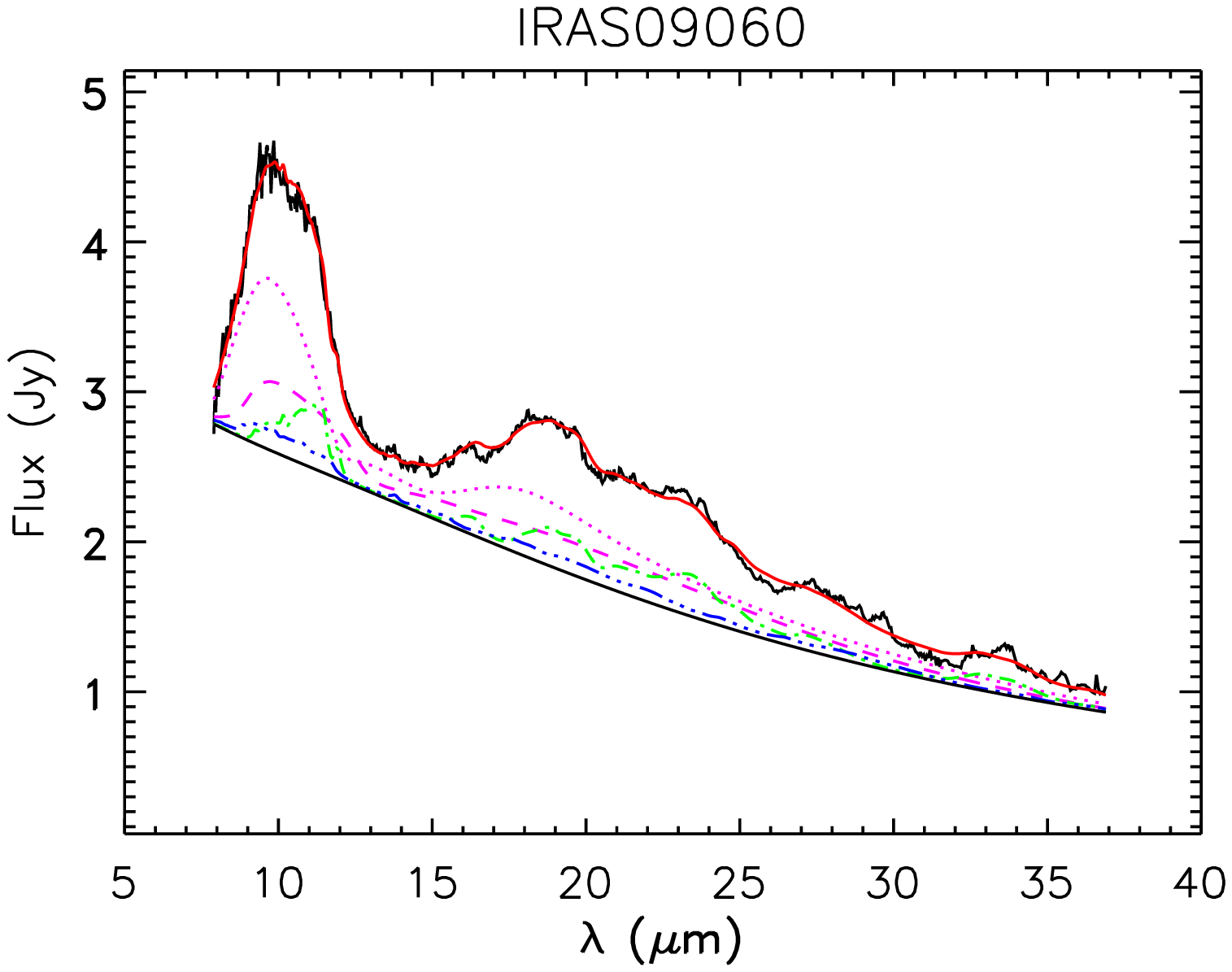}}
\caption{Best model fits for our sample stars, showing the contribution of the different dust species.
The observed spectrum (black curve) is plotted together with the best model fit (red curve) and the continuum (black solid line).
Forsterite is plotted in dash-dot lines (green) and enstatite in dash-dot-dotted lines (blue).
Small amorphous grains (2.0\,$\mu$m) are plotted as dotted lines (magenta) and large amorphous grains (4.0\,$\mu$m) as dashed lines (magenta).}
\label{fitting1}
\end{figure}

\begin{figure}
\vspace{0cm}
\hspace{0cm}
\resizebox{9cm}{!}{\includegraphics{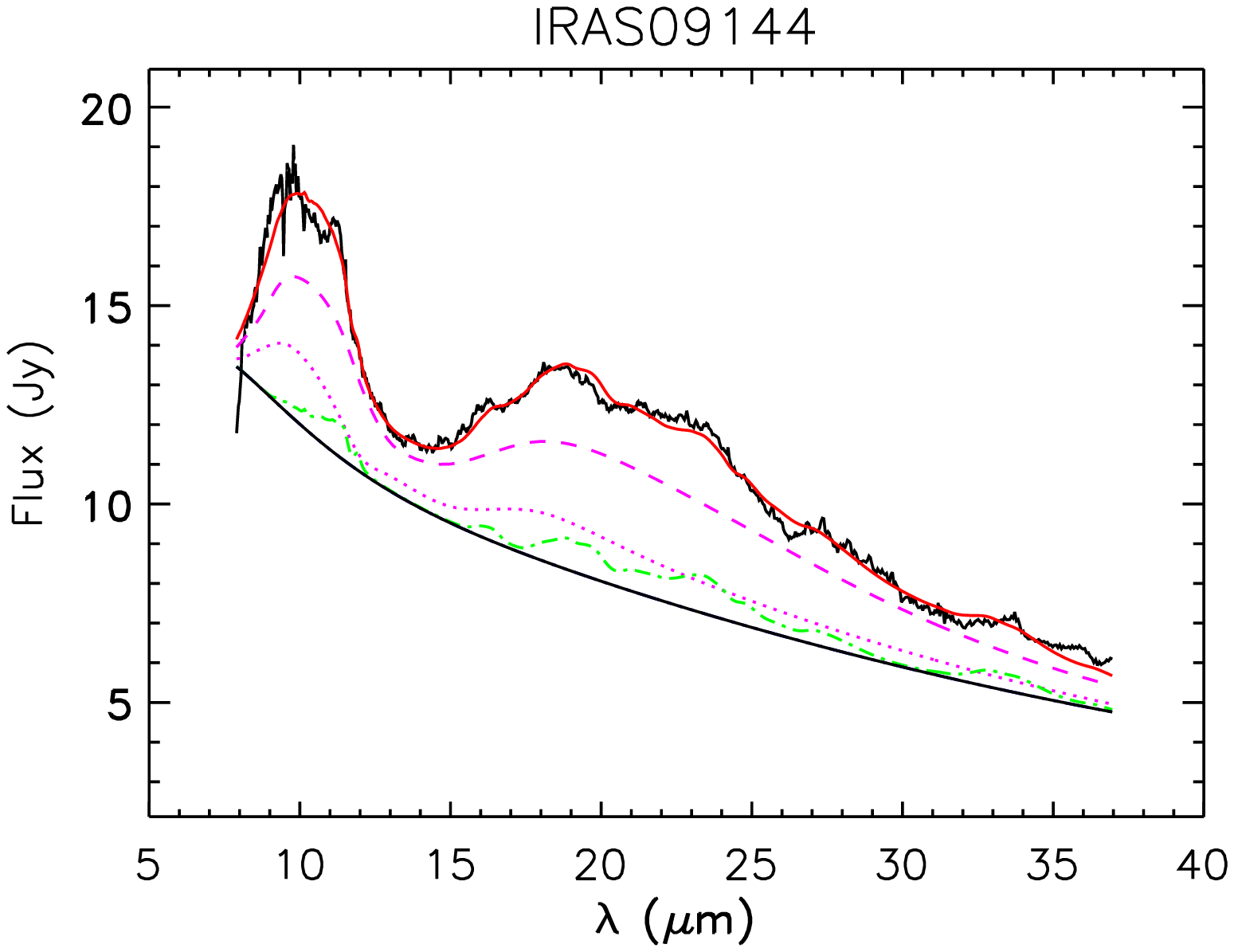}}
\resizebox{9cm}{!}{\includegraphics{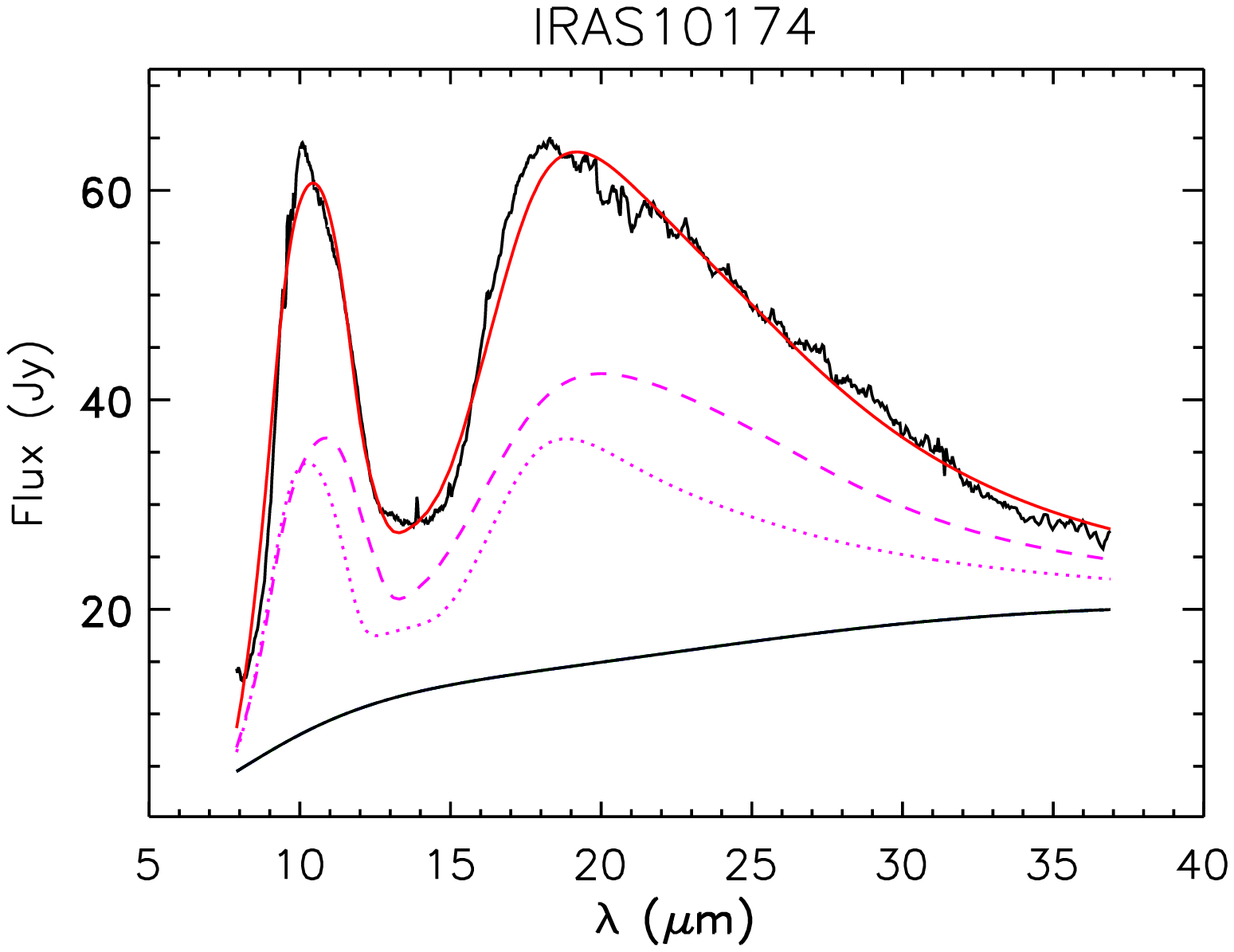}}
\resizebox{9cm}{!}{\includegraphics{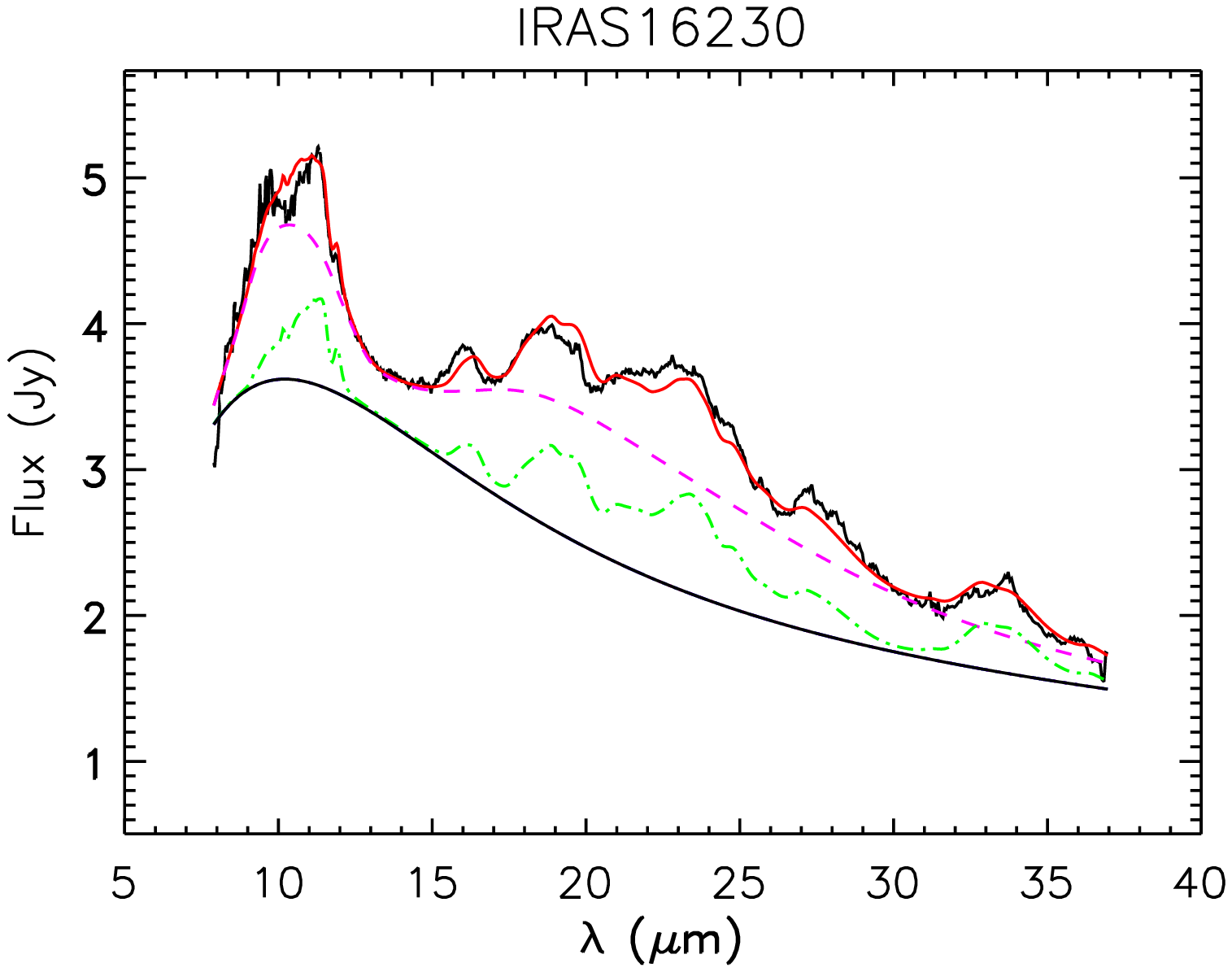}}
\resizebox{9cm}{!}{\includegraphics{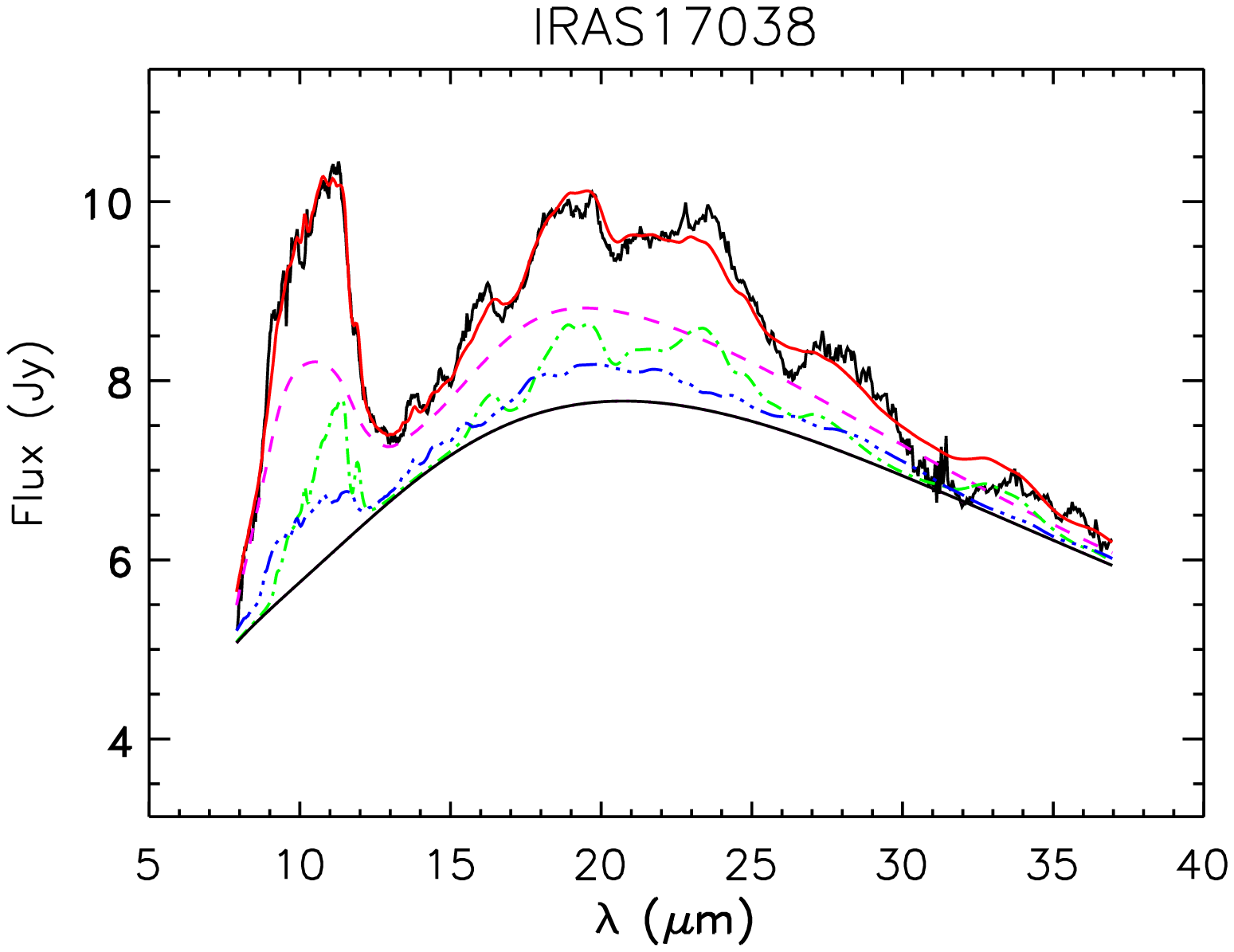}}
\resizebox{9cm}{!}{\includegraphics{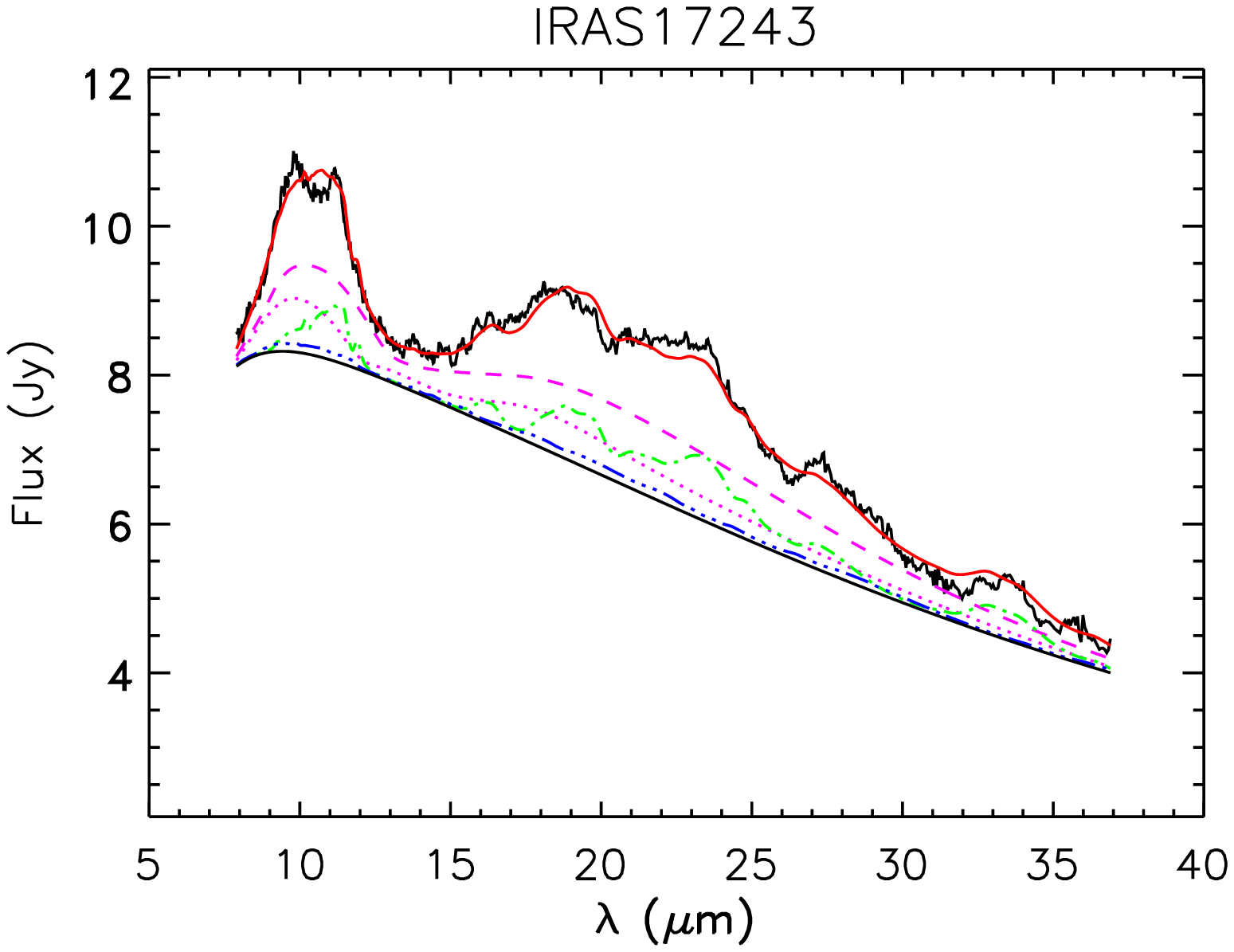}}
\resizebox{9cm}{!}{\includegraphics{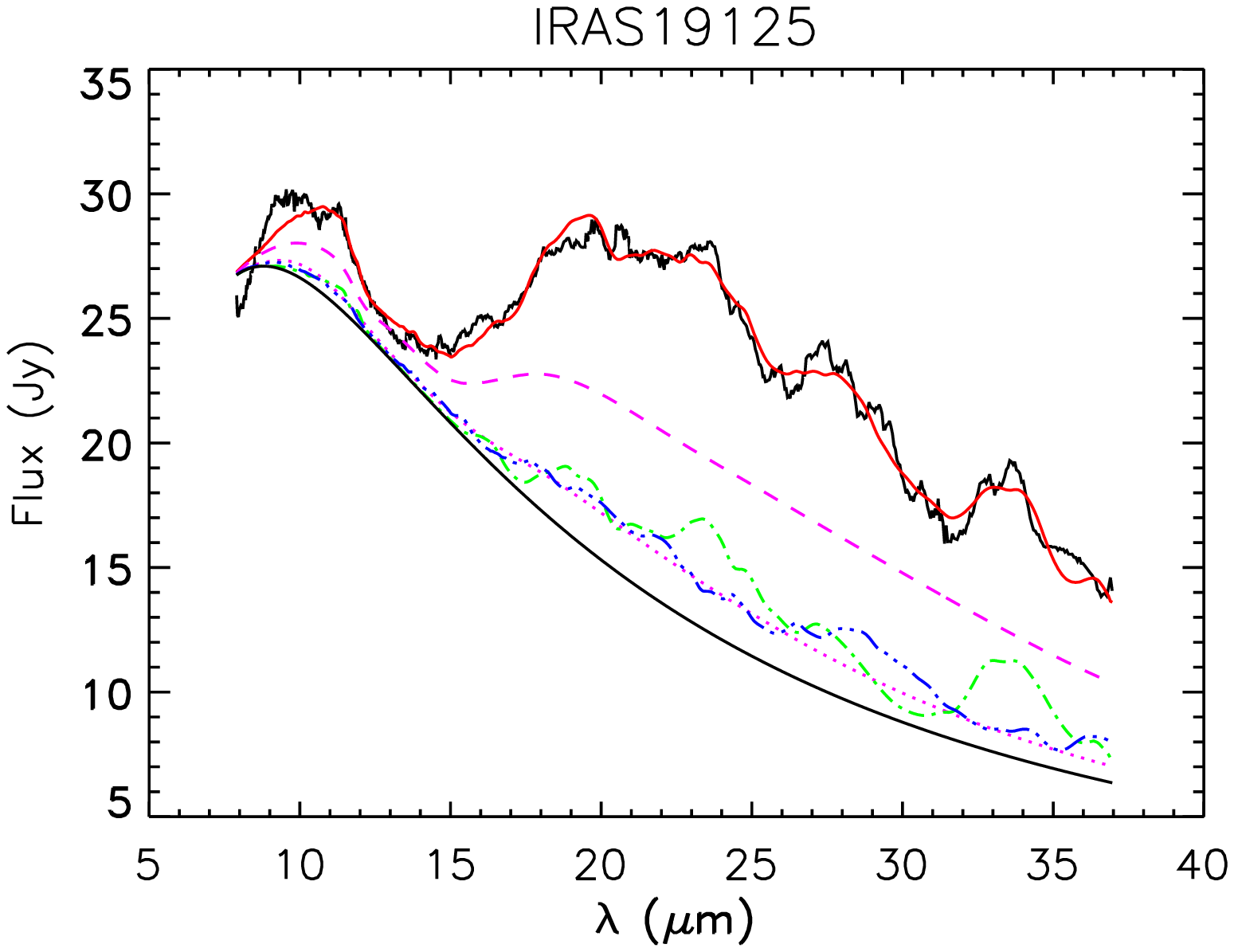}}
\caption{See previous caption.}
\label{fitting2}
\end{figure}
\begin{figure}
\vspace{0cm}
\hspace{0cm}
\resizebox{9cm}{!}{\includegraphics{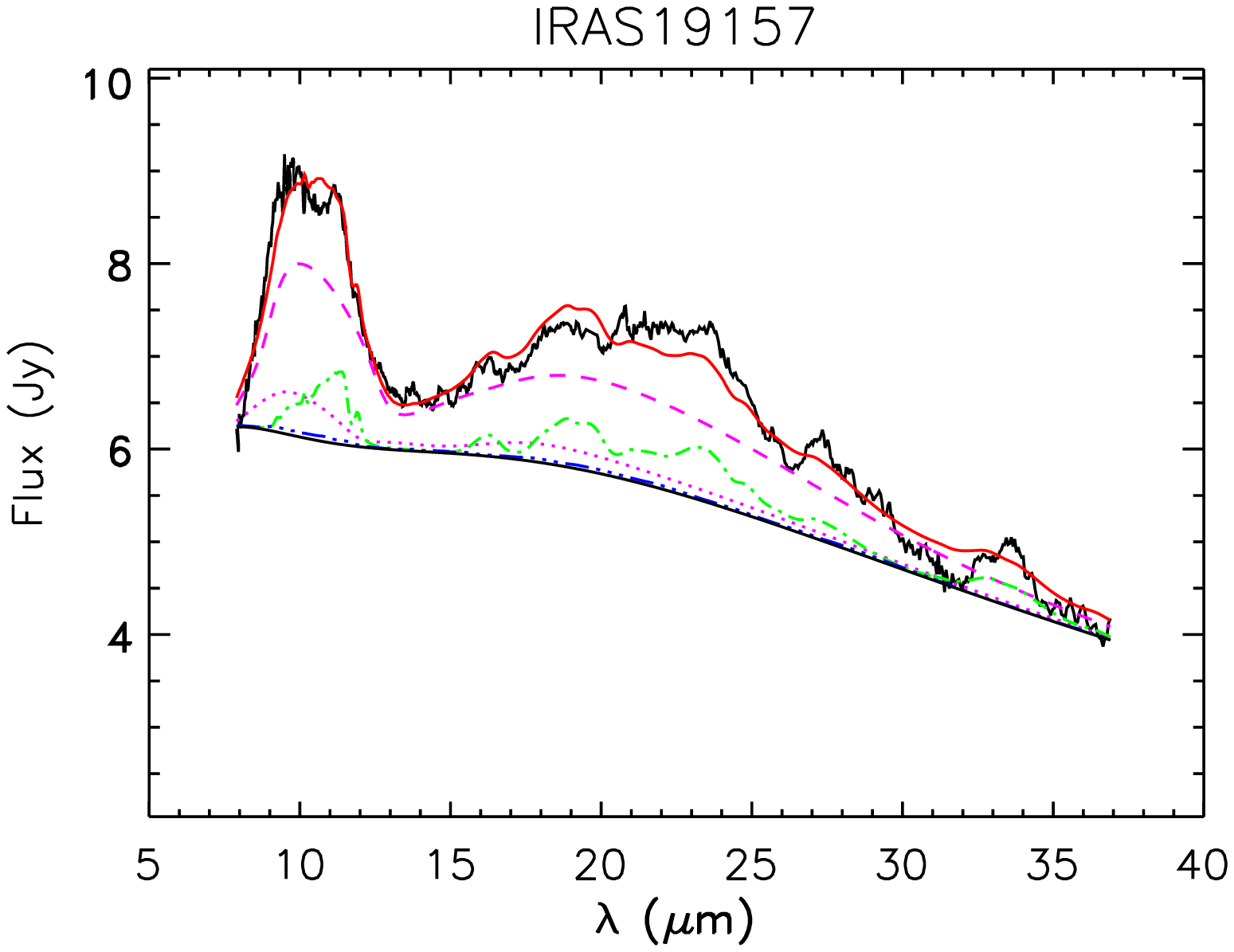}}
\resizebox{9cm}{!}{\includegraphics{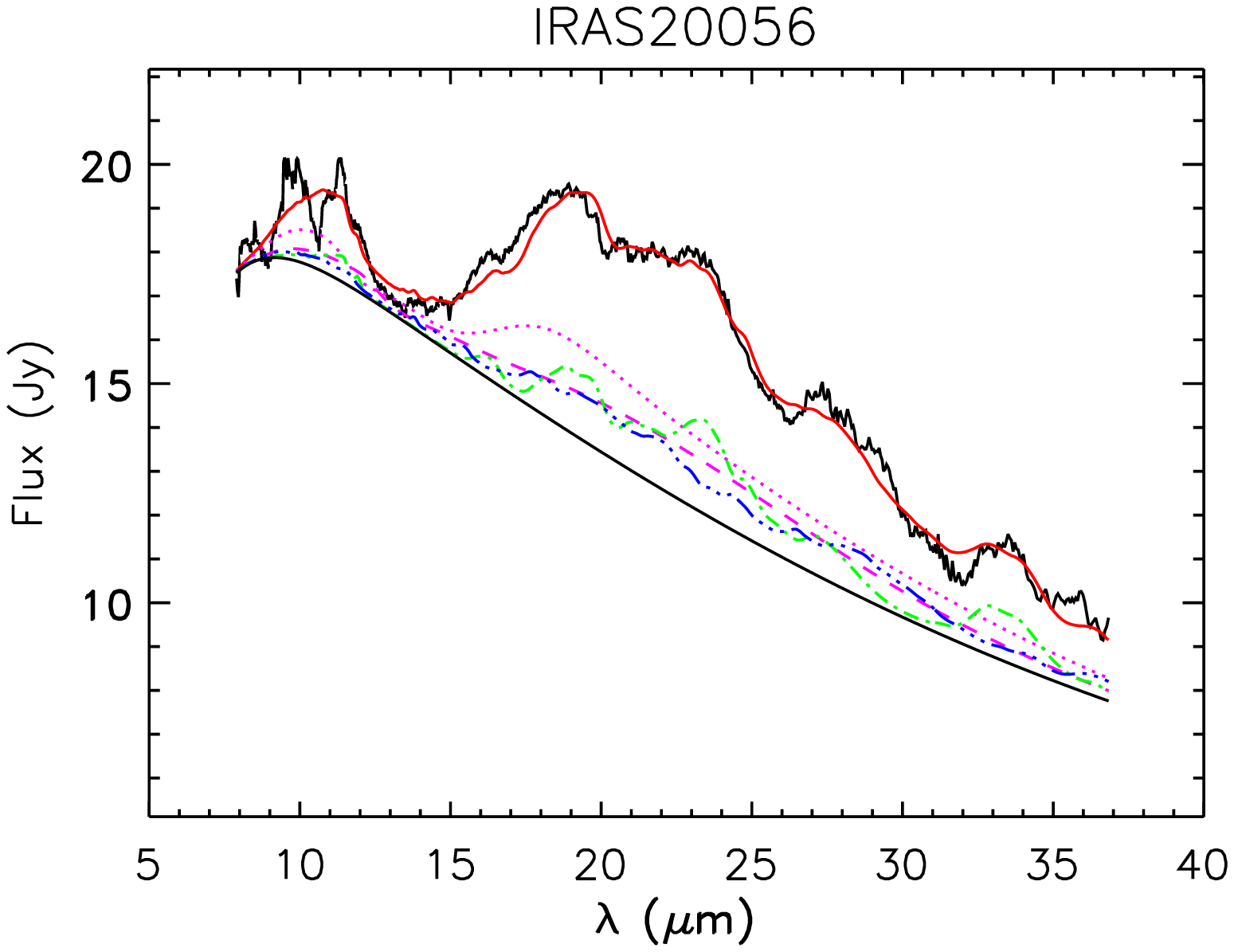}}
\resizebox{9cm}{!}{\includegraphics{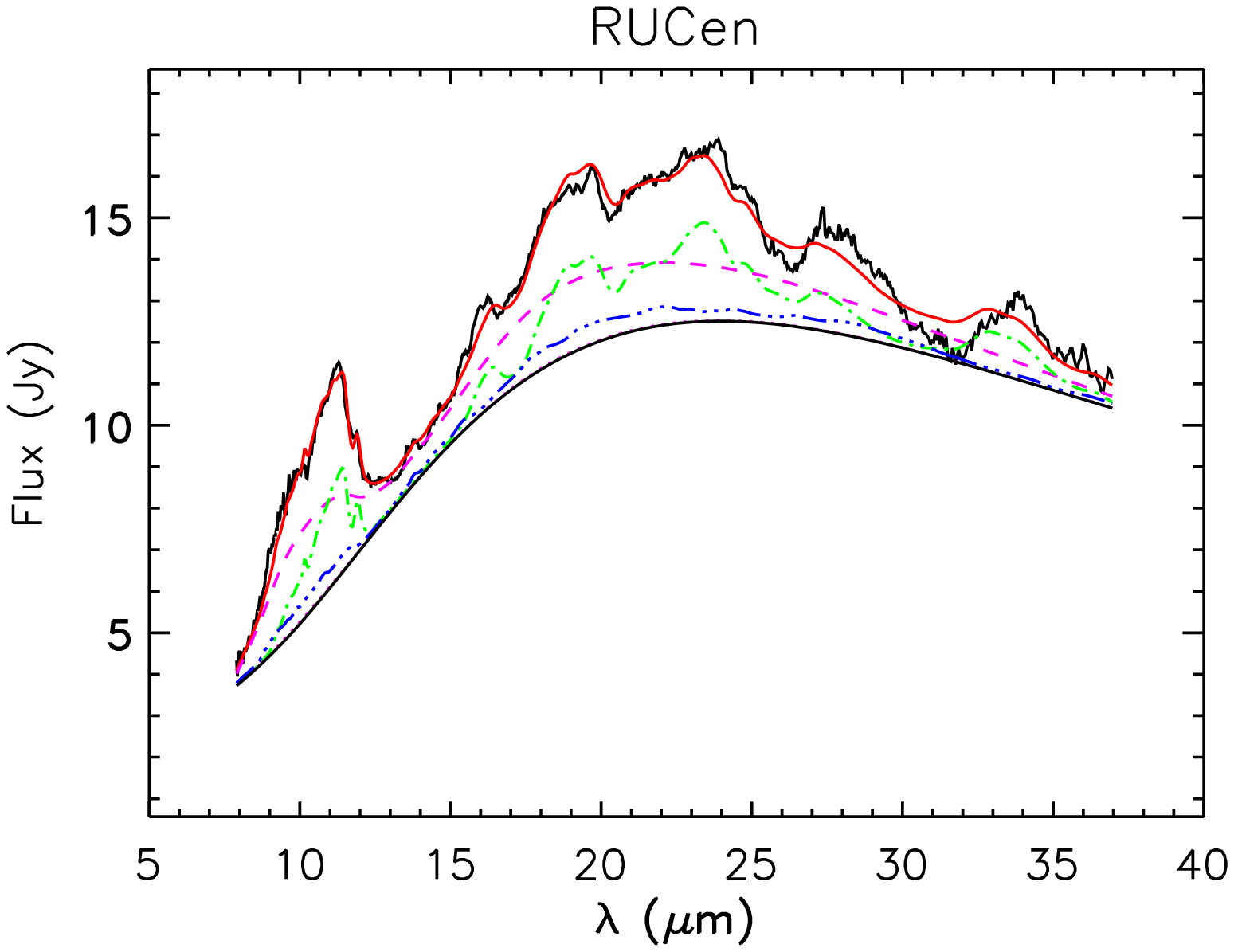}}
\resizebox{9cm}{!}{\includegraphics{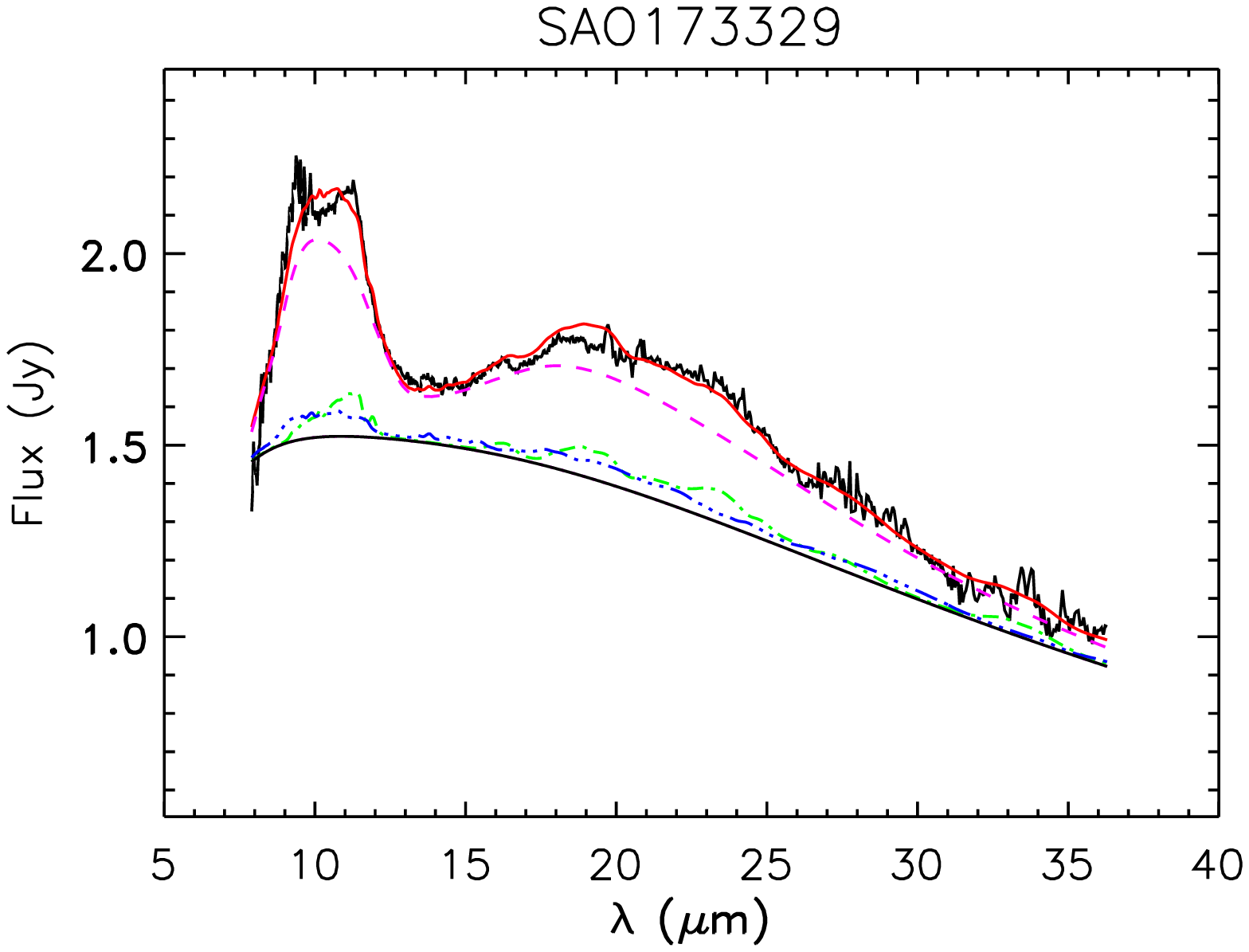}}
\resizebox{9cm}{!}{\includegraphics{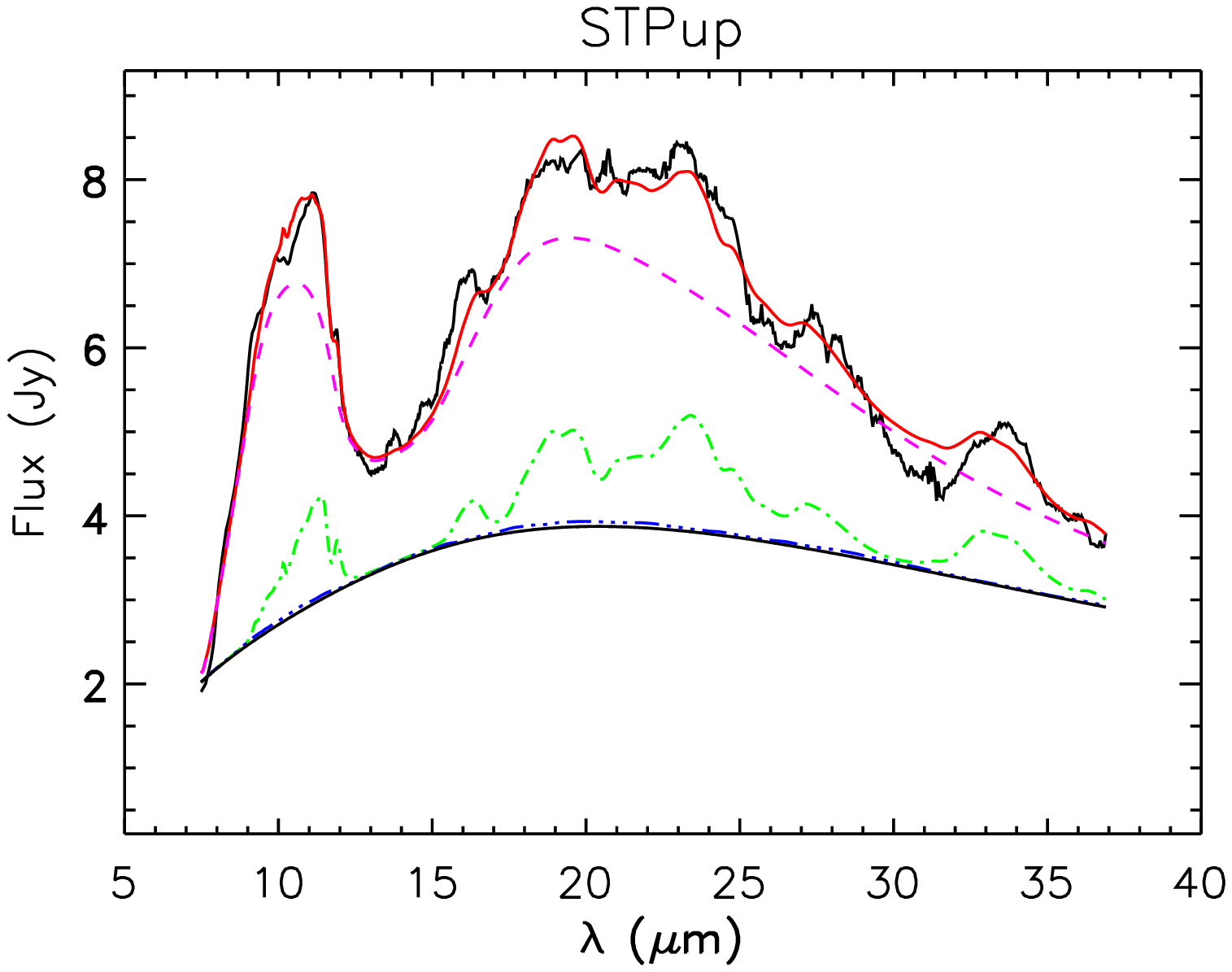}}
\resizebox{9cm}{!}{\includegraphics{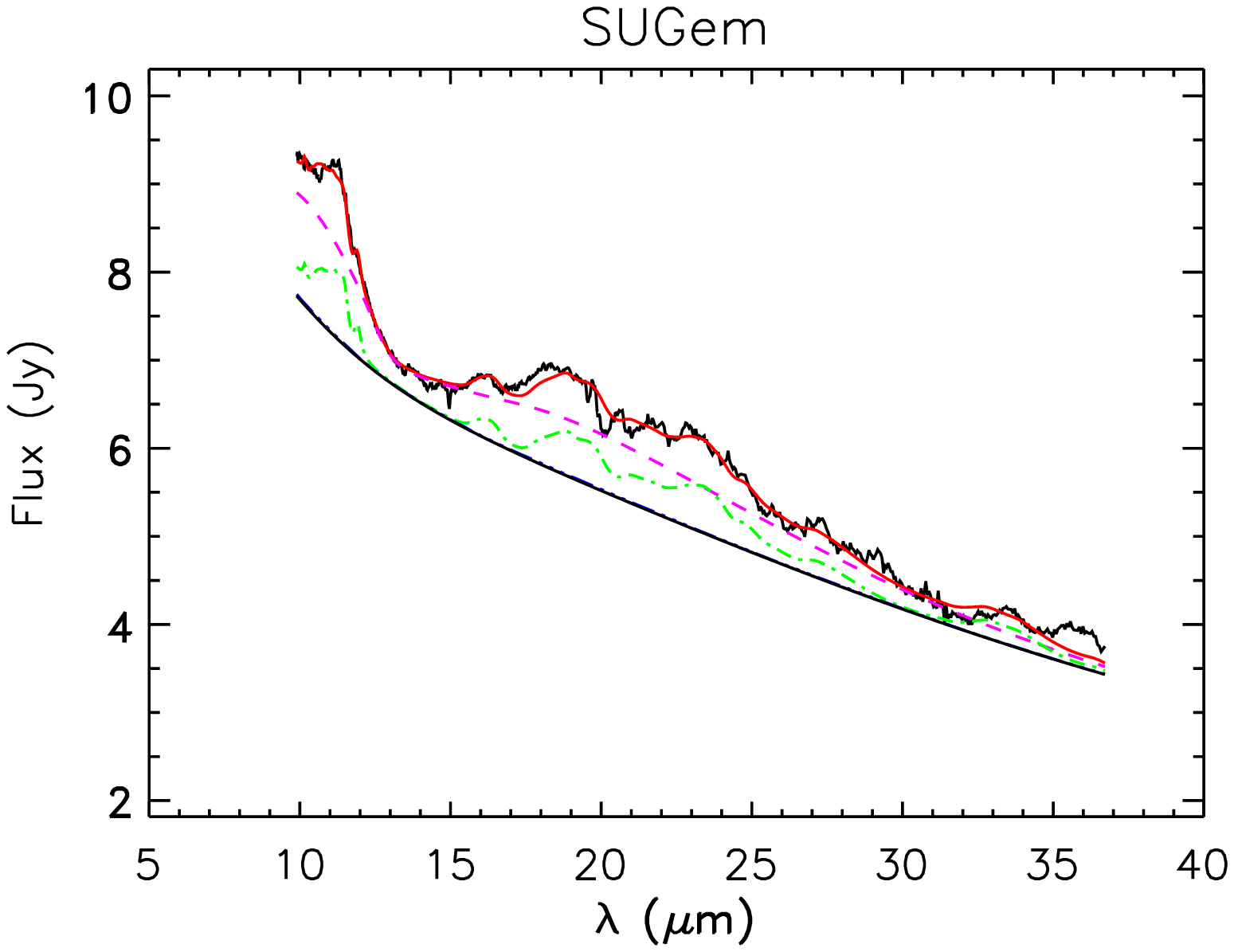}}
\caption{See previous caption.}
\label{fitting3}
\end{figure}
\begin{figure}
\vspace{0cm}
\hspace{0cm}
\resizebox{9cm}{!}{\includegraphics{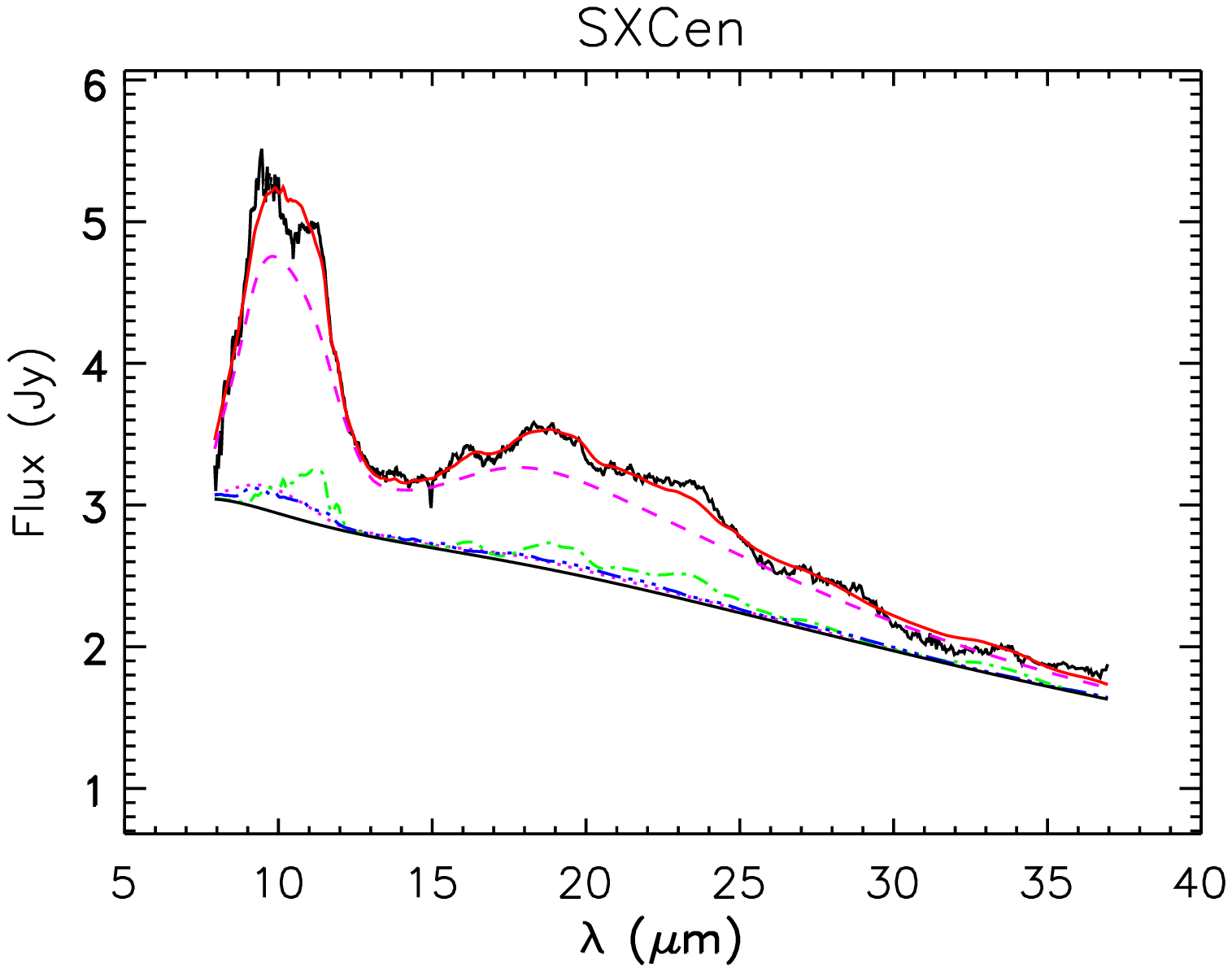}}
\resizebox{9cm}{!}{\includegraphics{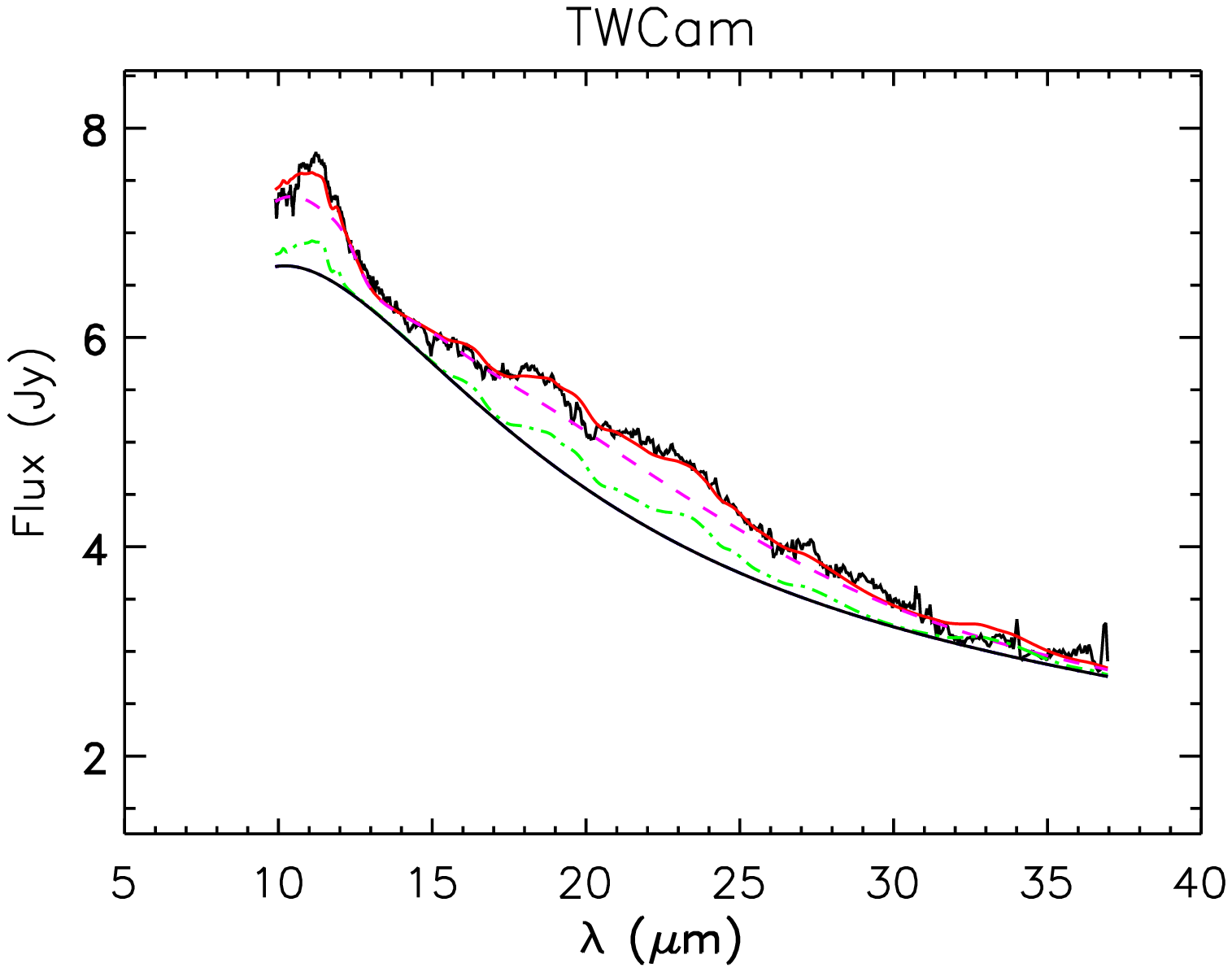}}
\resizebox{9cm}{!}{\includegraphics{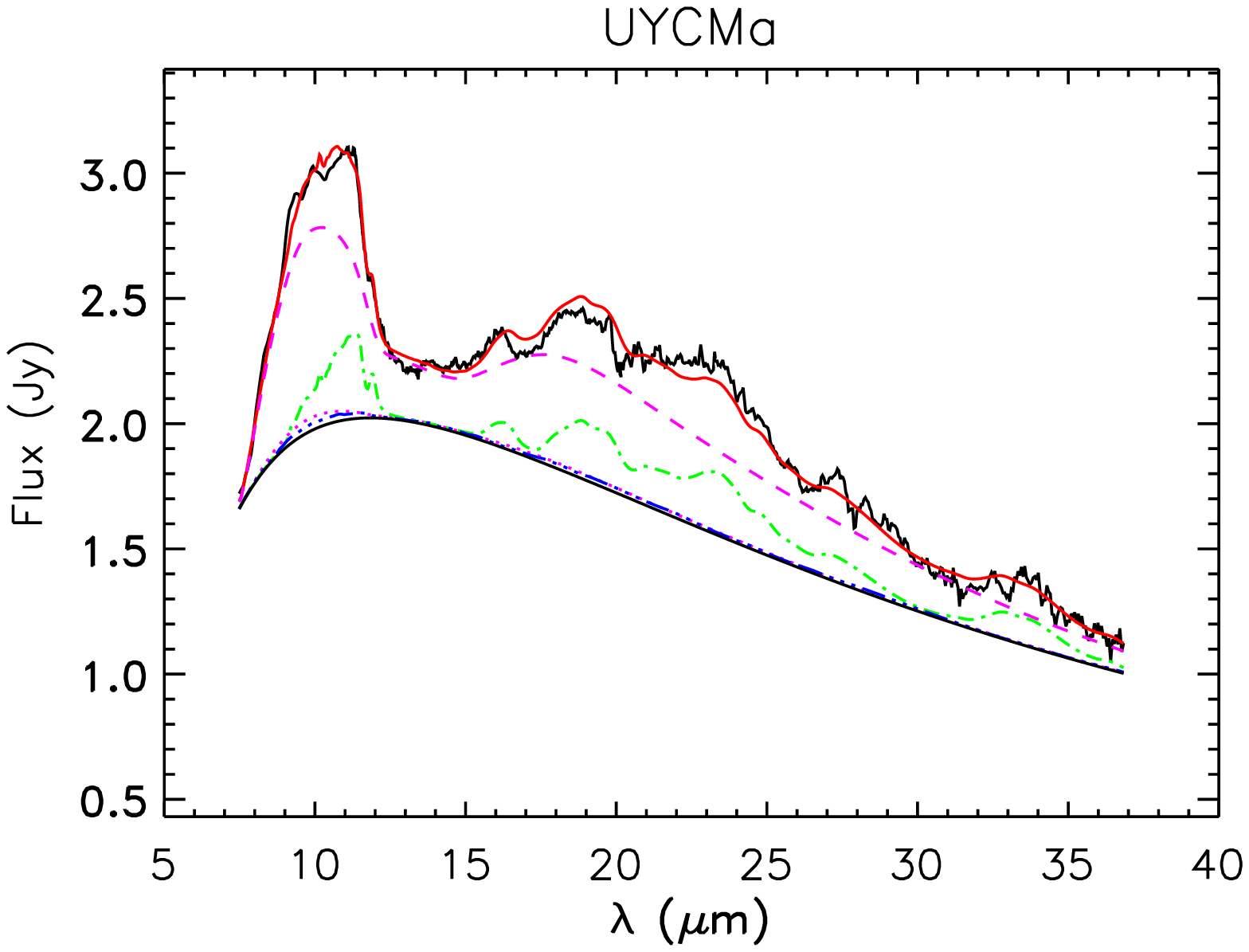}}
\caption{See previous caption.}
\label{fitting4}
\end{figure}

\end{appendix}
\end{document}